\documentclass[a4paper,11pt]{article}
\pdfoutput=1 

\usepackage{jcappub} 

\usepackage[T1]{fontenc} 
\usepackage{orcidlink}
\usepackage{newtxtext,newtxmath}
\usepackage{amsmath}
\usepackage{physics}
\usepackage{cleveref}
\usepackage{comment}
\usepackage{multicol,multirow}
\usepackage{lineno}
\usepackage{aas_macros}

\title{Impact and mitigation of spectroscopic systematics on DESI DR1 clustering measurements}

\author[1,2,3]{{A.~Krolewski},}
\author[4]{{J.~Yu}\orcidlink{0009-0001-7217-8006},}
\author[5,6,7]{{A.~J.~Ross}\orcidlink{0000-0002-7522-9083},}
\author[1]{{S.~Penmetsa},}
\author[1,2,3]{{W.~J.~Percival}\orcidlink{0000-0002-0644-5727},}
\author[8]{{R.~Zhou}\orcidlink{0000-0001-5381-4372},}
\author[9]{{M.~J.~Wilson},}
\author[10]{{J.~Hou},}
\author[8]{{J.~Aguilar},}
\author[11]{{S.~Ahlen}\orcidlink{0000-0001-6098-7247},}
\author[12]{{D.~Brooks},}
\author[8]{{E.~Chaussidon}\orcidlink{0000-0001-8996-4874},}
\author[8]{{T.~Claybaugh},}
\author[13]{{A.~de la Macorra}\orcidlink{0000-0002-1769-1640},}
\author[14]{{Biprateep~Dey}\orcidlink{0000-0002-5665-7912},}
\author[15,16]{{J.~E.~Forero-Romero}\orcidlink{0000-0002-2890-3725},}
\author[8]{{S.~Gontcho A Gontcho}\orcidlink{0000-0003-3142-233X},}
\author[8]{{J.~Guy}\orcidlink{0000-0001-9822-6793},}
\author[5,17,7]{{K.~Honscheid},}
\author[18]{{S.~Juneau},}
\author[19]{{D.~Kirkby}\orcidlink{0000-0002-8828-5463},}
\author[8]{{T.~Kisner}\orcidlink{0000-0003-3510-7134},}
\author[8]{{A.~Kremin}\orcidlink{0000-0001-6356-7424},}
\author[8]{{A.~Lambert},}
\author[20]{{L.~Le~Guillou}\orcidlink{0000-0001-7178-8868},}
\author[8]{{M.~E.~Levi}\orcidlink{0000-0003-1887-1018},}
\author[5,6,7]{{P.~Martini}\orcidlink{0000-0002-4279-4182},}
\author[18]{{A.~Meisner}\orcidlink{0000-0002-1125-7384},}
\author[21,22]{{R.~Miquel},}
\author[23]{{J.~Moustakas}\orcidlink{0000-0002-2733-4559},}
\author[24]{{A.~D.~Myers},}
\author[13]{{J.~A.~Newman}\orcidlink{0000-0001-8684-2222},}
\author[25,26]{{G.~Niz}\orcidlink{0000-0002-1544-8946},}
\author[27,8]{{N.~Palanque-Delabrouille}\orcidlink{0000-0003-3188-784X},}
\author[28]{{G.~Rossi},}
\author[29]{{E.~Sanchez}\orcidlink{0000-0002-9646-8198},}
\author[30]{{E.~F.~Schlafly}\orcidlink{0000-0002-3569-7421},}
\author[8]{{D.~Schlegel},}
\author[31,32]{{M.~Schubnell},}
\author[33]{{H.~Seo}\orcidlink{0000-0002-6588-3508},}
\author[18]{{D.~Sprayberry},}
\author[32]{{G.~Tarl\'{e}}\orcidlink{0000-0003-1704-0781},}
\author[18]{{B.~A.~Weaver},}
\author[34]{{C.~Zhao}\orcidlink{0000-0002-1991-7295}}

\affiliation[1]{Department of Physics and Astronomy, University of Waterloo, 200 University Ave W, Waterloo, ON N2L 3G1, Canada}
\affiliation[2]{Perimeter Institute for Theoretical Physics, 31 Caroline St. North, Waterloo, ON N2L 2Y5, Canada}
\affiliation[3]{Waterloo Centre for Astrophysics, University of Waterloo, 200 University Ave W, Waterloo, ON N2L 3G1, Canada}
\affiliation[4]{Institute of Physics, Laboratory of Astrophysics, \'Ecole Polytechnique F\'ed\'erale de Lausanne (EPFL), Observatoire de Sauverny, CH-1290 Versoix, Switzerland}
\affiliation[5]{Center for Cosmology and AstroParticle Physics, The Ohio State University, 191 West Woodruff Avenue, Columbus, OH 43210, USA}
\affiliation[6]{Department of Astronomy, The Ohio State University, 4055 McPherson Laboratory, 140 W 18th Avenue, Columbus, OH 43210, USA}
\affiliation[7]{The Ohio State University, Columbus, 43210 OH, USA}
\affiliation[8]{Lawrence Berkeley National Laboratory, 1 Cyclotron Road, Berkeley, CA 94720, USA}
\affiliation[9]{Institute for Computational Cosmology, Department of Physics, Durham University, South
Road, Durham DH1 3LE, UK}
\affiliation[10]{Department of Astronomy, University of Florida, 211 Bryant Space Science Center, Gainesville, FL 32611, USA}
\affiliation[11]{Physics Dept., Boston University, 590 Commonwealth Avenue, Boston, MA 02215, USA}
\affiliation[12]{Department of Physics \& Astronomy, University College London, Gower Street, London, WC1E 6BT, UK}
\affiliation[13]{Instituto de F\'{\i}sica, Universidad Nacional Aut\'{o}noma de M\'{e}xico,  Cd. de M\'{e}xico  C.P. 04510,  M\'{e}xico}
\affiliation[14]{Department of Physics \& Astronomy and Pittsburgh Particle Physics, Astrophysics, and Cosmology Center (PITT PACC), University of Pittsburgh, 3941 O'Hara Street, Pittsburgh, PA 15260, USA}
\affiliation[15]{Departamento de F\'isica, Universidad de los Andes, Cra. 1 No. 18A-10, Edificio Ip, CP 111711, Bogot\'a, Colombia}
\affiliation[16]{Observatorio Astron\'omico, Universidad de los Andes, Cra. 1 No. 18A-10, Edificio H, CP 111711 Bogot\'a, Colombia}
\affiliation[17]{Department of Physics, The Ohio State University, 191 West Woodruff Avenue, Columbus, OH 43210, USA}
\affiliation[18]{NSF NOIRLab, 950 N. Cherry Ave., Tucson, AZ 85719, USA}
\affiliation[19]{Department of Physics and Astronomy, University of California, Irvine, 92697, USA}
\affiliation[20]{Sorbonne Universit\'{e}, CNRS/IN2P3, Laboratoire de Physique Nucl\'{e}aire et de Hautes Energies (LPNHE), FR-75005 Paris, France}
\affiliation[21]{Instituci\'{o} Catalana de Recerca i Estudis Avan\c{c}ats, Passeig de Llu\'{\i}s Companys, 23, 08010 Barcelona, Spain}
\affiliation[22]{Institut de F\'{i}sica d’Altes Energies (IFAE), The Barcelona Institute of Science and Technology, Campus UAB, 08193 Bellaterra Barcelona, Spain}
\affiliation[23]{Department of Physics and Astronomy, Siena College, 515 Loudon Road, Loudonville, NY 12211, USA}
\affiliation[24]{Department of Physics \& Astronomy, University  of Wyoming, 1000 E. University, Dept.~3905, Laramie, WY 82071, USA}
\affiliation[25]{Departamento de F\'{i}sica, Universidad de Guanajuato - DCI, C.P. 37150, Leon, Guanajuato, M\'{e}xico}
\affiliation[26]{Instituto Avanzado de Cosmolog\'{\i}a A.~C., San Marcos 11 - Atenas 202. Magdalena Contreras, 10720. Ciudad de M\'{e}xico, M\'{e}xico}
\affiliation[27]{IRFU, CEA, Universit\'{e} Paris-Saclay, F-91191 Gif-sur-Yvette, France}
\affiliation[28]{Department of Physics and Astronomy, Sejong University, Seoul, 143-747, Korea}
\affiliation[29]{CIEMAT, Avenida Complutense 40, E-28040 Madrid, Spain}
\affiliation[30]{Space Telescope Science Institute, 3700 San Martin Drive, Baltimore, MD 21218, USA}
\affiliation[31]{Department of Physics, University of Michigan, Ann Arbor, MI 48109, USA}
\affiliation[32]{University of Michigan, Ann Arbor, MI 48109, USA}
\affiliation[33]{Department of Physics \& Astronomy, Ohio University, Athens, OH 45701, USA}
\affiliation[34]{Department of Astronomy, Tsinghua University, 30 Shuangqing Road, Haidian District, Beijing, China, 100190}

\emailAdd{akrolews@uwaterloo.ca}

\abstract{
The large scale structure catalogs within DESI Data Release 1 (DR1) use nearly 6 million galaxies and quasars as tracers of the large-scale structure of the universe to measure the expansion history with baryon acoustic oscillations and the growth of structure with redshift-space distortions.
In order to take advantage of DESI's unprecedented statistical power, we must ensure that the galaxy clustering measurements are unaffected by non-cosmological density fluctuations.
One source of spurious fluctuations comes from
variation in galaxy density with spectroscopic observing conditions, lowering the redshift efficiency (and thus galaxy density) in certain areas of the sky.
We measure the uniformity of the redshift success
rate for DESI luminous red galaxies (LRG), bright galaxies (BGS) and quasars (QSO), complementing the detailed discussion of emission line galaxy (ELG) systematics in a companion paper \cite{KP3s4-Yu}.
We find small but significant fluctuations of up to 3\% in redshift success rate with the effective spectroscopic signal-to-noise, and create and describe weights that remove these fluctuations.
We also describe the process to identify and remove data from certain poorly performing fibers from DESI DR1, and measure the stability of the redshift success rate with time.
Finally, we find small but significant correlations
of redshift success rate with position on the focal plane, survey speed, and number of exposures required, and show the impact of weights correcting these trends on the power spectrum multipoles and on cosmological parameters from BAO and RSD fits.
These corrections change the best-fit parameters by $<15\%$ of their statistical errors, and thus contribute negligibly to the overall DESI error budget.
}

\begin{document}
\maketitle
\flushbottom






\section{Introduction}

The Dark Energy Spectroscopic Instrument \cite{Levi2013,DESI2016a.Science,DESI2016b.Instr,DESI2023b.KP1.EDR,DESI2023a.KP1.SV} will obtain 40 million extragalactic spectra over its 5-year survey from 2021--2026, an order-of-magnitude increase in cosmological constraining power and sample size over previous samples from the Sloan Digital Sky Survey (SDSS).
DESI has 5000 robotically-controlled fiber positioners \cite{fiberoverview_claire} across a 7 deg$^2$ focal plane \cite{Silber23} at the 4-meter Mayall Telescope at Kitt Peak, Arizona, US \cite{DESI2022instrument}, with a wide field-of-view enabled by a new optical corrector \cite{Corrector.Miller.2023}.
DESI's goal is to obtain the tightest constraints to date on dark energy via
baryon acoustic oscillations and redshift-space distortions, in both cases
using galaxies and quasars as tracers of the dark matter field.
In its main survey, DESi is mapping the Universe using five tracers at $0.1 < z < 4.2$. During bright telescope time, DESI observes the Bright Galaxy Sample (BGS), which is used for cosmology at $0.1 < z < 0.4$ \cite{BGS.TS.Hahn.2023} (though the BGS targets span a slightly wider redshift range). During dark time, DESI observes the Luminous Red Galaxies (LRG) at $0.4 < z < 1.1$ \cite{LRG.TS.Zhou.2023}, Emission Line Galaxies (ELG) at $0.8 < z < 1.6$ \cite{ELG.TS.Raichoor.2023}, quasars (QSO) at $0.8 < z < 2.1$ \cite{QSO.TS.Chaussidon.2023}\footnote{The BAO and RSD analyses only use quasars up to $z=2.1$, but higher-redshift quasars are used in cross-correlation with the Ly$\alpha$ forest and to constrain primordial non-Gaussianity, up to $z=3.5$ \cite{DESI2024.VIII.KP7C}.}, and the Ly$\alpha$ forest at $z > 1.8$ \cite{DESI2024.IV.KP6}.

The first year of DESI observations will be made available as DESI DR1 \cite{DESI2024.I.DR1}.
This sample already includes 6 million galaxies used in the cosmological analysis (and another 4 million fainter BGS that were not used, since BGS is far denser than necessary to achieve a cosmic-variance limited measurement) \cite{DESI2024.II.KP3}.
The BAO measurements were presented in \cite{DESI2024.III.KP4,DESI2024.IV.KP6} and their cosmological interpretation given in \cite{DESI2024.VI.KP7A}.
The RSD measurements will come soon in \cite{DESI2024.V.KP5}, as well as their cosmological interpretation \cite{DESI2024.VII.KP7B} and primordial non-Gaussianity constraints \cite{DESI2024.VIII.KP7C}.

Making these percent-level measurements requires that the fluctuations in the galaxy field are cosmological rather than driven by systematics. The largest fluctuations are driven by inhomogeneities in the imaging used for targeting, and such mitigations are extensively described in \cite{DESI2024.II.KP3,KP3s2-Rosado,KP3s13-Kong,KP3s10-Chaussidon,KP3s14-Zhou}.

\begin{figure*}
    \includegraphics[width=0.7\linewidth]{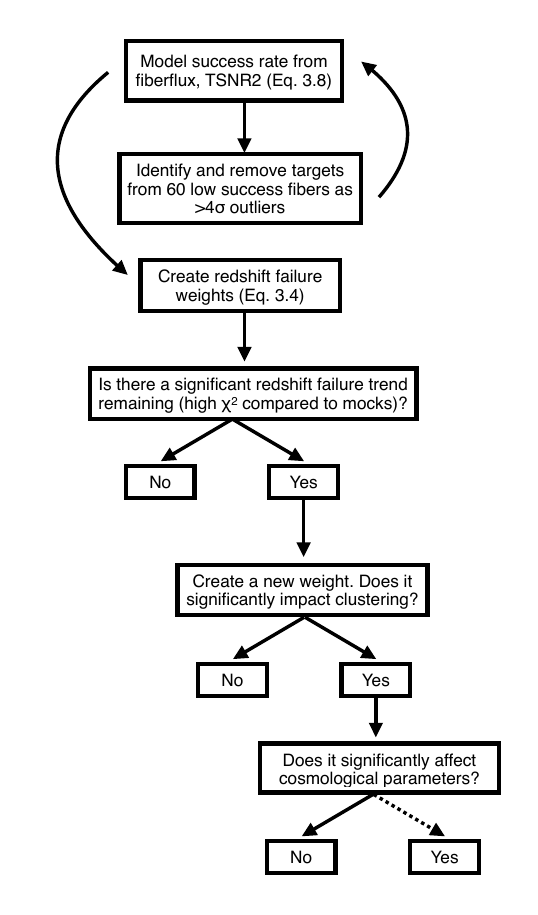}
    \caption{Flowchart describing the layout of the paper. A preliminary version of the redshift success model must be fit to the data to define the low success rate fibers; targets from these fibers are then removed from the final data and the fit is re-run. The flow-chart ends once each test passes. The impact on cosmological parameters is negligible for all tests; hence we show the final line as dashed to indicate that all tests pass and we never reach the final ``yes'' box.
 \label{fig:spec_syst_flowchart}}
\end{figure*}

This paper and the companion paper \cite{KP3s4-Yu} characterize the impact of fluctuations in spectroscopic observing conditions on large-scale structure measurements. Specifically, we aim to ensure that the effective spectroscopic signal-to-noise (or observing time) is as uniform as possible across all targets, and that its variations do not cause variations in redshift success rate and thus galaxy number density. 
This paper presents an overview of spectroscopic systematics for all tracers, while \cite{KP3s4-Yu} focuses specifically on ELG. ELG are both the largest sample in DESI and challenging to redshift accurately, owing to their faint fluxes and reliance on the narrow [OII] emission line, which at high redshift can be confused with the abundant skylines at the red end of the DESI wavelength range. Number density fluctuations driven by spectroscopic conditions are typically quite small (roughly 1\%, compared to the 
roughly 10\% fluctuations from imaging systematics), but can be detected at high significance due to the large sample sizes in DESI. Obtaining a measurement of the density field dominated by cosmological modes rather than systematics is of paramount importance for DESI's large-scale structure mission.

We focus on the discrete tracers (galaxies and quasars) since the Ly$\alpha$ forest has very different systematics.
Nevertheless, redshift errors and spectroscopic uniformity are also quite important for the Ly$\alpha$ analysis, and the impact of these systematics are described in \cite{KP6s4-Bault,KP6s5-Guy} (see also \cite{Napolitano24}).

We provide a flow chart describing the methods and logic of this paper in Fig.~\ref{fig:spec_syst_flowchart}. We describe the DR1 DESI data processing in Sec.~\ref{sec:data}, as well as the mocks used and the clustering measurements made.
We describe the model for redshift
failures and their corresponding weight, $w_{\textrm{zfail}}$, in Section~\ref{sec:zfail_wts}. Then we describe
the process to remove targets that were observed with low success-rate fibers in Sec~\ref{sec:badfibers}.
Finally, we validate the redshift failure weights in Section~\ref{sec:validate_zfail_wts}, showing that all trends of effective spectroscopic signal-to-noise with hardware and observational variables are greatly reduced and most are no longer significant after applying $w_{\textrm{zfail}}$.
For those trends that are significant,
we measure their impact on galaxy and quasar clustering in Section~\ref{sec:clustering}, and propagate their impact to cosmological parameters for those that change the observed power spectrum noticeably. We find $<20\%$ shifts
in parameters inferred from BAO and RSD fits. We conclude in Section~\ref{sec:conclusions}.

We use the DESI fiducial cosmology in this work where appropriate, a flat $\Lambda$CDM fiducial cosmology from the mean results of Planck \cite{Planck2018} \textsc{base\_plikHM\_TTTEEE\_lowl\_lowE\_lensing}  with $\omega_{\mathrm{b}} = 0.02237, \; \omega_{\mathrm{cdm}} = 0.12, \; h = 0.6736, \;  A_{\mathrm{s}} = 2.083 \cdot 10^{-9}, \; n_{\mathrm{s}} = 0.9649, \; \sigma_8=0.8079, \; N_{\mathrm{eff}} = 3.044, \; \sum m_{\nu} = 0.06 \; \mathrm{eV}$ (single massive neutrino). This cosmology is used to convert galaxy positions to Cartesian coordinates for computing two point clustering statistics.

\section{DESI Data}
\label{sec:data}

\begin{figure*}
\includegraphics[width=\textwidth]{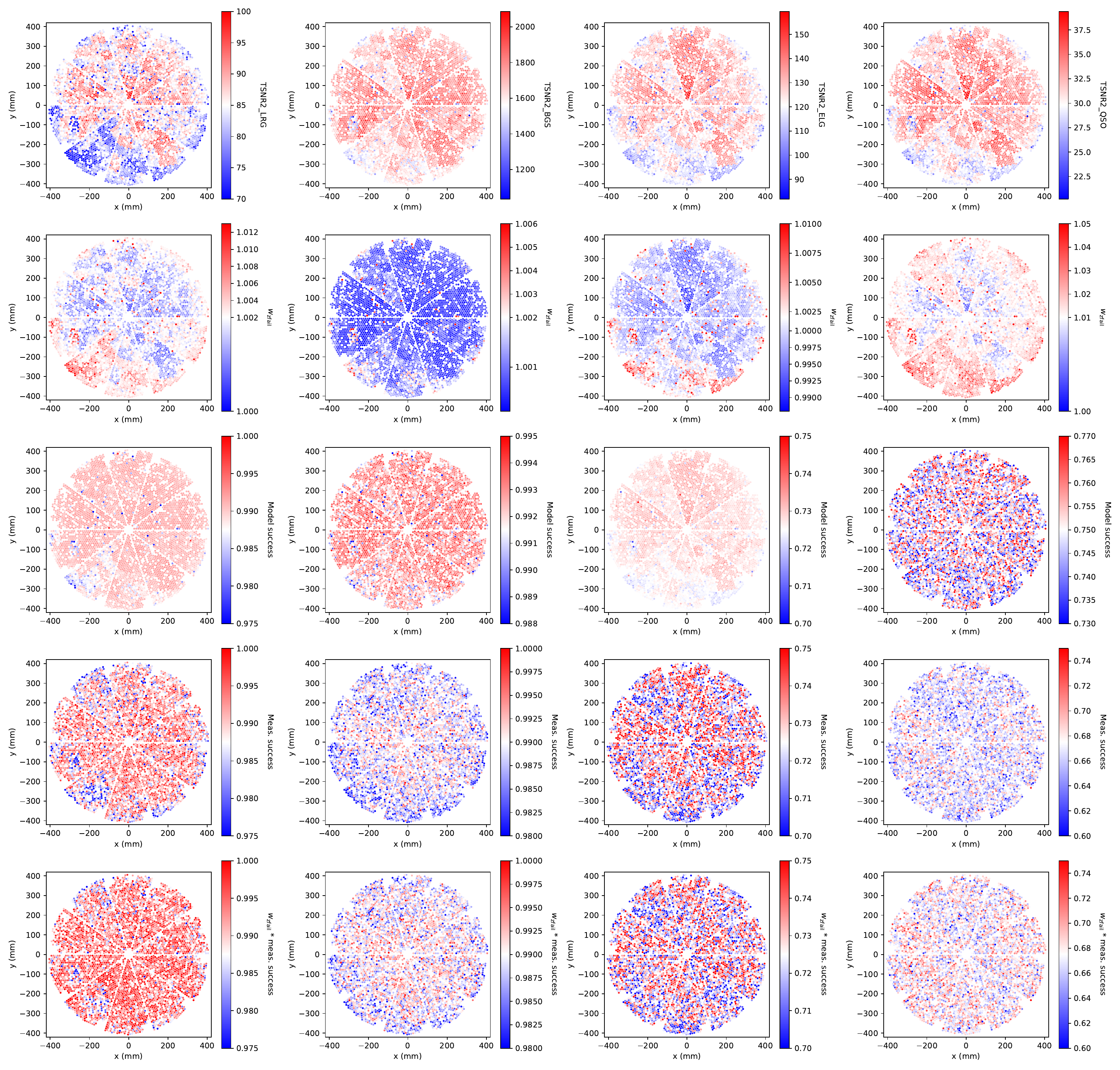}

\caption{Variation of TSNR2 (\textit{top}), modelled success rate (\textit{second row}), measured success rate (\textit{third row}), redshift failure weight (\textit{fourth row}), and measured success rate
weighted by the redshift failure weight (\textit{bottom row}).
The model does not necessarily match the normalization of the measured redshift success rate, especially for QSO, because it is designed to remove fluctuations rather than match the normalization. Comparison of the third row and the bottom row shows the model's performance: the $w_{\textrm{zfail}}$-weighted success rate is more uniform than the raw measured success rate.}
\label{fig:big_focal_plane}
\end{figure*}

\subsection{Observing strategy and catalog construction}
\label{sec:catalog_construction}

DESI DR1 contains data from the first year of the DESI Main Survey observed from May 2021 to June 2022 and covers over 7,400 deg$^2$ \cite{DESI2024.I.DR1,DESI2024.II.KP3}. DESI
targets are selected from the DESI Legacy Survey Imaging, consisting of the combination of the Beijing-Arizona Sky Survey (BASS) and the Mayall $z$-band Legacy Survey (MzLS) at $\delta > 32.375^\circ$ and the Dark Energy Camera Legacy Survey (DECaLS) at $\delta < 32.375^\circ$
\cite{BASS.Zou.2017,LS.Overview.Dey.2019,LS.dr9.Schegel.2024}.
Due to the differing image qualities of the surveys, a common systematic test is to split the data by region into BASS/MzLS (North) and DECaLS (South).
Targets are selected via multi-band color cuts, supplemented by infrared fluxes at 3.4 and 4.6 $\mu$m from the Wide-field Spectroscopic Explorer (WISE) for LRG and QSO, and GAIA $G$ band for a morphological proxy to reject stars from BGS. We also use flux cuts defined on fiber magnitudes, which measure the amount of predicted flux within a DESI fiber (1.5'' diameter) in 1'' Gaussian seeing.\footnote{See description here \url{https://www.legacysurvey.org/dr9/catalogs/}}
These fluxes are the most relevant for this work, since redshift success rate is connected to the total amount of light in a fiber, not the total amount of light in the entire galaxy.
The target selection pipeline is described in \cite{TS.Pipeline.Myers.2023}.
Galaxies are subsequently assigned to fibers 
\cite{FBA.Raichoor.2024}. Fiber assignment has a significant impact on small-scale galaxy clustering due to the 1.5 arcmin patrol radius of each fiber.
The impact of fiber assignment has been extensively studied
using both pairwise inverse probability weights \cite{KP3s7-Lasker} and ``fast fiberassign'' \cite{KP3s11-Sikandar} and its impact on two-point clustering is described in \cite{KP3s6-Bianchi}. Its impact on the RSD analysis is mitigated by removing all pairs with separation less than the patrol radius \cite{Kp3s5-Pinon}.

BGS targets are split into ``bright'' and ``faint'' subsamples
as described
in \cite{BGS.TS.Hahn.2023}.
The cosmology sample uses ``BGS\_BRIGHT'' with an additional restriction on the $r$-band absolute magnitude, $M_r < -21.5$ \cite{DESI2024.II.KP3}; this sample is called ``BGS\_BRIGHT-21.5.''
Hence there are three BGS samples for which large-scale structure (LSS) catalogs are constructed, ``BGS\_ANY'' (all BGS targets), ``BGS\_BRIGHT'' (without the absolute magnitude cut) and ``BGS\_BRIGHT-21.5'' (used for the cosmological analysis). In this paper we are primarily concerned with ``BGS\_BRIGHT-21.5'' but also briefly show the impact
of spectroscopic systematics on the other samples, which may be more relevant for other applications requiring higher number densities.
An important distinction between the samples is that only ``BGS\_ANY'' and ``BGS\_BRIGHT'' exist for the BGS \textit{targets}; defining ``BGS\_BRIGHT-21.5'' requires the galaxy's absolute magnitude (and thus its redshift), so ``BGS\_BRIGHT-21.5'' can only be defined  for successful redshifts.

DESI survey operations are designed to ensure maximum uniformity in the data's spectroscopic quality 
\cite{SurveyOps.Schlafly.2023}. The exposure time is dynamically adjusted based on observed conditions \cite{Expcalc.Kirkby.2024}.
Targets are observed from 5,000 fibers at once, with 4,326 available for the Y1 catalogs.\footnote{As explained in \cite{KP3s15-Ross}, the status of a fiber positioner can change from night to night (e.g.\ due to transient electronics issues). We therefore refer to the total number of fibers that were usable at any point during DESI Y1, which is 4,326.}
The astrometric performance and positioner accuracy are described in \cite{AstrometricCalib.Kent.2023}.
The fibers are divided into 10 petals, labelled with zero-based indexing (see Fig.~\ref{fig:big_focal_plane} in this paper and Fig.~1 in \cite{KP3s15-Ross} for a visualization of the focal plane). The fibers are read out continuously in fiber number in the CCD, so CCD issues will affect contiguous blocks of fibers. Amplifier issues are associated with half of the fibers within a petal (i.e.\ fiber numbers 0 to 250 or 251 to 500). Amplifier issues also cause flux offsets between
halves of spectra in the wavelength direction.

After nightly data is processed \cite{Spectro.Pipeline.Guy.2023}, QA is performed each morning, and tiles with at least 85\% of the target observing time can be approved, if the redshift success rate, redshift distribution, and sky fiber residuals are adequate
(as described in \cite{SurveyOps.Schlafly.2023}). The effective observing time for each target is parameterized by the ``template signal to noise ratio squared'' (TSNR2), as described in Section 4.14 of \cite{Spectro.Pipeline.Guy.2023}.

TSNR2, for a particular DESI target, is related to the signal-to-noise that a fixed-brightness galaxy template
would achieve on the DESI spectrograph, if it were observed
under exactly the same conditions as the target under consideration.
Hence despite having ``signal-to-noise'' in the name, it is \textit{not} related to the actual brightness of the target under consideration.
TSNR2 is calculated as follows:
\begin{equation}
    \textrm{TSNR2} = \sum_i T_i^2 \langle (\delta F)^2 \rangle_i / \sigma_i^2
\end{equation}
where $T_i$ is the throughput term, $\delta F$ is a signal term computed from a standard spectrum for each target class, and $\sigma_i$ is the noise.
For each target, five different versions of TSNR2 are calculated, corresponding to standard targets for BGS, LRG, ELG, QSO, and  Ly$\alpha$ forest.
For the signal part, an exponential surface brightness profile is assumed for ELG, LRG, and BGS (and a point source for QSO) with half-light radii of 0.45'', 1'', and 1.5'' respectively, independent of the properties of the target.
Likewise, the spectral dependence $\delta F$ varies between each target class, but is the same for each target type.
The spectral dependence is chosen so that TSNR2 is related to the inverse of the redshift measurement uncertainty for each target.
$\delta F$ is the rms of the difference between a standard spectrum $F$ and a 100 \AA\ median-filtered version of that same spectrum,
empirically found to be slightly more optimal than the derivative of the spectrum with respect to redshift. Thus the ``signal'' part of TSNR2 always comes from standard sources with no relationship to the target under consideration. 
In contrast, the throughput term $T_i$ and the noise $\sigma_i$ are taken from the actual observing conditions of the target, i.e.\ seeing, Galactic dust extinction, positioning error, plate scale, read noise, sky background, flat field correction, and transparency.

\begin{figure*}
    \includegraphics[width=0.8\linewidth]{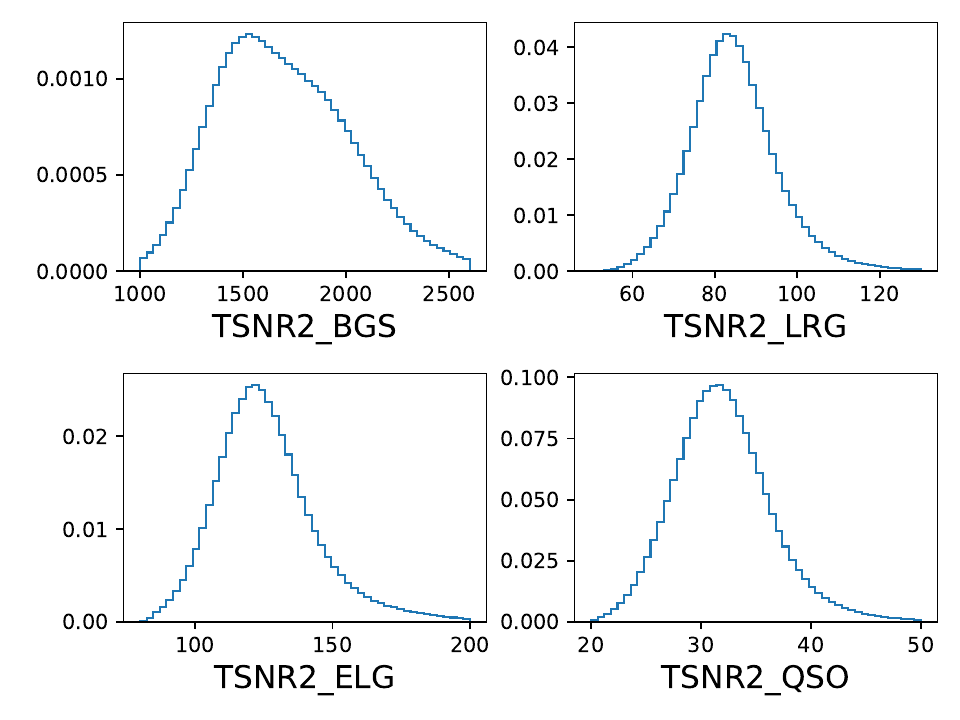}
    \caption{Normalized distribution of TSNR2 for targets in each tracer class. For BGS we show the distribution of TSNR2 for BGS\_BRIGHT targets; the distribution is nearly the same for BGS\_ANY targets.
    \label{fig:tsnr_distribution}}
\end{figure*}

TSNR2 is defined per fiber, and its average over an entire tile (specifically using TSNR2\_LRG for dark time and TSNR2\_BGS for bright time) is used to determine the tile's effective exposure time.
This is calculated via Eq.~22 in \cite{Spectro.Pipeline.Guy.2023}, assuming a sky-limited observation, relating TSNR2 to spectroscopic effective exposure time $T_{\textrm{spec}}$:
\begin{equation}
T_{\rm spec} = (12.15 \textrm{sec}) \times TSNR^2_{\textrm{LRG}}
\label{eqn:tsnr_lrg}
\end{equation}

TSNR2 for the other tracers is related to $T_{\textrm{spec}}$ in a similar
way but with different proportionality constants:
\begin{equation}
T_{\rm spec} = (0.135 \textrm{sec}) \times TSNR^2_{\textrm{BGS}}
\label{eqn:tsnr_bgs}
\end{equation}
\begin{equation}
T_{\rm spec} = (8.60 \textrm{sec}) \times TSNR^2_{\textrm{ELG}}
\label{eqn:tsnr_elg}
\end{equation}
\begin{equation}
T_{\rm spec} = (33.6 \textrm{sec}) \times TSNR^2_{\textrm{QSO}}
\label{eqn:tsnr_qso}
\end{equation}
If the effective exposure time is $<85\%$ of the goal effective exposure time ($0.85 \times 1000 = 850$ s for dark time and $0.85 \times 180 = 153$ s for bright time), another exposure is required. Another relevant variable is the survey speed, defined as the number of effective seconds observed per actual second of observing time, as computed using the Exposure Time Calculator and described in Section 5.3 of \cite{SurveyOps.Schlafly.2023}. If the survey speed is low
due to marginal observing conditions, leading to long exposures, the exposure may be split to facilitate cosmic-ray rejections. Thus tiles can be observed with up to 4 exposures in the Y1 sample, and these may be on different nights.
A small number of targets with low TSNR2 in the final data processing, TSNR2\_ELG < 80 or TSNR2\_BGS < 1000, are removed from the catalogs.

We show the normalized distributions of TSNR2 for each target class in Fig.~\ref{fig:tsnr_distribution}.
The TSNR2 distributions are very similar for all three dark time tracers, largely only differing by a scaling factor; on the other hand, BGS has quite a different distribution of TSNR2\_BGS. The distribution of TSNR2 values comes from both the distribution of effective exposure times between tiles, and variations in efficiency on the focal plane within each tile. While DESI uses a variable exposure time calculated by the exposure time calculator (ETC) \cite{Expcalc.Kirkby.2024} using real-time measurements of the sky background, seeing and transparency, this process is quite accurate but not perfect. Fig.~7 of \cite{SurveyOps.Schlafly.2023} shows
the distribution of actual effective exposure time relative to the goal exposure time; the dispersion is fairly tight for dark time, with a fractional standard deviation of 7.1\%, but much larger for bright time, with a fractional standard deviation of 16.6\%. This is due to varying color of the sky background from the Moon during bright time. The ETC only uses $r$-band sky brightness, which may not be representative of the actual wavelength-integrated sky brightness used to compuate TSNR2 and effective exposure time. We measure the standard deviation of TSNR2 for each tile (probing variations in efficiency on the focal plane), and the standard deviation of the mean TSNR2 across each tile (probing variation in observing conditions). For BGS, TSNR2\_BGS has a 17.5\% fractional standard deviation in the mean TSNR2 across each tile, and only a 9.6\% mean standard deviation within each tile. Thus the TSNR2\_BGS distribution is dominated by variations in observing conditions across tiles. On the other hand, TSNR2\_LRG has a 10.5\% fractional standard deviation in mean TSNR2 across each tile, and a 9.2\% standard deviation within each tile. Thus for LRG (and also other dark time tracers), variation on the focal plane is comparable to variation across tiles. The variation in TSNR2 within each tile is comparable to the 10\% variation shown in Fig.~26 of \cite{Spectro.Pipeline.Guy.2023} in relative transmission across the focal plane.

Once galaxies and quasars at $z < 2.1$ are observed to the specified effective exposure time, they are de-prioritized so that they can only be observed if no other science target is available for a given fiber.
Quasars at $z > 2.1$ are re-observed up to four times, to increase their signal-to-noise for the Ly$\alpha$ forest measurements.
Redshifts for the reduced and calibrated spectra are measured using a homogeneous processing of the full dataset; for Y1, this processing was internally referred to as ``Iron.'' Galaxies are redshifted using the \texttt{Redrock} software \cite{Redrock.Bailey.2024}, and quality is assessed by redshift warning flags \texttt{ZWARN} and $\Delta \chi^2$, the difference between the best-fit and the second best fit $\chi^2$.
Quasars are redshifted using updated DESI templates which separate low-redshift and high-redshift quasars \cite{RedrockQSO.Brodzeller.2023};
an additional fix to correctly account for the Ly$\alpha$ mean flux evolution \cite{KP6s4-Bault} was used for the Ly$\alpha$ analysis but not in the LSS 
catalogs--we have verified that applying this fix makes little difference in the LSS catalogs. The redshift success criterion follows that established for the DESI SV3 LSS catalogs \cite{DESI2023b.KP1.EDR}
\begin{itemize}
    \item BGS: \textsc{ZWARN}==0, $\Delta \chi^2 > 40$
    \item LRG: \textsc{ZWARN}==0, $\Delta \chi^2 > 15$
    \item ELG: $\log_{10}(S_{\textrm{[OII]}}) + 0.2 \log_{10}(\Delta \chi^2) > 0.9$
    \item QSO: Passes QSO criterion from \cite{QSO.TS.Chaussidon.2023}
\end{itemize}
The ELG and QSO targets partially overlap, and thus some of the ELG targets wind up as spectroscopically confirmed quasars. These targets are removed from all of the ELG LSS catalogs; the ELG sample is therefore internally referred to as ``ELG\_LOPnotqso.''

Both ELG and QSO require additional processing of the redrock outputs, referred to as an ``afterburner.'' For ELG, this is to measure the [OII] flux signal-to-noise ratio $S_{\textrm{[OII]}}$, and for QSO
the QuasarNet and MgII afterburners improve redshift accuracy and completeness.
These success criteria were extensively validated using the SV samples \cite{BGS.TS.Hahn.2023,LRG.TS.Zhou.2023,ELG.TS.Raichoor.2023,QSO.TS.Chaussidon.2023} and visually-inspected spectra from deep exposures \cite{VIGalaxies.Lan.2023,VIQSO.Alexander.2023}.
The cosmological analysis also uses a volume-limited subset of BGS with absolute magnitude $M_r < -21.5$, as determined using \textsc{fastspecfit} \cite{Fastspecfit}. Since $M_r$ can only be defined for galaxies with a successful redshift, we only apply this cut to successful BGS targets, not all BGS targets.

Finally, LSS catalogs are created from the successful targets, with appropriate weights. The process is described in detail in \cite{DESI2024.II.KP3,KP3s15-Ross}. In this paper, we describe how the redshift failure weights are defined, which upweight galaxies observed with less effective exposure time to account for the lack of uniformity due to redshift failures.

\subsection{Redshift success rate}
\label{sec:redshift_success_rate}

The redshift failure weights are applied to ensure that the redshift success rate (averaged over all redshift) is constant as a function of observing conditions (i.e.\ TSNR2). Identically to \cite{KP3s4-Yu}, we always consider the normalized redshift success rate, i.e.\ the spectroscopic success rate within a particular tracer, observing condition, and/or redshift bin, divided by the overall redshift success rate:
\begin{equation}
    \textrm{SSR} (x) = N_{\rm goodz}(x)/N_{\rm obs}(x)
    \label{eq:ssr}
\end{equation}
\begin{equation}
    f_{\textrm{goodz}}(x) = \frac{\textrm{SSR}(x)}{\sum_x N_{\rm goodz}(x)/\sum_x N_{\rm obs}(x)}, 
    \label{eq:fgoodz}
\end{equation}
As a result of the normalization by overall redshift success
rate, $f_{\textrm{goodz}}$ always fluctuates around one; hence it does not characterize the actual fraction of good redshifts,
but rather the uniformity of the good redshift fraction
as a function of spectroscopic observing condition.

Within Eqs.~\ref{eq:ssr} and~\ref{eq:fgoodz}, ``goodz'' denotes targets
that have obtained a successful
redshift following the classification in Sec.~\ref{sec:catalog_construction}.
Note that in some cases, this means that the ``failures'' might actually have a robustly classified redshift, but as the wrong type of target (e.g.\ QSO targets that can be spectroscopically confirmed as stars or galaxies).

We differentiate here betweeen two defintions of ``goodz.''
In the first, used to \textit{fit} the redshift failure weights,
a good redshift is merely any target that passes the successful redshift criterion of Sec.~\ref{sec:catalog_construction}, without any reference to its redshift.
In contrast, to \textit{validate} the redshift failure weights for the LSS catalogs,
we are only concerned with the ipact of spectroscopic systematics
on tracers within the specified redshift ranges.
Thus, in the second definition of ``goodz'' used for validation, we also require that the targets
lie in the fiducial redshift range to be classified as ``goodz'':
$0.1 < z < 0.4$ for BGS, $0.4 < z < 1.1$ for LRG, 
$0.8 < z < 1.6$ for ELG, and $0.8 < z < 2.1$ for QSO
(and we also validate using wider redshift ranges in Appendix~\ref{sec:validation_other_z_ranges}).
We also consider narrower redshift ranges, where the successful
redshifts must lie in $\Delta z = 0.1$ bins.
For BGS, the \textit{validation} definition requires that ``goodz'' passes
the $M_r < -21.5$ cut (but not necessarily the target), so the overall redshift success rate passing this cut is quite low. 
The $M_r$  cut is not applied to targets because the $M_r < -21.5$ cut is not well defined for galaxies with poorly measured redshift, as determining $M_r$ relies on the galaxy's redshift.
In contrast, the \textit{fitting} definition of ``goodz'' for BGS does not include the $M_r$ cut, i.e.\ the redshift failure weights are fit to ``BGS\_BRIGHT'' and not ``BGS\_BRIGHT-21.5.''

The overall ``fitting'' redshift success rates (with no extra redshift cuts) are 98.9\% for LRG, 98.8\% for BGS\_BRIGHT (before the $M_r < -21.5$ cut), 72.4\% for ELG, and 66.2\% for QSO.
For LRG and BGS, failures are mainly faint galaxies with no discernable emission features, with a small number of stellar interlopers. For ELG, failures are mainly galaxies with insufficient [OII] emission and minimal continuum detected, as well as ELGs at $z > 1.6$ where [OII] redshifts out of the DESI wavelength coverage. 

For QSO, there are more possiblities. Based on visual inspection
of deep observations (deep VI), 71\% of quasar targets are quasars, 16\% are galaxies, 6\% are stars, and 7\% are inconclusive \cite{QSO.TS.Chaussidon.2023}.\footnote{The fraction of quasars in deep VI is higher than the DR1 pipeline classification fraction because the pipeline misses some true quasars, and also due to sample variance from the small size of the visually inspected dataset.}
This is a crude treatment of the failures, since some of these spectra can be confidently classified, particularly the stars.
However, it is harder to devise a high-purity classification
for the quasar targets that are truly galaxies. 
A past study aimed at constraining primordial non-Gaussianity from DESI quasar target and Planck CMB lensing  \cite{Krolewski23}
created a spectroscopic success criterion for galaxies targeted as quasars, in order to measure the
redshift distribution of the entire quasar targeting sample, including galaxies.
This criterion is a hybrid of the ELG and LRG
spectroscopic success criterion but only achieves a 90\% 
purity, and the rest of the spectroscopically-classified galaxies are true quasars.
Moreover, the stellar classification has only 80\% completeness (the remaining 20\% are ``unclassified'' by spectroscopic criteria), and, even worse,
we have no truth classification for the 7\% of spectra
that are unclassified even in the deep VI exposures.
To sidestep some of these difficulties, we will also compare spectroscopic success rate trends in the full quasar sample to those in a ``high-purity'' sample defined in \cite{Krolewski23}, which removes much of the contaminating
stars and galaxies, though at the price of reducing the number density by half.
Overall, we continue to use the simple and crude classification where all non-quasar quasar targets are classified as failures. We discuss the impact of this choice on our results below and plan to revisit this decision for future data releases.

\subsection{Mocks}
\label{sec:mocks}

\begin{figure*}
    \includegraphics[width=0.8\linewidth]{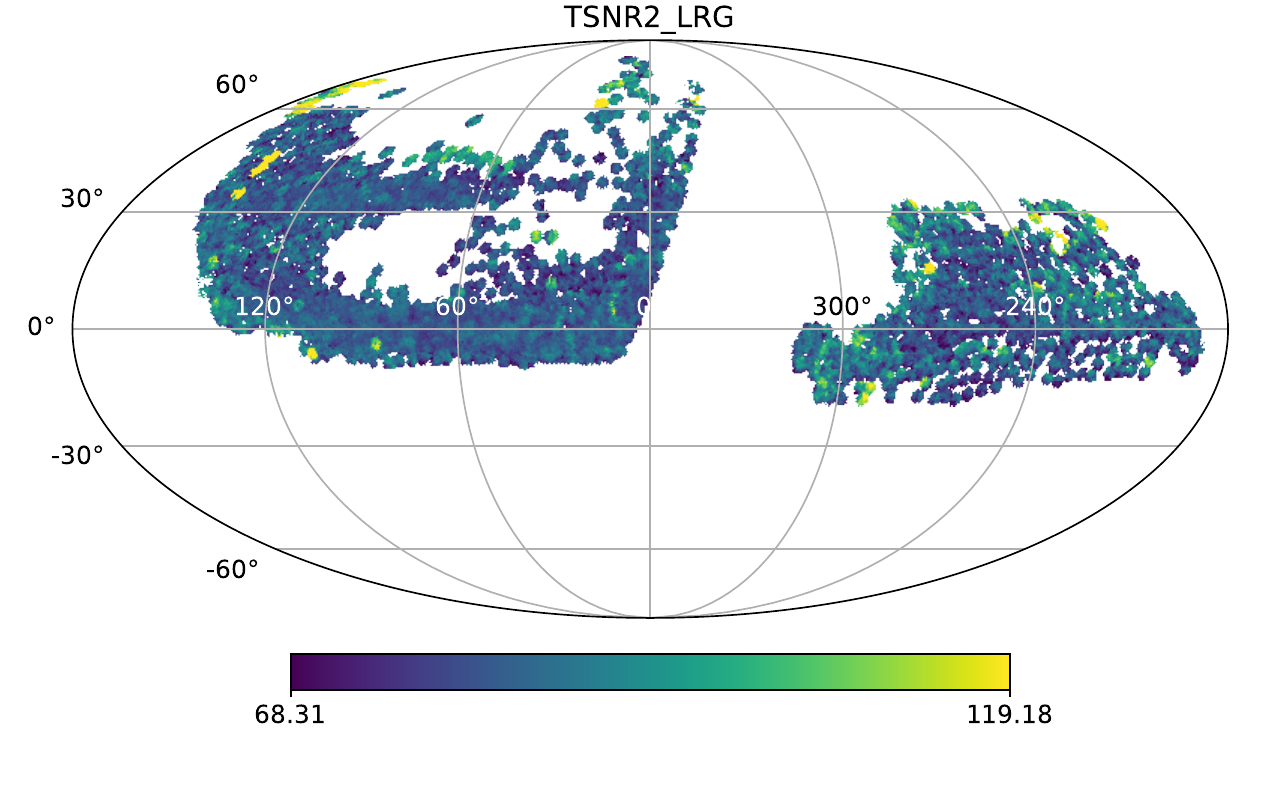}
    \includegraphics[width=0.8\linewidth]{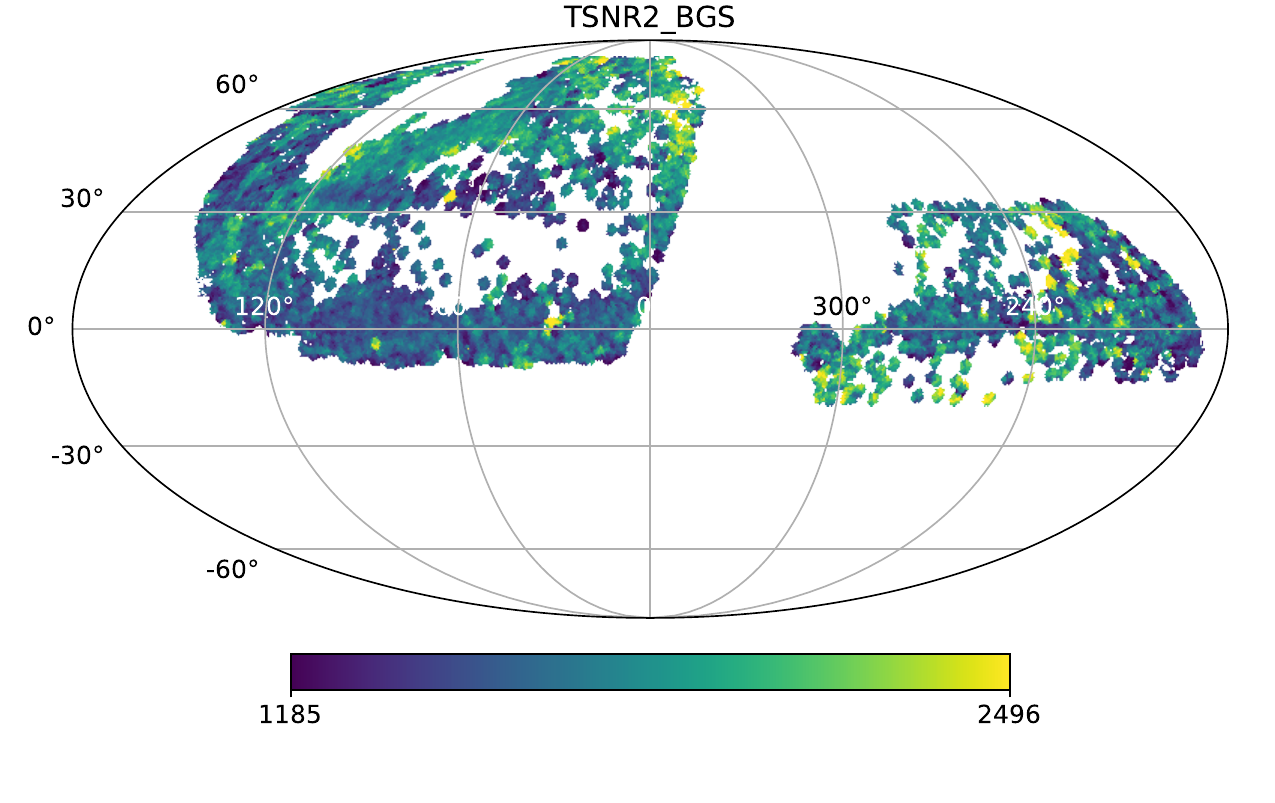}
    \caption{\textit{Top:} average TSNR2\_LRG for the LRG sample, binned in
    nside=128 HEALPix pixels. Other dark time tracers show a similar pattern, since they were observed in the same conditions in the same patches of sky.
    \textit{Bottom:} Same plot using TSNR2\_BGS for the BGS\_BRIGHT sample.
    \label{fig:tsnr2_map}}
\end{figure*}

We use a suite of 25 AbacusSummit mocks
\cite{AbacusSummit,kp3s8-Zhao} to characterize the noise on some of our redshift
success statistics. These mocks have been passed through the ``fast fiberassign'' \cite{KP3s11-Sikandar} emulator to select DESI targets, and also to give each mock target a per fiber and tile TSNR2 matched to the sky distribution in the data. We refer to these mocks as the ``Abacus FFA'' mocks.
The likelihood of the redshift success rate within TSNR2 bins is complicated.
Certain areas of the survey preferentially have higher or lower
TSNR2 (Fig.~\ref{fig:tsnr2_map}).
As a result, the probability of success for a single observation does not follow a 
 binomial distribution that would
be expected if all galaxies were independent, but contains
a contribution from cosmic variance as well.
The cosmic variance contribution is akin to the super-sample covariance effect and is inversely proportional to the volume of the patches (area of the TSNR2 patch times depth of the redshift bin used).
In other words, if a $\Delta z = 0.1$ bin in TSNR2 just happened
to contain a few more clusters than expected, its $f_{\textrm{goodz}}$ would fluctuate higher than expected based on binomial statistics alone.
These fluctuations are possible for redshift bins of any depth, but are suppressed as $\Delta z$ gets larger.
Therefore, to assess the significance of the fluctuations
in redshift success rate in the narrow redshift bins, 
we measure $\chi^2$ from the deviation between the observed redshift success rate and unity, using binomial errors:
\begin{equation}
    \chi^2 = \sum_{\textrm{TSNR2}} \frac{(f_{\textrm{goodz}}(\textrm{TSNR2}) - 1)^2}{\sigma_{\textrm{binomial}}^2}
    \label{eq:chi2}
\end{equation}
Rather than interpreting this statistic with the usual $\chi^2$ distribution, we directly
compare the distribution of $\chi^2$ between data and mocks with randomly-assigned redshift failures, but TSNR2 assigned following the observed sky distribution. The mocks themselves do not have redshift failures, so we randomly downsample the mocks by the observed mean success rate, 99\% for LRG, 97\% for BGS\_BRIGHT, and 66.5\% for QSO.\footnote{While these are slightly different from the success rate on data, the impact of the difference on the TSNR2 trends in the mocks is negligible.} This simple model will under-estimate the scatter in success rate at low TSNR2, where the success rate is lower than the mean (recalling that in the limit as the failure rate $(1-p)$ goes to zero, the binomial error goes to $\sqrt{(1-p)}$). This effect is most severe for the LRG and BGS, where the success rate is close to 100\%, and, at low flux, the failure rate may double at low TSNR2. However, we primarily use the mocks to assess the significance of TSNR2 trends in narrow redshift bins, where the ``success rate'' is typically low, due to the redshift cut applied to the definition of a ``success'' (Eq.~\ref{eq:ssr}). As a result, small changes in the success rate do not impact its error much.

\subsection{Clustering measurements}


We test the impact of significant residual redshift success trends on the observed power spectrum, quantifying their impact using $\chi_{\textrm{sys}}^2$
\begin{equation}
\chi_{\textrm{sys}}^2 = (P_{\ell}^{\textrm{extra wt.}}(k_i) - P_{\ell}^{\textrm{std}}(k_i)) (\boldsymbol{C}^{-1})^{ij} (P_{\ell}^{\textrm{extra wt.}}(k_j) - P_{\ell}^{\textrm{std}}(k_j))
\end{equation}
where $P_\ell^{\textrm{std}}$ are the power spectrum multipoles using the standard weights (Eq.~\ref{eq:wtot}), $P_\ell^{\textrm{extra wt.}}$
are the multipoles using the additional weight under consideration, and $\boldsymbol{C}$ is the power spectrum covariance.
Shifts in derived parameters (i.e.\ cosmological, nuisance or BAO shift parameters $\alpha_{\textrm{iso}}$ or $\alpha_{\textrm{AP}}$) from adding the additional weights are at most $\sqrt{\chi_{\textrm{sys}}^2}$, if the impact
of the extra weight and the parameter under consideration are perfectly degenerate. This sets the systematic uncertainty from residual inhomogeneities.
If $\chi_{\textrm{sys}}^2 << 1$, we conclude that the systematic error does not measurably affect parameters inferred from galaxy clustering, without actually having to fit for those parameters.
If $\chi_{\textrm{sys}}^2 \gtrsim 1$, we test the impact of weights by inferring cosmological parameters from both the default power spectrum and the extra-weight power spectrum, to test how much of the change projects onto cosmological parameters of interest, as opposed to nuisance parameters.

We measure the power spectrum multipoles $P_0$, $P_2$, and $P_4$ using \textsc{pypower}\footnote{\url{https://github.com/cosmodesi/pypower}} \cite{pypower2017}, measured between $k = 0$ and 0.4 $h$ Mpc$^{-1}$, with bins of width $\Delta k = 0.005$ $h$ Mpc$^{-1}$.
The multipoles are calculated from the Fourier transform of the galaxy overdensity field, using the Yamamoto estimator \cite{yamamoto2006}
\begin{equation}
P_{\ell}(k) = \frac{2 \ell + 1}{A N_{k}} \sum_{\boldsymbol{k}} F_{0}(\boldsymbol{k}) F_{\ell}(-\boldsymbol{k}) - \mathcal{SN}_{\ell}
\label{eq:power_spectrum_multipoles}
\end{equation}
where $N_k$ is the number of modes in the $k$ bin, $A$ is a normalization (defined below), and $F_{\ell}(k)$ is the Legendre-weighted Fourier transform of the overdensity field:
\begin{equation}
F(\boldsymbol{r}) = n_{d}(\boldsymbol{r}) - \alpha n_{r}(\boldsymbol{r}).
\label{eq:fkp}
\end{equation} 

\begin{equation}
F_{\ell}(\boldsymbol{k}) = \sum_{\boldsymbol{r}} F(\boldsymbol{r}) \mathcal{L}_{\ell}(\mu) e^{i \boldsymbol{k} \cdot \boldsymbol{r}}.
\label{eq:fkp_multipoles}
\end{equation}
where $n_d$ and $n_r$ are the real-space densities of the data and randoms,\footnote{The randoms are created for the imaging survey by \texttt{DESITARGET} \cite{TS.Pipeline.Myers.2023} and their processing for 3D clustering is described in \cite{KP3s15-Ross}.} on a Cartesian grid with cell size 6 $h^{-1}$ Mpc, and $\mu$ is the angle to the line of sight.
The galaxies and randoms are weighted to correct
for incompleteness (due to the limited number of DESI passes in the Y1 dataset) $w_{\rm comp}$ and observational systematics ($w_{\rm sys}$ and $w_{\rm zfail}$):
\begin{equation}
    w_{\rm tot} = w_{\rm comp}w_{\rm sys}w_{\rm zfail},
    \label{eq:wtot}
\end{equation}
In addition, the FKP weight \cite{FKP1994} is applied to optimally combine clustering measurements at different redshifts:
\begin{equation}
    w_{\rm FKP} = \frac{1}{1+n_\mathrm{local}(z, N_{\rm tile})P_0},
\end{equation}
where $n_\mathrm{local}(z, N_{\rm tile})$ is the local number density for a given tracer, accounting for completeness variations as described in Section 7.3 of \cite{KP3s15-Ross}, and $P_0$ is the amplitude of the observed power spectrum at $k \approx 0.15$ $h$ Mpc$^{-1}$, as given in Section 8.2 of \cite{DESI2024.II.KP3}.
The systematics weights include imaging systematics weights $w_{\rm sys}$, designed to null the dependence of galaxy density on non-cosmological fluctuations (described in Section 6 of \cite{DESI2024.II.KP3}) and redshift failure weights $w_{\rm zfail}$, defined in Section~\ref{sec:zfail_wts} below and designed to null the dependence of galaxy density on spectroscopic observing conditions.
Both data and randoms are weighted by $w_{\rm tot}$.

The weighted field $F_{\vec{r}}$ is normalized to have mean zero: 
$\alpha=\sum^{N_d}_{i=1}w_{\rm tot,i(d)}/\sum^{N_r}_{i=1}w_{\rm tot,i(r)}$.
The normalization of the power spectrum $A=\alpha\sum^k n_{d,k}n_{r,k}$ is summed over a cell with 10 $h^{-1}$ Mpc in size. The shot noise is only nonzero for the monopole:
\begin{equation}
\mathcal{SN}_{0} = \frac{1}{A} \left[\sum^{N_d}_{i=1}w_{\rm tot,i(d)}^{2}+\alpha^2\sum^{N_r}_{i=1}w_{\rm tot,i(r)}^{2} \right].
\end{equation}

We use analytic covariance matrices calculated by \textsc{THECOV}\footnote{\url{https://github.com/cosmodesi/thecov}} \cite{KP4s8-Alves}, based on the perturbative covariance matrix code \textsc{CovaPT}\footnote{\url{https://github.com/JayWadekar/CovaPT/}} \cite{CovaPT_Wadekar:2019rdu}.
The covariance matrices are validated for DESI Y1 in \cite{KP4s6-Forero-Sanchez}.

\section{Redshift failure model and weights}
\label{sec:zfail_wts}

We define a set of redshift failure weights to remove density fluctuations due to spectroscopic observing conditions
for each of DESI's four tracers.\footnote{This is implemented in the DESI clustering code in \url{https://github.com/desihub/LSS/blob/v1.4-DR1/py/LSS/ssr_tools_new.py}.} 
The process is very similar for LRG, BGS, and QSO, and quite different for ELG, whose redshift failure weights are described in detail in Section 4.1 of \cite{KP3s4-Yu}.

For the LRG, BGS, and QSO weights, 
we start by defining a model for the overall spectroscopic success rate (SSR), i.e.\ not normalized by the mean spectroscopic success rate of each tracer.
We fit a simple model to the overall spectroscopic success rate as a function of TSNR2, allowing 
for an exponentially rising failure rate at low TSNR2.

In the model, the spectroscopic success rate is defined by:
\begin{equation}
    \textrm{SSR} = 1 - \textrm{FR}_{\textrm{clip}}
\end{equation}

\begin{equation}
  \textrm{FR}_{\textrm{clip}} = 
  \begin{cases}
    0 & \textrm{FR} < 0 \\
    \textrm{FR} \equiv \exp{(-(\textrm{TSNR2} + a)/b)} + c & 0 < \textrm{FR} < 1 \\
    1 & \textrm{FR} > 1 \\
  \end{cases}
  \label{eqn:fr_clip}
\end{equation}
where the clipping ensures that the simple function does not go below zero or above one. The model is always constrained to ensure that failure rate cannot decrease with increasing TSNR2.
While overall cuts to the sample are defined in terms of TSNR2\_LRG, the fits are done in terms of the tracer-specific TSNR2.

We define 10 bins in TSNR2. For LRG, these are defined between 41.1 (= 500/12.15, corresponding to 500 ``effective seconds'') and 139.9 (= 1700/12.15).
For QSO, the bins are defined between 13.3 (= 450/33.72, using the conversion between effective time and TSNR2\_QSO) and 53.4 (= 1800/33.72).
For BGS, the bins are defined between 889 (=120/0.135, using the conversion between effective time and TSNR2\_BGS) and 2222 (=300/0.135).
In the regression we fit against the median TSNR2 in each bin.
When defining the weights and validating their performance, we only use the first observation for high redshift quasars with multiple observations, otherwise the high-TSNR2 bin for quasars would be highly skewed towards these high redshift quasars.

\begin{figure*}
    \includegraphics[width=\linewidth]{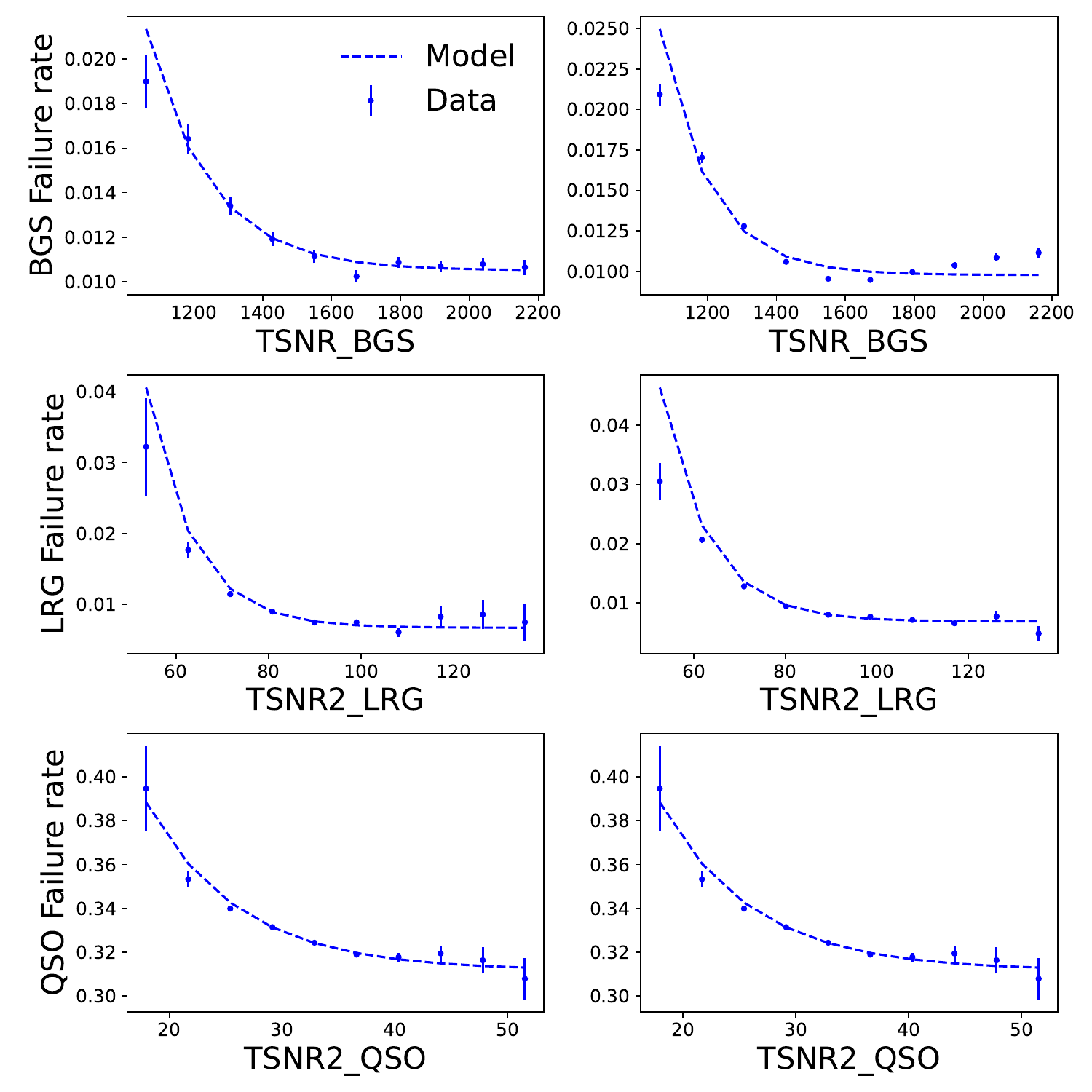}
    \caption{Overall failure rate model for tracers in the North (left column) and South (right column), for BGS\_BRIGHT (top), LRG (middle) and QSO (bottom).
    \label{fig:fit_quality}}
\end{figure*}

This model is shown in Fig.~\ref{fig:fit_quality} and the parameters $a$, $b$ and $c$ are given in Table~\ref{tab:my_label}. The model fits the observed trends
reasonably well in both hemispheres, particularly at the low TSNR2 end. At moderate to high TSNR2, we see difficulties in BGS\_BRIGHT (South region) and QSO (North region).  These are further discussed below, but are believed to result from chance alignments between TSNR2 and other properties controlling failure rate. For quasars, high TSNR2 happens to lie at low galactic latitude, particularly in the North (see Fig.~\ref{fig:tsnr2_map}), and for BGS, high TSNR2 happens to coincide with slightly fainter targets, which have a much higher failure rate.

Next, we account for the fiberflux dependence of the success rate. Faint galaxies are considerably less likely
to have successful redshift measurements. This is particularly important because flux is correlated with redshift,
and having the correct flux dependence allows the redshift-failure weights to properly account for spectroscopic observing fluctuations at all redshifts. 
The fiberflux dependence of each tracer uses the band used for the target selection fiber magnitude cut:
$r$ for QSO and BGS, $g$ for ELG, and $z$ for LRG.
For BGS and QSO, we allow for a linear relationship between success rate and fiberflux $F$, using two new parameters, $f$ and $F_{\textrm{piv}}$. The weights are defined as follows:
\begin{equation}
    \textrm{SSR}_{\textrm{rel}} = \frac{\textrm{SSR}}{\textrm{max(SSR)}}
\end{equation}

\begin{equation}
    w_{\textrm{zfail}} = 
    \begin{cases}
    1 & w_{\textrm{raw}} < 1 \\
     w_{\textrm{raw}} \equiv
    \left[f \left( 1  - \frac{F}{F_{\textrm{piv}}}\right) + 1\right] \left(\frac{1}{\textrm{SSR}_{\textrm{rel}}} - 1\right) + 1 & w_{\textrm{raw}} > 1
    \end{cases}
\label{eqn:model_flux_tsnr}
\end{equation}
The free parameters $f$ and $F_{\textrm{piv}}$ allow us to smoothly
change success rate with flux, and to change between the regime where success rate strongly
depends on flux and where it does not depend on flux.
If $f = 0$, then the spectroscopic weight does not depend on flux, and is simply the inverse of the global spectroscopic success rate fit.
If $f = 1$, flux maximally affects success rate.
If $F = 0$ (i.e.\ a faint galaxy), then $w_{\textrm{zfail}}-1$ for this galaxy is upweighted by $(f + 1)$ relative to $w_{\textrm{zfail}}-1$ for a galaxy with flux at $F_{\textrm{piv}}$.
For LRG, we find that allowing for a piecewise linear relationship between success rate and flux, with a third parameter $C$, fits the data considerably better:
\begin{equation}
    w_{\textrm{zfail}} = 
    \begin{cases}
    1 & w_{\textrm{raw}} < 1 \\
     w_{\textrm{raw}} \equiv
    \left[G(F; F_{\textrm{piv}}, f, C)\right] \left(\frac{1}{\textrm{SSR}_{\textrm{rel}}} - 1\right) + 1 & w_{\textrm{raw}} > 1
    \end{cases}
    \label{eqn:model_flux_tsnr_lrg}
\end{equation}
where $G$ is a function of fiberflux $F$ with free parameters $F_{\textrm{piv}}$, $f$, and $C$
\begin{equation}
    G = 
    \begin{cases}
    f \left( 1  - \frac{F}{F_{\textrm{piv}}}\right) + 1 & F < F_{\rm break} \\
    C + D F & F \geq F_{\rm break}
    \end{cases}
\end{equation}
where $D$ is defined to ensure contnuity of $G(F)$
\begin{equation}
D = \left(f + 1 -  \frac{f F_{\rm break}}{F_{\rm piv} - C}\right)\frac{1}{F_{\rm break}}
\end{equation}
The break flux $F_{\rm break}$ is fixed to 3 nanomaggies;\footnote{This is the 30th percentile in extinction-corrected $z$-band fiberflux.} we do not find significant improvements if we vary it.

The fit parameters are determined by applying the weights as a continuous function to each galaxy as a function of its TSNR2 and fiberflux, and then minimizing the difference between the weighted histogram and a uniform distribution. The galaxies are binned in a histogram with the same 10 bins in TSNR2 and 5 bins in flux, defined using the 20th, 40th, 60th, and 80th percentiles.
The fit parameters and $\chi^2$ are given in Table~\ref{tab:my_label}, both for the overall TSNR2 fit (with 10 bins in TSNR2) and the flux-dependent fit.

The model success rate $f_{\textrm{suc}}$ is then 
directly related to $w_{\textrm{zfail}}$
\begin{equation}
    f_{\textrm{suc}} = \textrm{SSRF}(F)/w_{\textrm{zfail}}
\end{equation}
where the function $\textrm{SSRF}(F)$
is a fit to the flux-dependent success rate, with the following model
\begin{equation}
    \textrm{SSRF}(F) = a_F \erf{((b_F + F)/c_F)}
\end{equation}
with $a_F$, $b_F$, and $c_F$ fit to the
success rate in 20 equally spaced flux bins.
For this fit, the success rate
is weighted by $w_{\textrm{zfail}}$.
The function $\textrm{SSRF}(F)$ accounts for the fact that the success rate can vary strongly
with flux, but this is normalized
out in the redshift failure weights, since we do not want to over-weight
faint galaxies with potentially
very low success rates.

The ELG redshift success model is significantly different from what is described above. Like the LRG, BGS, and QSO model, the ELG model is designed to account for variation in redshift success rate with both spectroscopic observing condition and redshift. However, for ELG the dominant galaxy property controlling redshift success is not fiberflux, but rather [OII] line emission strength. This is discussed extensively in Section 4.1 of \cite{KP3s4-Yu}.








\begin{table*}
    \centering
    \begin{tabular}{c|cccc|cccc}
       Tracer  & $a$ & $b$ & $c$ & TSNR2-fit $\chi^2$ & $f$ & $F_{\textrm{piv}}$ & C & Flux-fit $\chi^2$ \\
       & & & & & & (nmgy) & & \\
       \hline
         LRG N & -19.68 & 10.00 & 0.00668 & 4.92 & 14.72 & 2.94 & 1.50 & 42.76 \\
         LRG S &  -18.89 & 10.42 & 0.00683 & 10.60 & 15.42 & 2.96 & 2.63 & 85.16 \\
         BGS\_BRIGHT N & -238.15 & 181.71 & 0.0105 & 12.82 & 10.86 & 5.41 & & 58.30 \\
        BGS\_BRIGHT S & -469.99 & 141.72 & 0.00975 & 127.68 & 8.56 & 5.54 & &  192.72 \\
        BGS\_ANY N & -277.55 & 188.96 & 0.00999 & 15.86 & 3.52 & 5.42 & & 94.50 \\
        BGS\_ANY S & -464.11 & 151.49 & 0.00941 & 163.81 & 3.02 & 5.43 & & 350.23 \\
        QSO N & -11.24 & 2.25 & 0.36 & 50.99 & 25.58 & 1.24 & & 73.44 \\
        QSO S & -3.10 & 6.33 & 0.33 & 9.35 & 4.27 & 1.58 & & 85.07 \\
    \end{tabular}
    \caption{Best-fit parameters for the overall TSNR2 fits (first set of columns; $a$, $b$ and $c$ referring to parameters in Eq.~\ref{eqn:fr_clip}) and the flux-dependent fits (second set; $f$, $c$, and $F_{\textrm{piv}}$ referring to Eq.~\ref{eqn:model_flux_tsnr} and~\ref{eqn:model_flux_tsnr_lrg}). We use 10 bins for the TSNR2 fits and 50 bins for the flux-dependent fits.
    \label{tab:my_label}}
\end{table*}

\section{Identifying and removing targets from low success rate fibers}
\label{sec:badfibers}

We find that certain fibers have a systematically low success rate, perhaps indicative of instrumental issues in the fiber or CCD, even beyond the success rate expected from the average TSNR2 of observations in that fiber. We remove the fibers with the worst success rates, rather than attempting to up-weight these observations. Specifically, we remove any fiber that we deem to be more than a 4$\sigma$ outlier, based on the observation that we expect less than 1 fiber out of the $\sim$4000 usable fibers to be a 4$\sigma$ outlier just by random chance.

Determining whether to remove these fibers requires an accurate model for the PDF of success rates.
In this section, we neglect the cosmic variance contribution to the success rate, since each fiber has observed galaxies spread over the DESI Y1 footprint, and therefore has negligible cosmological correlations.
The success rate of each observation is a binomial distribution, with probability mass function
\begin{equation}
    P(k) = \binom{n}{k} p^k (1-p)^{n-k}
\label{eqn:binomdist}
\end{equation}
where $p$ is the success rate and $P(k)$ is the probability of having exactly $k$ successes in $n$ independent Bernoulli trials.
Many galaxies have been observed with each fiber, at different times and locations on the sky.
Each observation has its own spectroscopic observing conditions, and thus its own TSNR2.
This means that the PDF of success rate for each fiber is the sum of many binomial distributions with different success rates $p$. There exists an analytic formula for the variance of $Z = \sum_{i=1}^N X_i$, the sum of $N$ binomial random variables $X_i$ with different success rates \cite{Drezner93}
\begin{equation}
    \mathrm{Var}(Z) = N \bar{p} ( 1 - \bar{p}) - \sum_{i=1}^N (p_i - \bar{p})^2
\label{eqn:sum_binomial_var}
\end{equation}
where $\bar{p}$ is the average success rate. As $N \rightarrow \infty$, the central limit theorem guarantees
that the PDF of $Z$ is Gaussian and fully specified by its mean $\bar{p}$ and variance (Eq.~\ref{eqn:sum_binomial_var}). 

\begin{figure*}
    \includegraphics[width=0.8\linewidth]{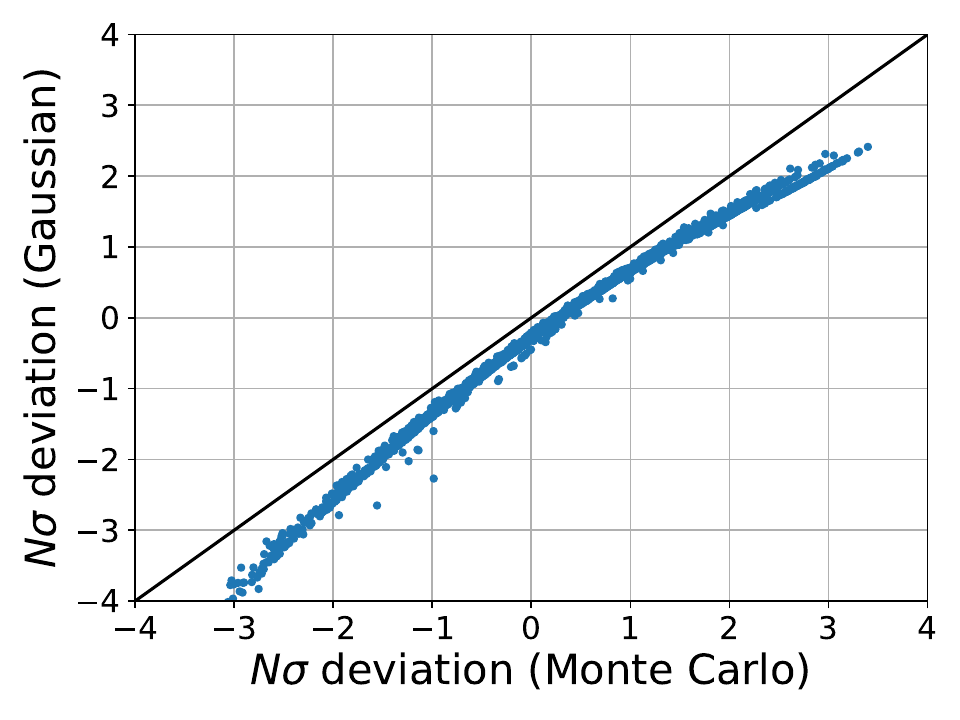}
    \caption{Comparison between Monte Carlo simulation and analytic (Gaussian) approximation for the mean and variance of the average success rate per fiber. The analytic approximation is a sum over binomial random variables with different success rates for each galaxy observation with a different TSNR2. We compute the likelihood of the observed success rate for each fiber, and convert probabilites to a number of $\sigma$ offset (using a Gaussian PDF). The line $y= x$ is shown in black. The probabilites shown are computed for the LRG sample. The offset points to the upper right are all fibers with zero failures; they are offset from all the fibers with exactly one failure, just above.
    \label{fig:analytic_vs_sim}}
\end{figure*}

The number of observations per fiber is typically large ($N \sim 600$ for LRG, for instance).
However, we want to measure probabilities deeply in the non-Gaussian tail of the distribution
to accurately determine whether a given fiber is a 4$\sigma$ outlier. In the tail, the number of observations
required for the Gaussian approximation to hold is much larger than typically needed for probabilities close to the core of the distribution.
In particular, the Gaussian approximation with Eq.~\ref{eqn:sum_binomial_var} dramatically overstates the number of 4$\sigma$ outliers as compared to a Monte Carlo simulation (Fig.~\ref{fig:analytic_vs_sim}).

As a result, we perform a Monte Carlo simulation to determine the likelihood of the observed success rate for each fiber, within each tracer class, following the Monte Carlo simulation for LRG in \cite{LRG.TS.Zhou.2023}.
For each fiber, we generated 3 million samples from the binomial PDF for each observation, using the modelled success rate for each observation. We re-normalize the modelled success rate to ensure that the mean success rate in the simulation matches the observed mean success rate (note that in the model as defined in Sec.~\ref{sec:zfail_wts}, such a match is not guaranteed because we care about removing fluctuations from the sample rather than accurately determining the true total number of any given target). The large sample size is necessary to confidently determine whether a given fiber is indeed a 4$\sigma$ outlier. We then directly determined the probability of observing at least as many successes as were seen in the actual data. This gives a $p$ value for each fiber, which was then converted into a Gaussian ``n-$\sigma$'' deviation to put the numbers on a familiar scale:
\begin{equation}
n\sigma = -\sqrt{2} \times \text{erfc}^{-1}(2 \times p\text{-value})
\label{eqn:nsig}
\end{equation}
where
the complementary error function, erfc(x), is defined as:
\begin{equation}
\text{erfc}(x) = 1 - \frac{2}{\sqrt{\pi}} \int_0^x e^{-t^2} dt
\label{eqn:erfc}
\end{equation}

\begin{figure*}
    \includegraphics[width=1.05\linewidth]{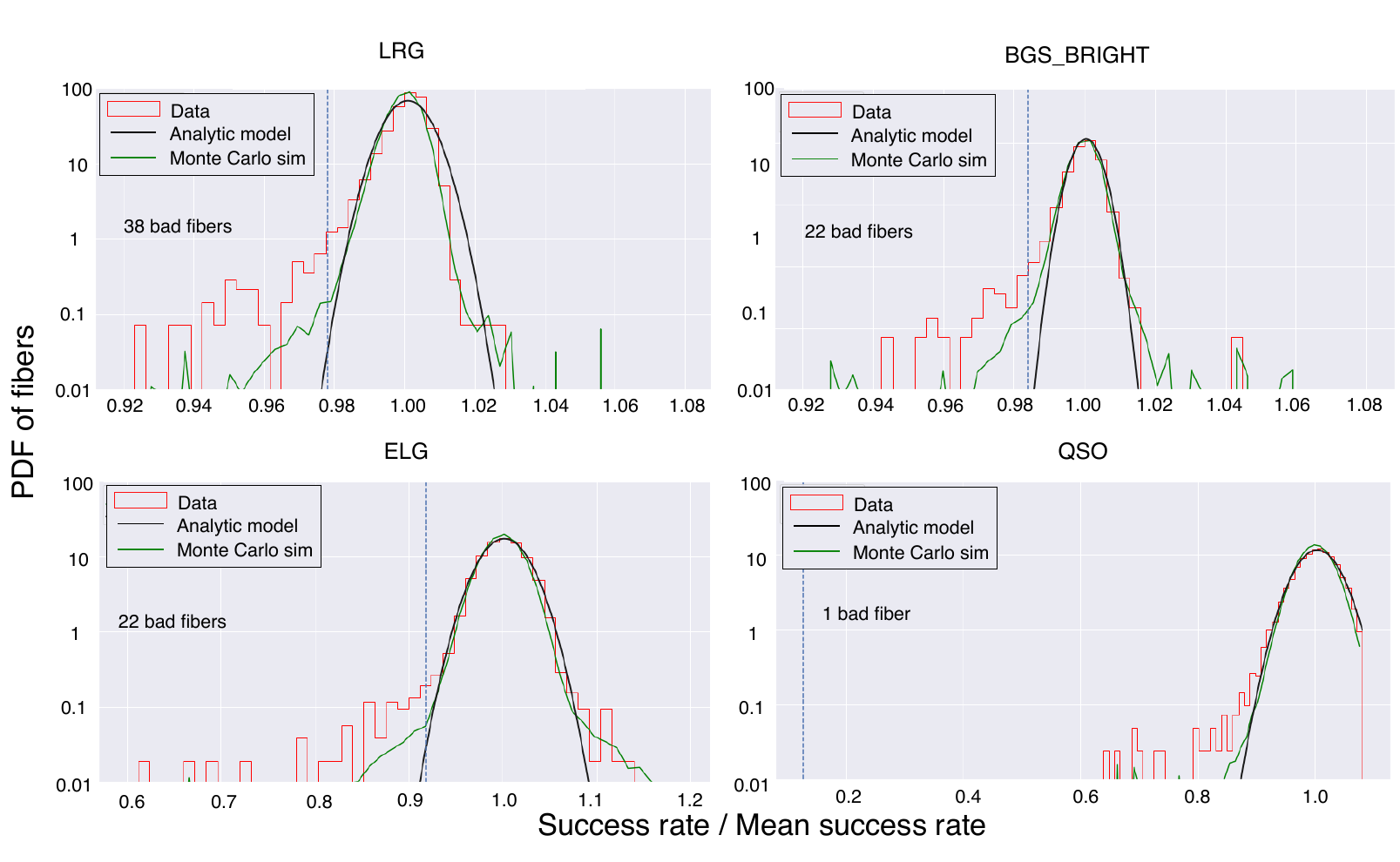}
    \caption{Comparison of the success rate PDF between the observed data (red), the Gaussian model using Eq.~\ref{eqn:sum_binomial_var} (black) and the Monte Carlo simulations (green), in both cases using the TSNR2 of each observation to model the success rate. The dashed blue line indicates the cutoff for 4$\sigma$ outliers in the data, as compared to the Monte Carlo simulations. The 4$\sigma$ cutoff was chosen so that the expected number of outliers, just by chance, was $<1$ for the $\sim$4000 active DESI fibers. On this plot, the $x$ position of a ``4$\sigma$'' outlier can change for each fiber, since it depends on the number of galaxies contributing to that fiber. As a result, we cannot define a single ``4$\sigma$'' cutoff line, but only show the maximum normalized success rate of the fibers that we remove.
    This is why the blue line is at success rate $\sim$15\% for the quasars, since there is only one bad fiber identified with success rate $\sim$15\%.
    \label{fig:badfiber}}
\end{figure*}

In Fig.~\ref{fig:badfiber}, we show the observed distribution of success rates, compared to the Monte Carlo simulation based on TSNR2-varying success rates and the analytic approximation based on the central limit theorem.
We see that there is a significant excess of fibers with low success rate for LRG, BGS\_BRIGHT, and ELG.
These fibers may have other issues impacting the success rate that are not folded into the TSNR2, or the model
linking TSNR2 to success rate may break down for these fibers (if they preferentially sample regions in fiberflux and TSNR2 where the model does not represent the data well). In either case, the success
rate for these fibers cannot be modelled well, and so we remove them from further analysis.

While low success rate fibers are identified separately for each tracer, we ultimately merge them into a single list of 55 unique low success rate fibers. We remove all targets observed with these fibers from the LSS catalogs. We also remove five additional fibers by hand, which are adjacent to low success rate fibers already identified in petal 5. This leads to removing targets from the contiguous range of fibers from 2675 to 2691.\footnote{Further work has identified a time-variable, non-linear, saturated feature in the 2D spectra that leads to the poor redshift performance, as discussed here: \url{https://github.com/desihub/desispec/issues/2193\#issuecomment-2004984304}.} Overall, we remove 60 fibers from further analysis. This removes 1.2\% of the total redshifts from the DR1 analysis.

\begin{figure*}
    \includegraphics[width=1.05\linewidth]{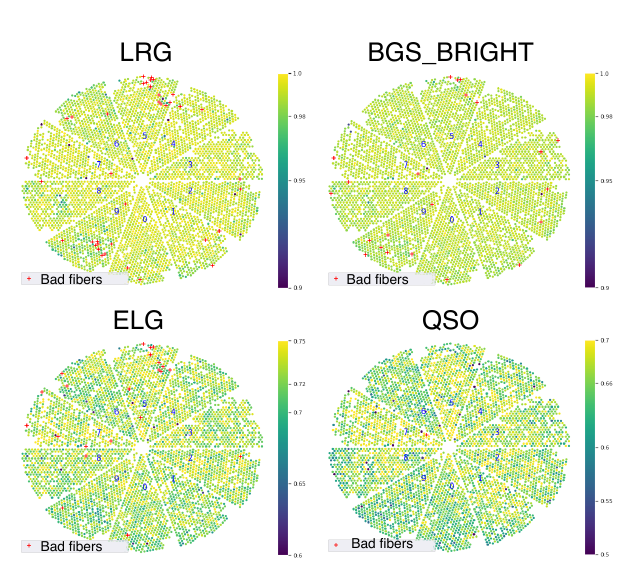}
    \caption{Success rate for all tracers on the focal plane, with low success rate fibers identified with red crosses. Targets on these fibers are removed from large-scale structure catalogs.
    \label{fig:badfiber_focalplane}}
\end{figure*}

\begin{figure*}
    \includegraphics[width= \linewidth]{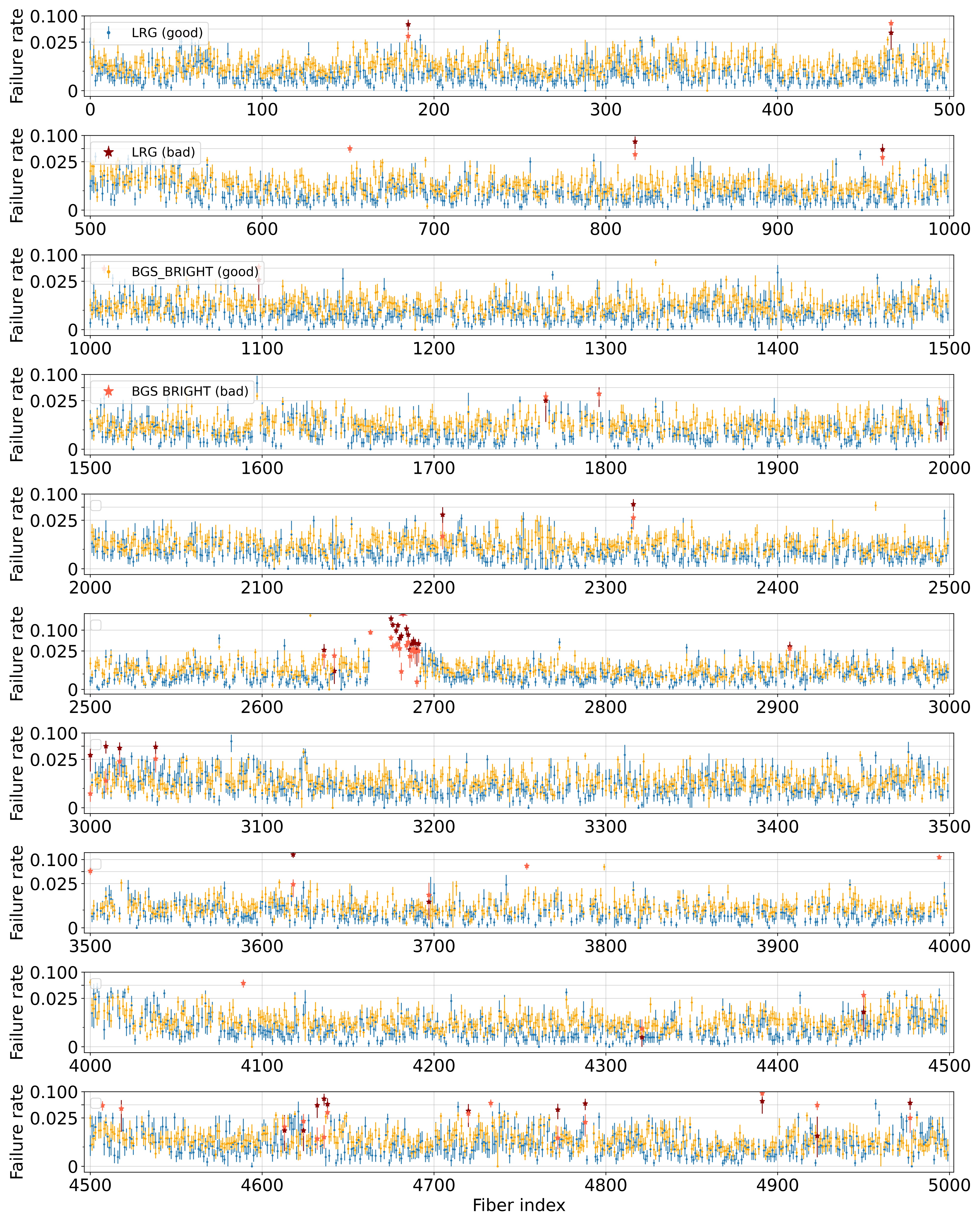}
    \caption{Per-fiber failure rate of LRG (blue) and BGS\_BRIGHT (orange), with low success rate fibers marked in dark red (LRG) and light red (BGS\_BRIGHT). We use a linear scale below failure rate of 0.025 and a logarithmic scale at higher failure rate. Plots are arranged by petal from 0 to 9 (top to bottom).
    \label{fig:badfiber_lrgbgs}}
\end{figure*}

\begin{figure*}
    \includegraphics[width=\linewidth]{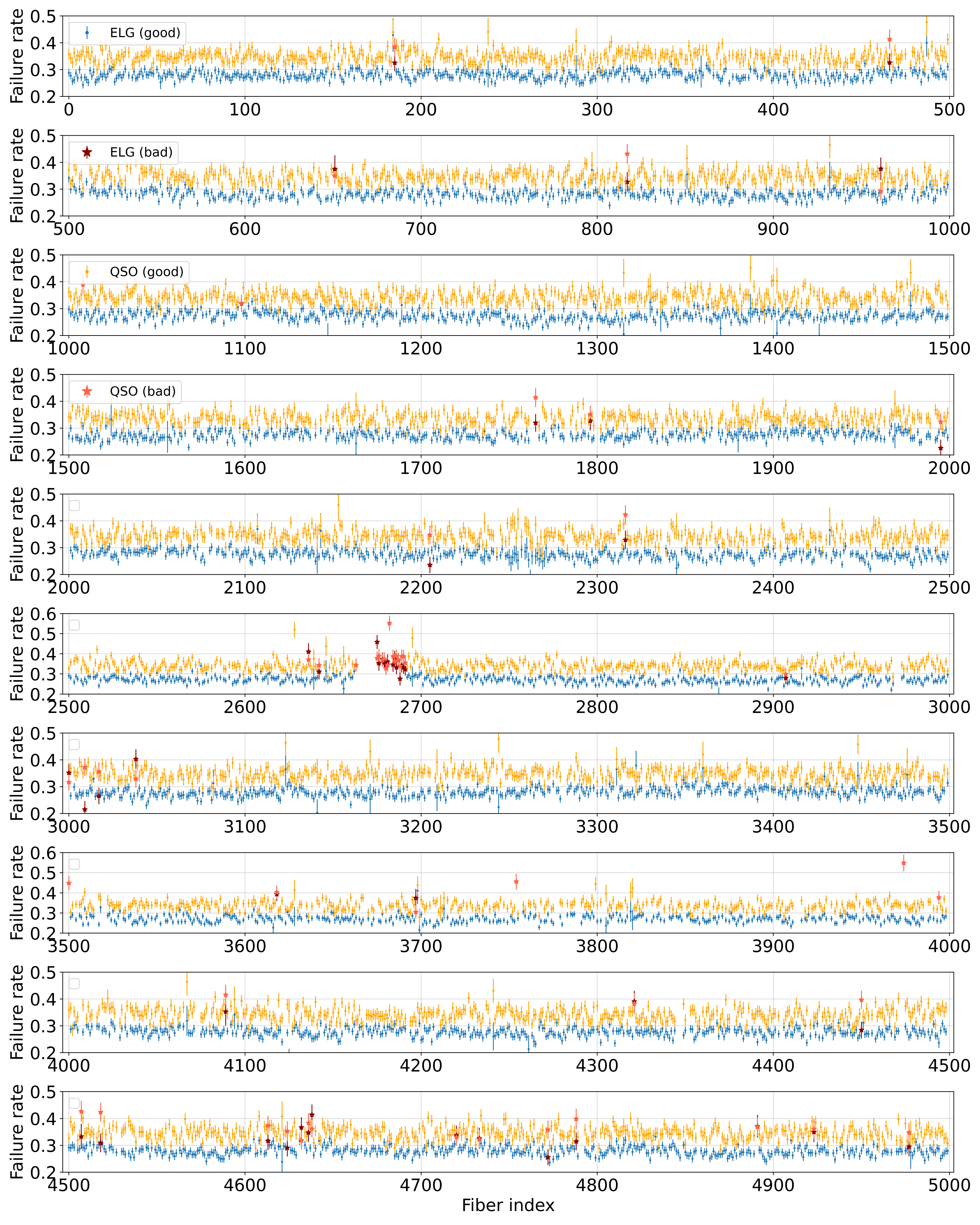}
    \caption{Per-fiber failure rate of ELG (blue) and QSO (orange), with low success rate fibers marked in dark red (ELG) and light red (QSO). Plots are arranged by petal from 0 to 9 (top to bottom).
    \label{fig:badfiber_elgqso}}
\end{figure*}

We show the position of these low success rate fibers on the focal plane in Fig.~\ref{fig:badfiber_focalplane}. We also unroll these plots, to show the fiber-by-fiber success rate in Figs.~\ref{fig:badfiber_lrgbgs} and~\ref{fig:badfiber_elgqso}. The union of the per-tracer low success rate fibers are shown in these plots, so the failure rate of the fibers is not necessarily higher than average, if the fiber was identified as a low success rate fiber in a different tracer.

By visual inspection of the $n(z)$ per fiber,\footnote{Plots and redshift distributions are available in the Zenodo repository linked in the Data Availability section.}
we find that several fibers in petal 2 have a significant upward fluctuation in the ELG $n(z)$ at $z = 1.5$: fibers 1261, 1269, 1295, 1296, and 1307. In these fibers, the  spurious $z = 1.5$ galaxies were observed between Oct 1, 2021, and the end of DR1; they are not restricted to observations at a particular time.
This is roughly equivalent to a 0.1\% impact in the catastrophic redshift failure rate.
The impact of the ELG catastrophic rate, and its robustness to variations
in the assumed rate, is explored in Figs.~9 and 10 and Section 5.3 in \cite{KP3s4-Yu}.

We show further exploration of the interaction between the redshift distribution and redshift success rate in Appendix~\ref{sec:meanz_vs_success_rate}. We find weak evidence for relationships between success rate deviation
and mean redshift for some of the tracers, but with differing signs. Due to the smallness of the observed shifts, we do not correct for these trends in DR1, but flag them for further study in future DESI data releases.

To robustly model the redshift success rate,
we first remove targets from low success rate fibers, and then fit the model described in Section~\ref{sec:zfail_wts}. However, identifying low success rate fibers
requires a success rate model to determine whether a given fiber is a 4$\sigma$ outlier. We therefore proceed iteratively, first fitting a model with a fiducial ``by-eye'' low success rate fiber list, then creating the systematic low success rate fiber list described in this section, and then fitting the redshift success rate model and creating $w_{\textrm{zfail}}$.

\section{Validation of the redshift failure weights}
\label{sec:validate_zfail_wts}

We test that applying the redshift failure weights leads to a uniform distribution of success rate within
the redshift ranges of interest, as well as in narrow redshift bins to make sure that we are not missing trends at any particular redshift. 

\subsection{Redshift failure trends in the fiducial redshift ranges}
In Fig.~\ref{fig:validate_overall_rate}, we test the uniformity of the success rate against TSNR2, using each tracer within the default redshift range: $0.4 < z < 1.1$ for LRG, $0.1 < z < 0.4$ for BGS (and additionally applying the $M_r < -21.5$ cut, which can only be applied to successful redshifts), $0.8 < z < 1.6$ for ELG, and $0.8 < z < 2.1$ for QSO.
This is in contrast to the fits, which use the redshift success rate defined over the entire redshift range (as we make no reference to redshift when defining the model and weights, since failed observations do not have a good redshift by definition).
We measure the significance of trends using the $\chi^2$ difference from unity in all bins, using a binomial error for each bin. As discussed in Sec.~\ref{sec:mocks}, this is not a good model for the covariance in general, as it neglects the cosmic variance contribution, but in the limit of a large redshift depth, the cosmic variance contribution is suppressed.
In this figure and all like it, we always split the data by imaging region (North and South) to ensure that our results are robust to variation in target properties between the different imaging surveys.

We show several extensions to these plots in Appendix~\ref{sec:validation_other_z_ranges}: success rate trends using the entire redshift range of each tracer (Fig.~\ref{fig:validate_overall_rate_full_z_range}); success rate trends for the extended quasar redshift range of $0.8 < z < 3.5$ (Fig.~\ref{fig:validate_overall_rate_full_z_range}); and success rate trends for BGS without the ``BRIGHT'' or absolute-magnitude cuts (Fig.~\ref{fig:validate_overall_rate_other_bgs}).

Similar plots for ELG are shown in Fig.~4 of \cite{KP3s4-Yu}. However, in those plots, there is no redshift limit applied to the definition of a successful redshift (not even the looser limits that we apply in Appendix~\ref{sec:validation_other_z_ranges} and Fig.~\ref{fig:validate_overall_rate_full_z_range}); this accounts for the slightly different
trends and $\chi^2$ seen in \cite{KP3s4-Yu}.

For LRG, ELG, and QSO in the South, there is a significant relationship between success rate and TSNR2 that is removed by applying the weights.
Note that the unweighted trends, while
statistically significant due to the large sample sizes and thus small errors, are quite small in magnitude, $<3\%$.
For BGS\_BRIGHT-21.5, there is no significant relationship, and applying the weights does not measurably change the $\chi^2$. In contrast, Fig.~\ref{fig:fit_quality}
shows a clear increase in BGS\_BRIGHT failure rate at low TSNR2.
This difference is driven by galaxies outside the fiducial redshift range, most likely at $z > 0.4$; referring to Fig.~\ref{fig:validate_overall_rate_other_bgs} in Appendix~\ref{sec:validation_other_z_ranges}
shows that BGS\_BRIGHT and BGS\_BRIGHT-21.5 have similarly mild success rate trends at low TSNR2.


\subsubsection{Anomalous trends in the quasars}
\label{sec:qso_anomaly}

\begin{figure*}
    \includegraphics[width=0.5\linewidth]{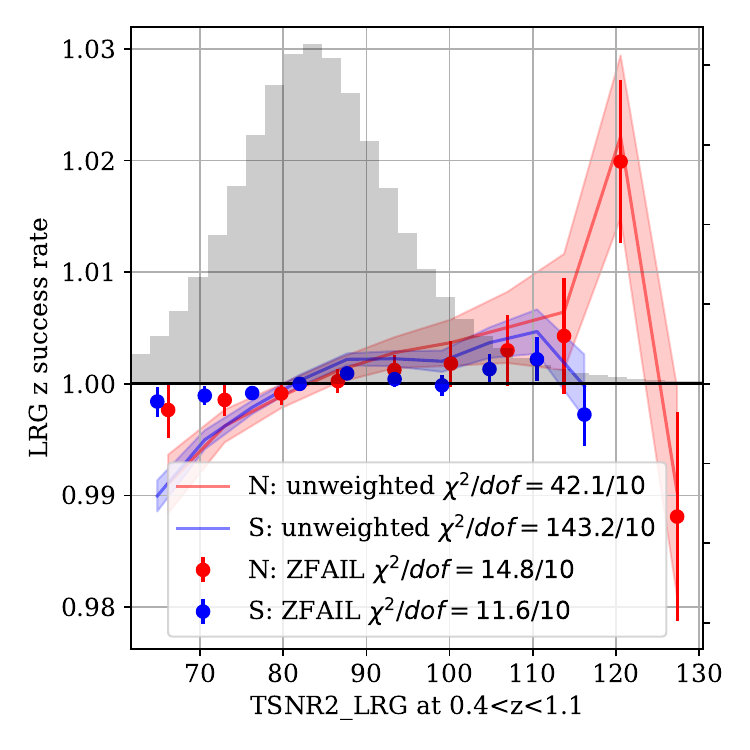}
     \includegraphics[width=0.5\linewidth]{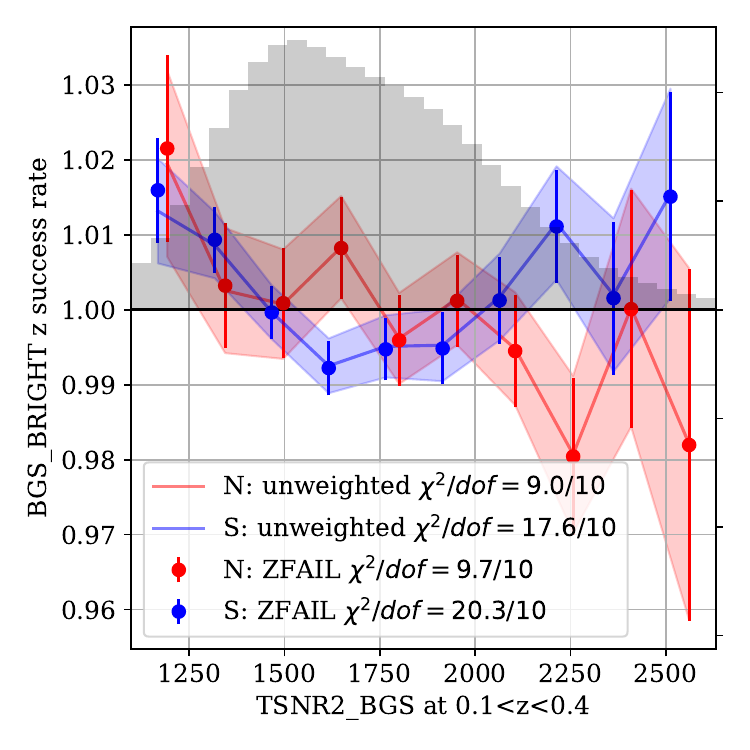}
     \includegraphics[width=0.5\linewidth]{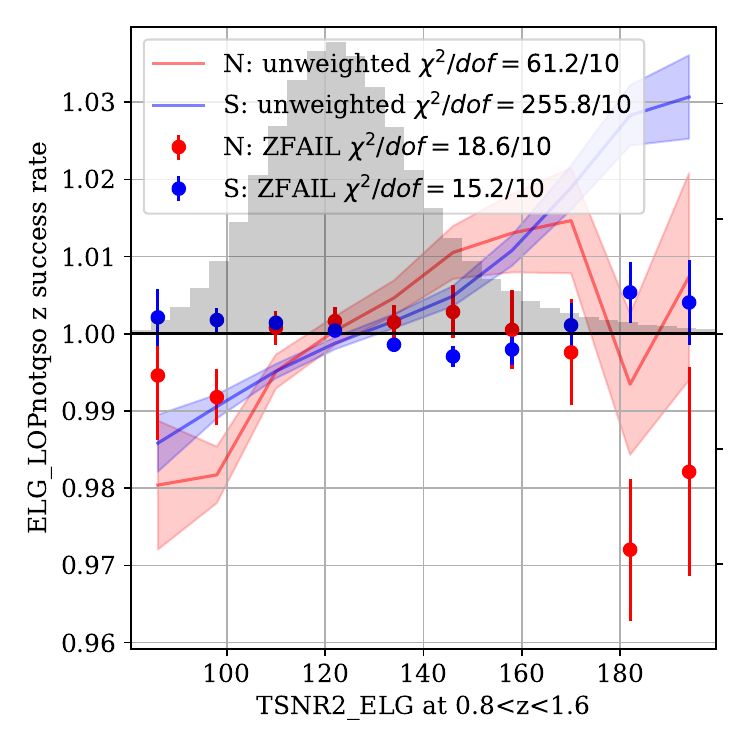}
     \includegraphics[width=0.5\linewidth]{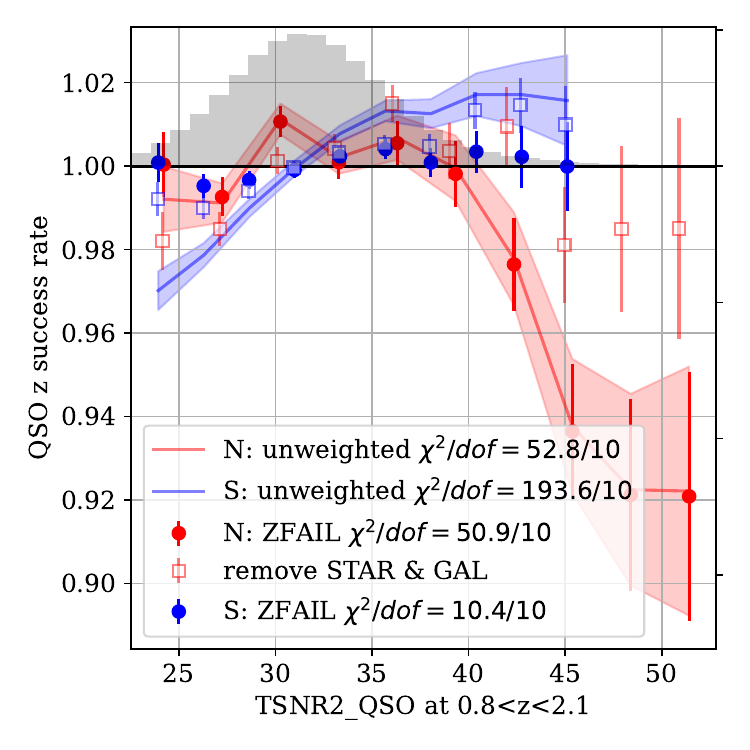}
    \caption{Validation of the redshift failure weights for LRG, BGS, ELG, and QSO in their fiducial redshift ranges. 
    Background histogram shows the distribution of TSNR2 for each tracer. Once redshift failure weights are applied (points with errorbars), there are no significant trends between TSNR2 and success rate, removing the significant trends seen without the weights (solid lines with shaded errorband). The $\chi^2$/dof are measured from the deviation between the measured TSNR2 trend and a flat success rate at unity in all TSNR2 bins. There remains a significant deviation
    from a flat success rate trend in QSO N. This is because in the North imaging, high-TSNR2 regions of the sky lie at lower Galactic latitude, where quasar targets are more likely to be spectroscopically classified as galaxies. If we remove galaxies and stars from the numerator and denominator of the quasar redshift success rate, the dip at high TSNR2\_QSO in the N is eliminated (light points). However, removing stars and galaxies is complicated due to quasar contamination in the galaxy classification and uncertainty in the classification of the residual ``unknown'' spectra: see Section~\ref{sec:qso_anomaly} for details.
    \label{fig:validate_overall_rate}}
\end{figure*}

QSO N shows a minimal relationship between TSNR2\_QSO and success rate at TSNR2\_QSO $< 40$, and then a significant 8\% drop in the success rate at high TSNR2\_QSO (Fig.~\ref{fig:validate_overall_rate}).
The $\chi^2$ indicates a significant trend that is not mitigated by the weights, since they
are unable to account for declining success rate with increasing TSNR2\_QSO.
The drop in TSNR2\_QSO contrasts with our intuitive expectation that as we increase the signal-to-noise, the success rate should increase. Using the classifications of \cite{Krolewski23}, we find that at TSNR2\_QSO $>40$ and in the northern hemisphere, 62.9\% of quasar targets are classified as quasars, 24.0\% are galaxies, 9.4\% are bad, and 3.7\% are stars. At TSNR2\_QSO $<40$, 65.1\% of quasar targets are quasars, 21.6\% are galaxies, 10.2\% are bad, and 3.1\% are stars.\footnote{We point out that in the North, a higher fraction of quasars targets are spectroscopic galaxies due to the worse PSF in the BASS $g$ and $r$ bands \cite{Chaussidon22}, compared to the 16\% galaxy in deep VI tiles from SV \cite{QSO.TS.Chaussidon.2023}. The larger PSF size makes it more difficult to distinguish stars and galaxies, leading to a higher galaxy contamination fraction.} The high TSNR2\_QSO quasar targets lie at lower Galactic latitude, although the only notable difference in their average imaging properties is a 20\% increase in Milky Way star density. It therefore appears that the drop in success rate at high TSNR2\_QSO is due to a targeting fluctuation, not a spectroscopic fluctuation: these targets happen to lie at lower Galactic latitude where a higher fraction of quasar targets are spectroscopic galaxies.

If we re-define the quasar success rate so that spectroscopic
stars and galaxies are counted as neither successes nor failures,  the anomalous trend in the North disappears (lower left panel of Fig.~\ref{fig:validate_overall_rate}).  While the relative success rate is still $<1$ for the three highest TSNR2\_QSO bins at TSNR2\_QSO $>45$, none of these deviations are statistically significant.
This supports our argument that the anomalous trend
at high TSNR2\_QSO is due to a chance alignment
between TSNR2\_QSO and regions at low Galactic latitude with a higher fraction of galaxies in the quasar targets. 

This result suggests that we should use the classifications of \cite{Krolewski23} and exclude successfully
classified stars and galaxies from both the numerator and the denominator of the quasar success rate.
Redshift success rate should be defined as the fraction of all true quasars that are successfully classified as quasars by the DESI pipeline. While the stars and galaxies are quasar targets, if they can be confidently classified as stars or galaxies, they are clearly not true quasars. However, there is a problem with this proposal. The galaxy classification of \cite{Krolewski23} is only 90\% pure; the remaining 10\% are almost all true quasars (using the deep VI data as truth).
Furthermore, it is not clear that all of the ``unclassified'' spectra are true quasars, rather than stars or galaxies. Of the 40\% of the unclassified spectra with a VI classification, 54\% are quasars, 27\% are stars, and 19\% are galaxies.
When multiplying by the overall classified fraction (67\% quasars, 16\% galaxies, 12\% unclassified, 5\% stars in the South region), this means that there are nearly as many true quasars classified as galaxies (1.6\% of the total) as those classified as ``unknown'' (2.6\%). Hence, without a higher-purity classification of the galaxies, and a better understanding of the unclassified spectra (7\% of even the deep VI spectra cannot be confidently classified \cite{QSO.TS.Chaussidon.2023}), it is difficult to properly modify the quasar success rate by removing stars and galaxies from the calculation.
As a result, we retain the simple classification where all non-quasar targets are designated ``failures'' in the success rate calculation, and defer re-considerations of this definition to future DESI data releases.
These choices only affect the tracer QSO clustering and do not affect the Ly$\alpha$ forest, which depends
on the observed Ly$\alpha$ absorption along the line of sight to each quasar.



As a further validation of the quasar
spectroscopic success rate definition,
we also measure the success rate for two subsets
of the quasar target sample with higher quasar purity, defined in \cite{Krolewski23}.
The main quasar target sample has relatively low purity, with only 60\% of targets spectroscopically classified as quasars, 18\% unclassified, 16\% galaxies, and 5\% stars (using the \cite{Krolewski23} classifications on the Y1 data rather than the SV or VI measurements). We also consider a ``no star'' sample, where we use the photometric star probability of \cite{Duncan22}, removing any quasar target with $P_{\rm star} > 0.01$.
This sample is 66\% quasars, 16\% unclassified, 17\% galaxies, and 1\% stars. We also consider a ``high purity'' sample, where we remove roughly the faintest half of the quasar sample using cuts in W2 and $r$ magnitudes.
This sample has half the number density of the main quasar targets, but is 90\% quasars, 4\% unclassified, 4\% galaxies, and 2\% stars.

We then measure the failure rate as a function of TSNR2\_QSO for the main quasar sample and the ``no star'' and ``high purity'' subsets.
We find consistent behavior between the 
``full,'' ``no star'' and ``high purity'' subsamples. This allows us to
conclude that the measured quasar redshift success rate is representative of the ``true'' redshift success rate, defined as the ratio of spectroscopically-classified quasars to true quasars within the target sample, and is robust to the presence of interlopers (stars and galaxies) within the quasar sample.
We provide more details in Appendix~\ref{sec:qso_full_nostar_high_purity}.



\subsection{Redshift failure trends in narrow redshift bins}

We divide each sample into redshift bin subsets and test whether the redshift failure weights
successfully remove any redshift failure trends. We use bins of $\Delta z = 0.1$ for LRG and BGS, and $\Delta z = 0.3$ for QSO (except for the final QSO bin, which extends from $1.7 < z < 2.1$). The same analysis for ELG is shown  in Fig.~2 of \cite{KP3s4-Yu}. That paper finds no significant redshift-dependent trends for ELG once the redshift failure weights are applied.

\subsubsection{LRG}

\begin{figure*}
    \includegraphics[width=\linewidth]{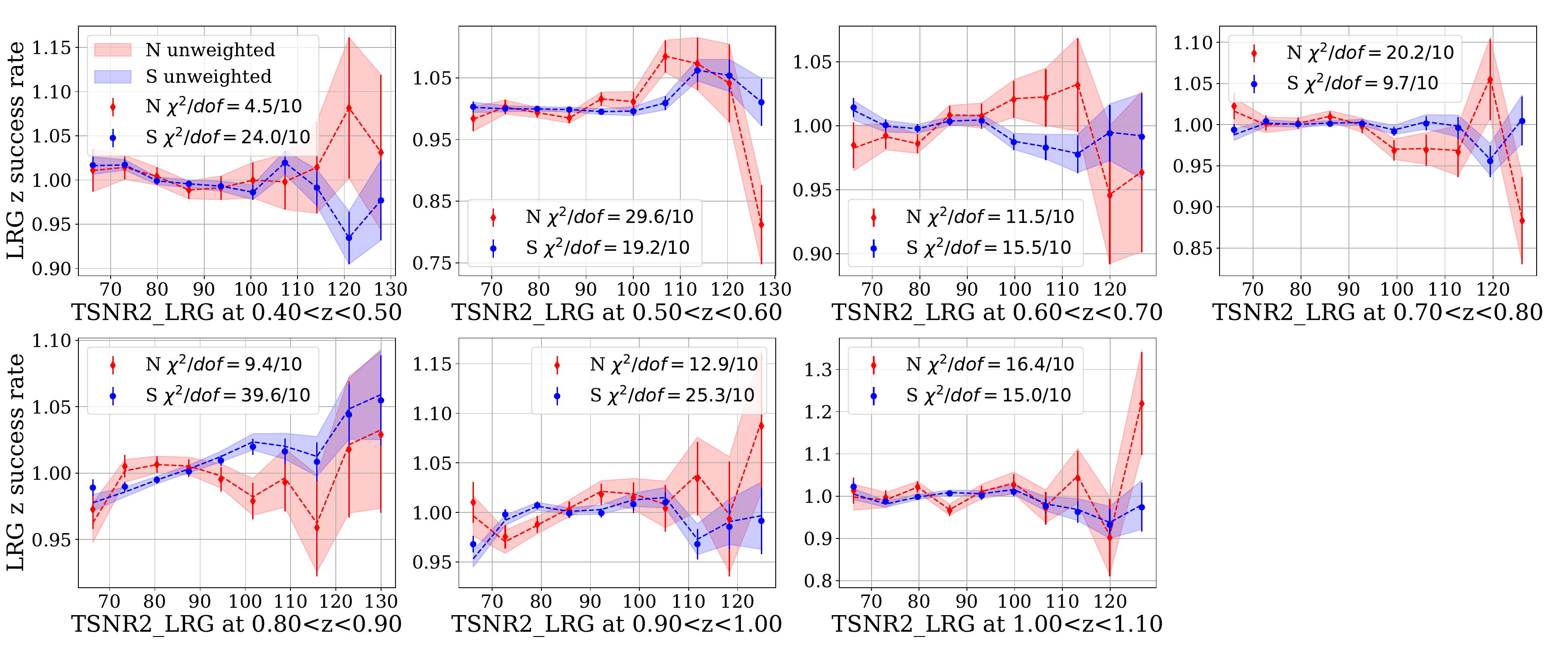}
    \caption{Relationship between success rate and TSNR2 in $\Delta z = 0.1$ bins for LRG, both without redshift failure weights (lines and shaded regions) and with them (points with errorbars). The $\chi^2$ uses binomial errors in each TSNR2 bin and requires comparison to mocks to interpret the significance, due to the contribution of cosmic variance to the redshift failure errorbar within each bin.
    \label{fig:validate_by_redshift_bins_lrg}} 
\end{figure*}

The trends for LRG are shown in Fig.~\ref{fig:validate_by_redshift_bins_lrg}. The largest outlying trend is in the South region at $0.8 < z < 0.9$, which shows a  trend of increasing success rate as TSNR2 increases.
If the $\chi^2$ statistic follows a $\chi^2$ distribution with the specified number of degrees of freedom, this trend is significant at the equivalent of 4.1 Gaussian $\sigma$.
The only other trend significant at $>3\sigma$ is in the North region at $0.5 < z < 0.6$, which is significant at 3.1$\sigma$ and is largely driven by the point at highest TSNR2. The total $\chi^2$ is 104.5 in N and 148.3 in S, across 70 bins each. This is significant at a Gaussian equivalent of 2.6$\sigma$ in N and 5.1$\sigma$ in S.

\begin{figure*}
    \includegraphics[width=\linewidth]{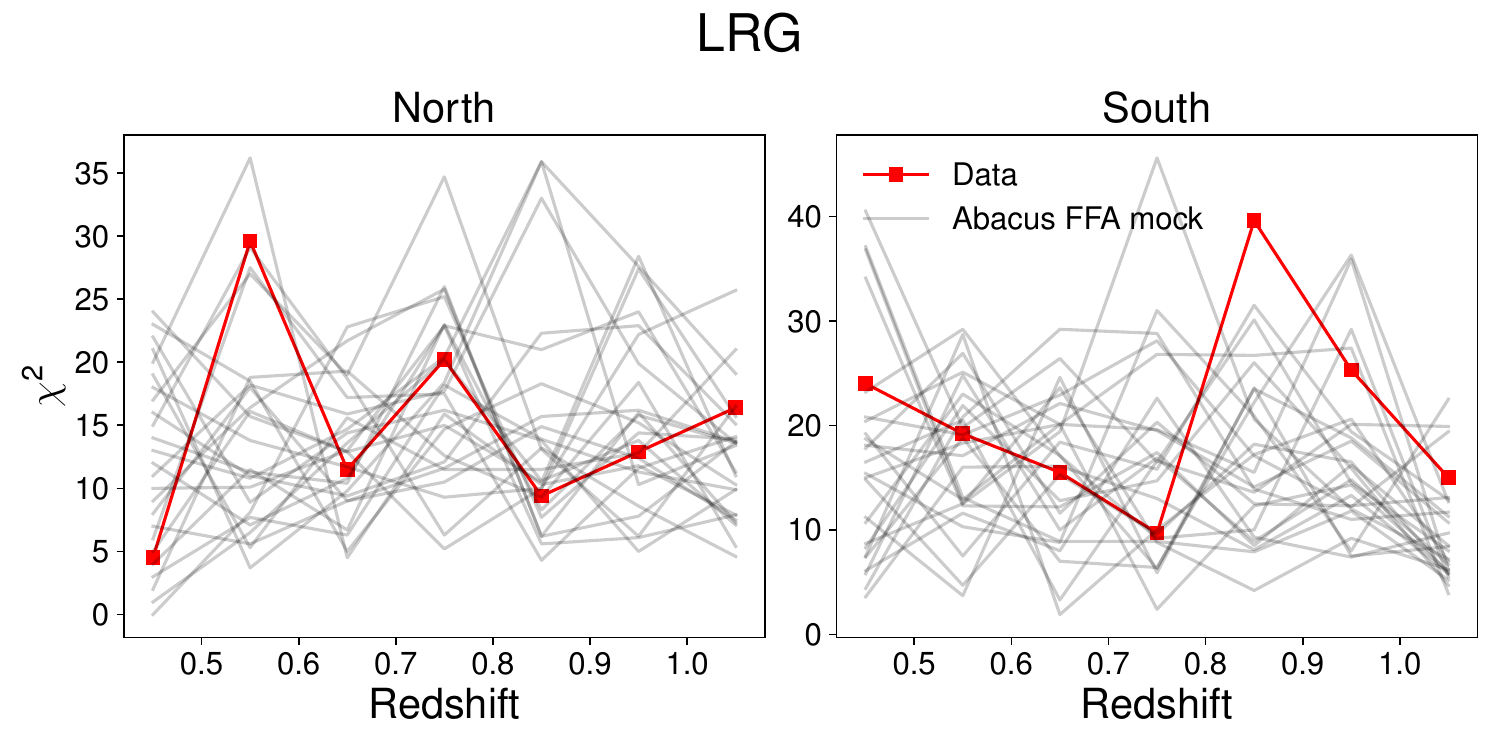}
    \caption{Distribution of $\chi^2$ within each redshift bin for the 25 Abacus FFA mocks (gray lines) and the data (red) for LRG. The data lies within the fluctuations of the mocks, except for the $0.8 < z < 0.9$ bin in the South.
    \label{fig:data_vs_abacus_ffa_lrg}}
\end{figure*}

The $\chi^2$ calculation uses binomial errors, but as discussed above, the contribution from cosmic variance may be non-negligible, particularly in these relatively narrow bins.
As a result, we also compute $\chi^2$ for each redshift bin and imaging region using the Abacus FFA mocks (Fig.~\ref{fig:data_vs_abacus_ffa_lrg}). We find that the mock mean and standard deviation (summing over all seven redshift bins) $\chi^2 = 110 \pm 23$ in the North and $96 \pm 13$ in the South. Thus we estimate that the North data is perfectly consistent with the mocks (with $p = 0.56$ since 14 of the 25 mocks have higher $\chi^2$ than data) while the South is roughly a 4.0$\sigma$ outlier (none of the mocks have higher $\chi^2$ than data). The most outlying bin in the South is the $0.8 < z < 0.9$ bin, which is a 3.2$\sigma$ outlier based on the mock mean and standard deviation; none of the mocks have a higher $\chi^2$ than the data.
The combination of North and South has $p = 0.08$, so it is possible that this is simply a fluctuation in the $0.8 < z < 0.9$ bin driving the high $\chi^2$. However, given that the shape of the trend matches what we expect from issues in the redshift failure model, we test the application of an extra set of weights to the South LRG at $0.8 < z < 0.9$ which explicitly nulls this trend. We describe the impact of these weights on the observed power spectrum multipoles in Section~\ref{sec:clustering}.

\subsubsection{BGS}

\begin{figure*}
    \includegraphics[width=\linewidth]{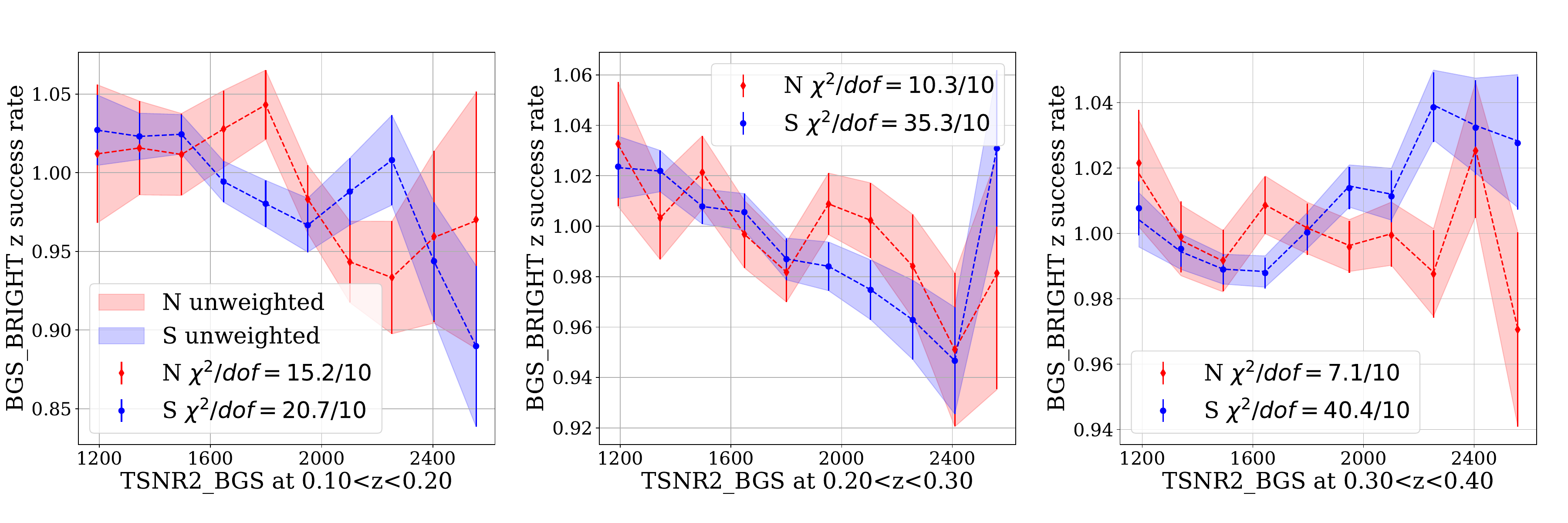}
    \caption{Same as Fig.~\ref{fig:validate_by_redshift_bins_lrg}, but for BGS\_BRIGHT-21.5.
    \label{fig:validate_by_redshift_bins_bgs}} 
\end{figure*}

\begin{figure*}
    \includegraphics[width=\linewidth]{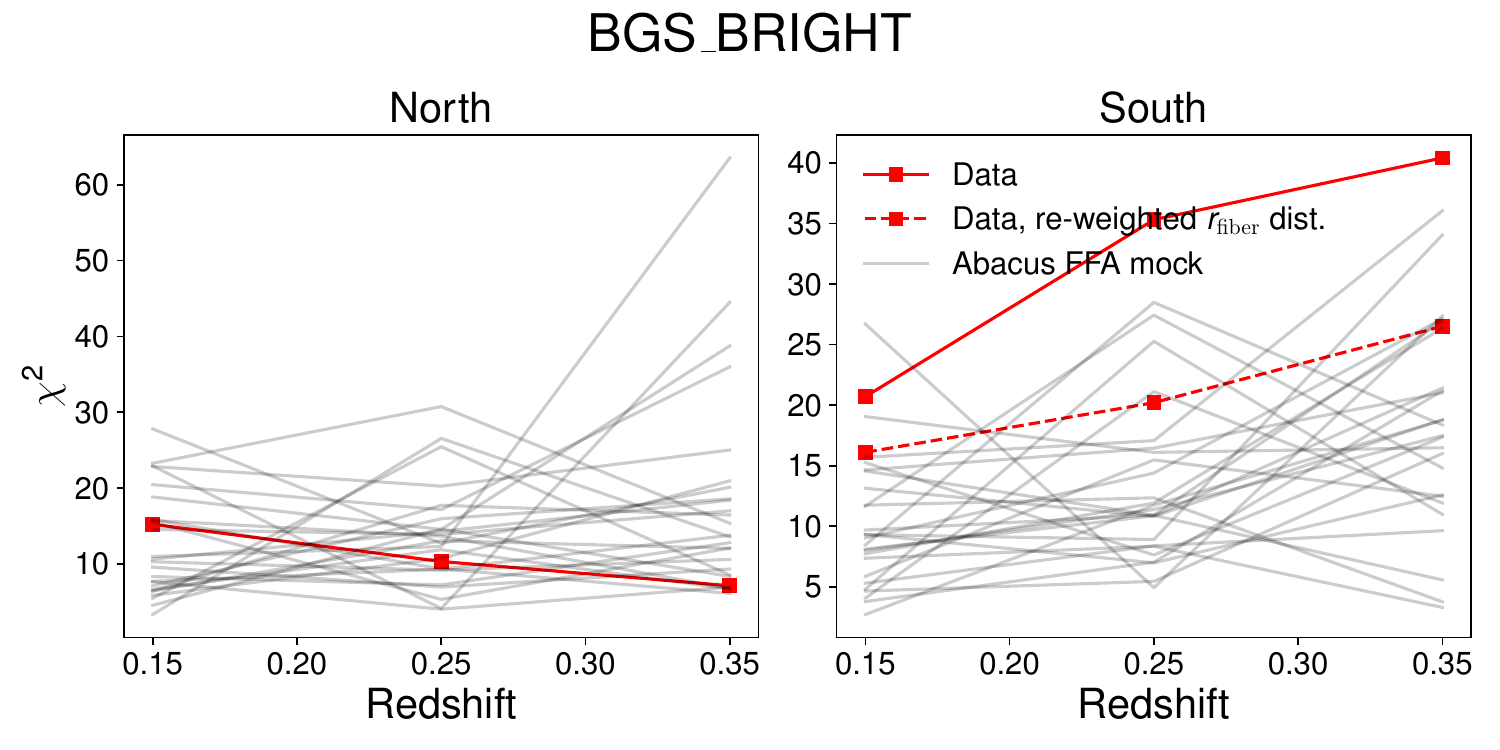}
    \caption{Same as Fig.~\ref{fig:data_vs_abacus_ffa_lrg}, but for BGS\_BRIGHT-21.5. The data lies within the fluctuations of the mocks for North, but  not for South.
    This is due to slightly fainter $r_{\textrm{fiber}}$ targets in the high TSNR2\_BGS bins. Re-weighting each TSNR2\_BGS bin to match the $r_{\textrm{fiber}}$ distribution in the first TSNR2\_BGS bin greatly reduces the discrepancy between data and mocks (dashed red curve).
    \label{fig:data_vs_abacus_ffa_bgs}}
\end{figure*}

For BGS, we find significant trends in the South, especially in the $0.2 < z < 0.3$ and $0.3 < z < 0.4$ bins (Fig.~\ref{fig:validate_by_redshift_bins_bgs}, where none of the mocks have
higher $\chi^2$ than observed in data (Fig.~\ref{fig:data_vs_abacus_ffa_bgs}). We find $\chi^2$ in data is high by 3.4$\sigma$ at $0.2 < z < 0.3$ and 2.2$\sigma$ at $0.3 < z < 0.4$, and the combination in the South is high by 4.2$\sigma$ (all compared to mocks). However, the shapes of the observed trends are puzzling, as the most significant trend is the declining trend with increasing TSNR2\_BGS in the $0.2 < z < 0.3$ bin, rather than the more expected increasing trend.

\begin{figure*}
    \includegraphics[width=\linewidth]{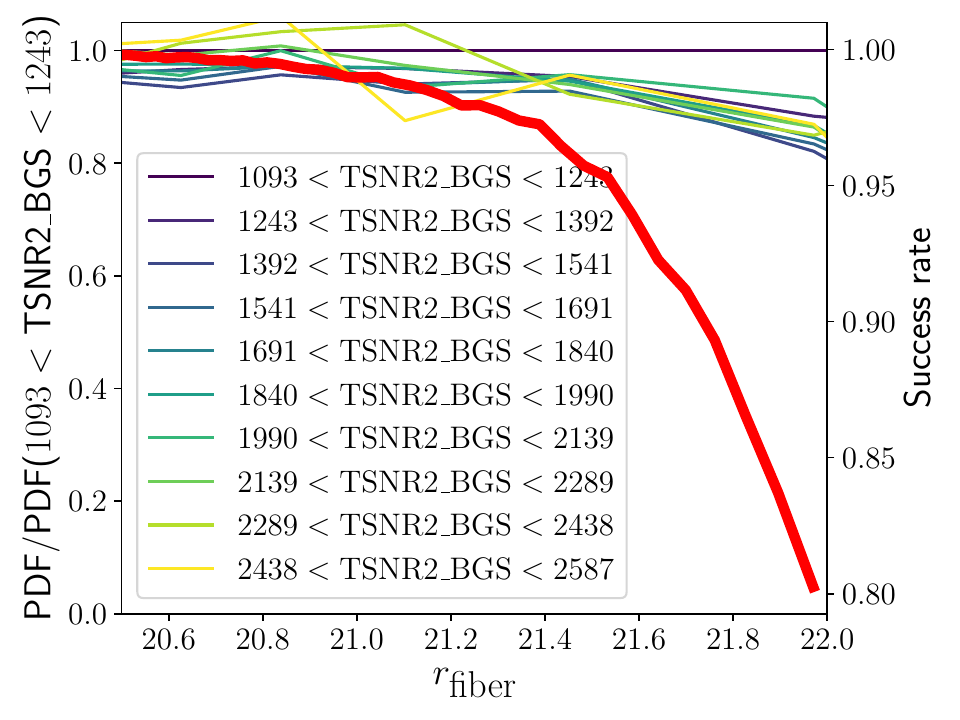}
    \caption{PDF of $r_{\textrm{fiber}}$ in bins of TSNR2\_BGS, normalized by the PDF in the lowest TSNR2\_BGS bin. The right-hand $y$-axis shows the success rate for BGS\_BRIGHT (without the absolute magnitude cut); the success rate is plotted in the thick red line as a function of $r_{\textrm{fiber}}$.
    \label{fig:bgs_bright_fiberflux_r_distribution}}
\end{figure*}

This trend is likely due to slight differences in the $r_{\textrm{fiber}}$ distribution in the different TSNR2\_BGS bins, i.e.\ again due to targeting fluctuations that happen to align with regions with higher TSNR2\_BGS.
In Fig.~\ref{fig:bgs_bright_fiberflux_r_distribution}, we show that targets in the higher TSNR2\_BGS bins are generally slightly fainter than those in the lower TSNR2\_BGS bins. Combined with the steep drop in spectroscopic success as a function of $r_{\textrm{fiber}}$, this leads to a lower success rate in the higher TSNR2\_BGS bins. We therefore divide the BGS targets into 5 bins in $r_{\textrm{fiber}}$, matching the 5 bins used in the flux-dependent redshift success fit.
We then re-weight each TSNR2\_BGS bin to match the $r_{\textrm{fiber}}$ distribution of the first bin, and measure $\chi^2$ for the re-weighted sample. We find that $\chi^2$ (with 10 degrees of fredom each) drops in the $0.1 < z < 0.2$ bin from 19.8 to 16.1, in the $0.2 < z < 0.3$ bin from 34.4 to 20.2, and in the $0.3 < z < 0.4$ bin from 35.5 to 26.5. This drops the data from a 3.4$\sigma$ deviation to a 1.1$\sigma$ deviation at $0.2 < z < 0.3$, from 2.2$\sigma$ to 1.0$\sigma$ at $0.3 < z < 0.4$, and from 4.2$\sigma$ to 1.8$\sigma$ considering the combination of all three bins. This implies that the TSNR2\_BGS trends are driven by the slight change in $r_{\textrm{fiber}}$ distribution with TSNR2\_BGS, and are not indicative of underlying spectroscopic uniformity issues.

\subsubsection{QSO}

\begin{figure*}
    \includegraphics[width=\linewidth]{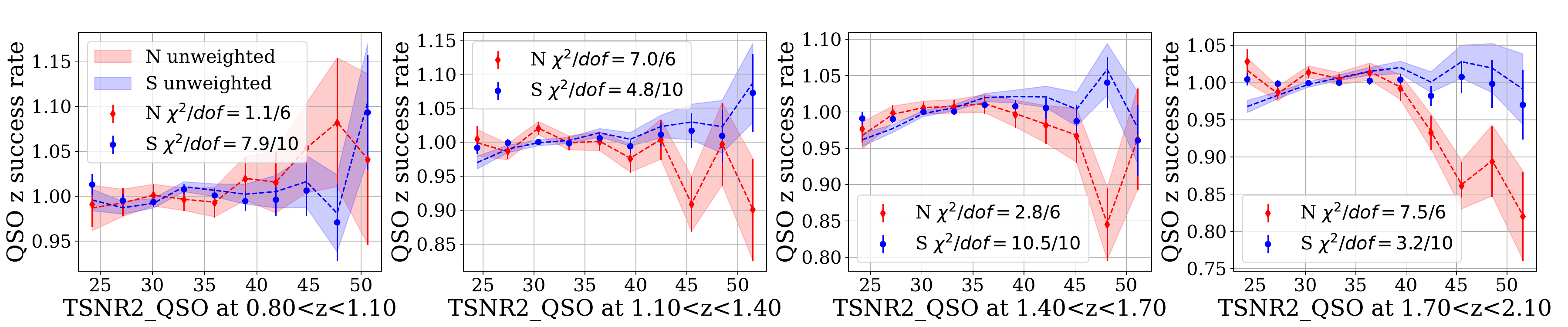 }
    \caption{Same as Fig.~\ref{fig:validate_by_redshift_bins_lrg}, but for QSO. While we show the success rate at TSNR2\_QSO $>40$ in the North, we do not include those points in the $\chi^2$ calculation, since they are affected by the targeting fluctuation leading to more spectroscopic galaxies in regions of high TSNR2\_QSO (lower right panel of Fig.~\ref{fig:validate_overall_rate}).
    \label{fig:validate_by_redshift_bins_qso}}
\end{figure*}

\begin{figure*}
    \includegraphics[width=\linewidth]{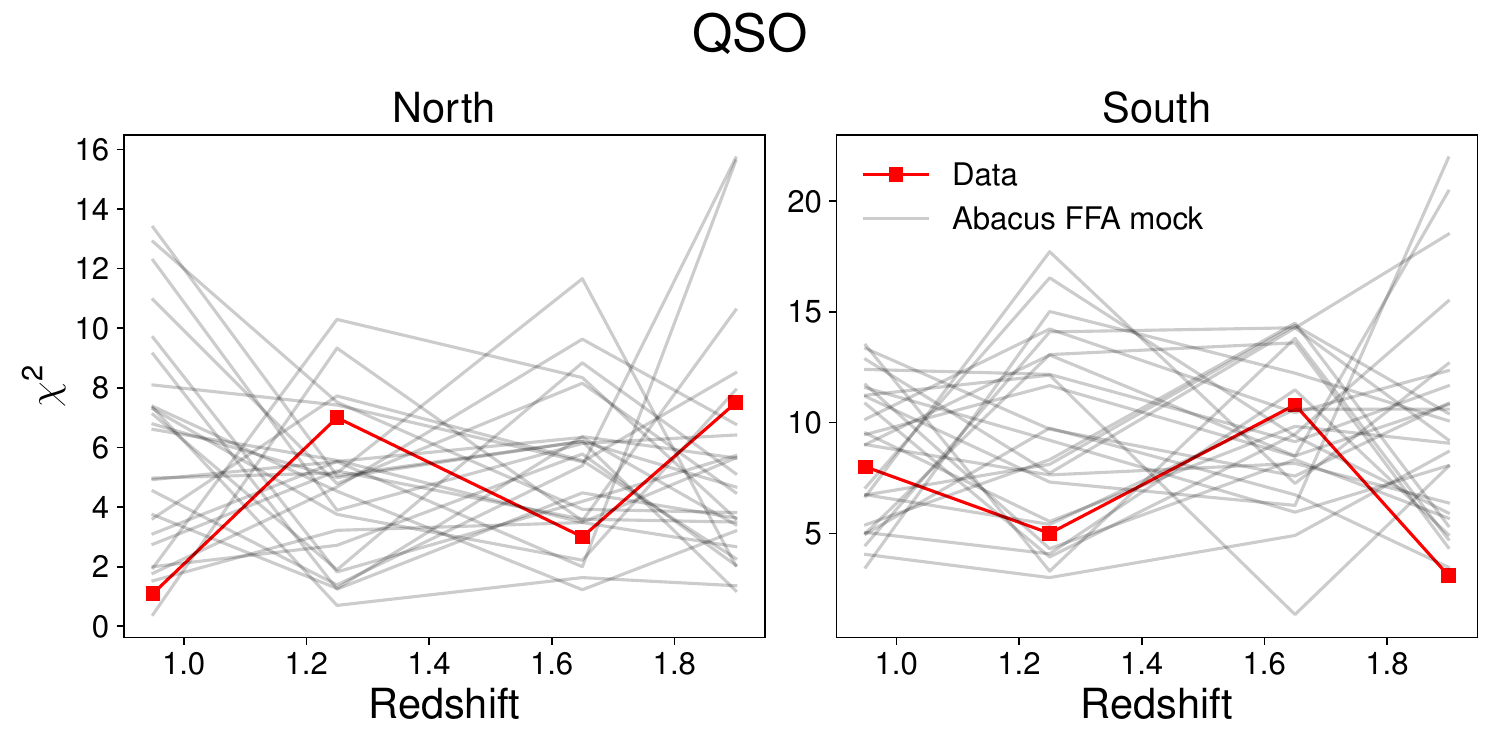}
    \caption{Same as Fig.~\ref{fig:data_vs_abacus_ffa_lrg}, but for QSO.
    \label{fig:data_vs_abacus_ffa_qso}}
\end{figure*}

We perform the same test for QSO as for LRG and BGS (Fig.~\ref{fig:validate_by_redshift_bins_qso}). Due to the fluctuation in the targeting sample leading to more galaxies at higher TSNR2\_QSO in the North (which has a particularly strong impact on the success rates in the $1.7 < z < 2.1$ bin), we only use the first six points with TSNR2\_QSO $< 40$ when computing $\chi^2$ in the North. Summing over the redshift bins, we find a total significance of $\chi^2 = 21.6 \pm 5.3$ in North and $37.8 \pm 7.8$ in South. The data values are very consistent with the mocks in all redshift bins (Fig.~\ref{fig:data_vs_abacus_ffa_qso}). Summing over all redshift bins in the data, $\chi^2 = 18.4$ (24 dof) and 26.4 (40 dof) in North and South, with $p$ values of 0.68 and 0.88, respectively.

\subsection{Trends with focal plane radius}

Even after applying the redshift failure weights,
we find a residual significant, though sub-percent, relationship between redshift
success rate and the target's position on the focal plane.
We show the relationship between redshift success
and focal plane radius on each of the ten petals for LRG (where the trends are strongest) in Fig.~\ref{fig:focal_plane_trends_lrg}.
The trends are weaker for other tracers, and we show them in Appendix~\ref{sec:focal_plane_trends}.
In these figures, errorbars are estimated by jackknife resampling, sequentially leaving out all galaxies within each nside=32 HEALPix pixel.

Trends are generally much more significant when defined using the entire redshift range of a tracer rather than the clustering redshift range.
These trends are only mildly reduced if we fit a linear relationship between focal plane radius and success in each petal and apply it as a weight.

Using the entire redshift range, we find significant trends for LRG in all petals except petal 7. Applying linear focal plane radius weights reduces the significance of all trends, but the $\chi^2$ are still too high for all petals except petal 2. 

For LRG, we attempted to model these focal plane variations (and additional angular variations apparent in Fig.~\ref{fig:big_focal_plane}) with Zernike polynomials defined on the entire focal plane. However, we found that it took a very large number of polynomials to even roughly fit the observed trends.
Instead, we consider the impact of these trends on clustering statistics by applying a set of weights that fully nulls all fiber-to-fiber variation in the success rate, $\eta_{\textrm{zfail}}$, defined similarly for ELG in \cite{KP3s4-Yu}. These ``focal plane weights'' 
 are prone to overfitting, since they will null the Poisson fluctuations to the per-fiber success rate as well as the true fluctuations.
We
measure the impact of the focal plane weights on two-point clustering
statistics in Sec.~\ref{sec:clustering} to provide an upper bound for the impact of these residual
variations in spectroscopic success rate on the focal plane.

\begin{figure*}
    \includegraphics[width=0.8\linewidth]{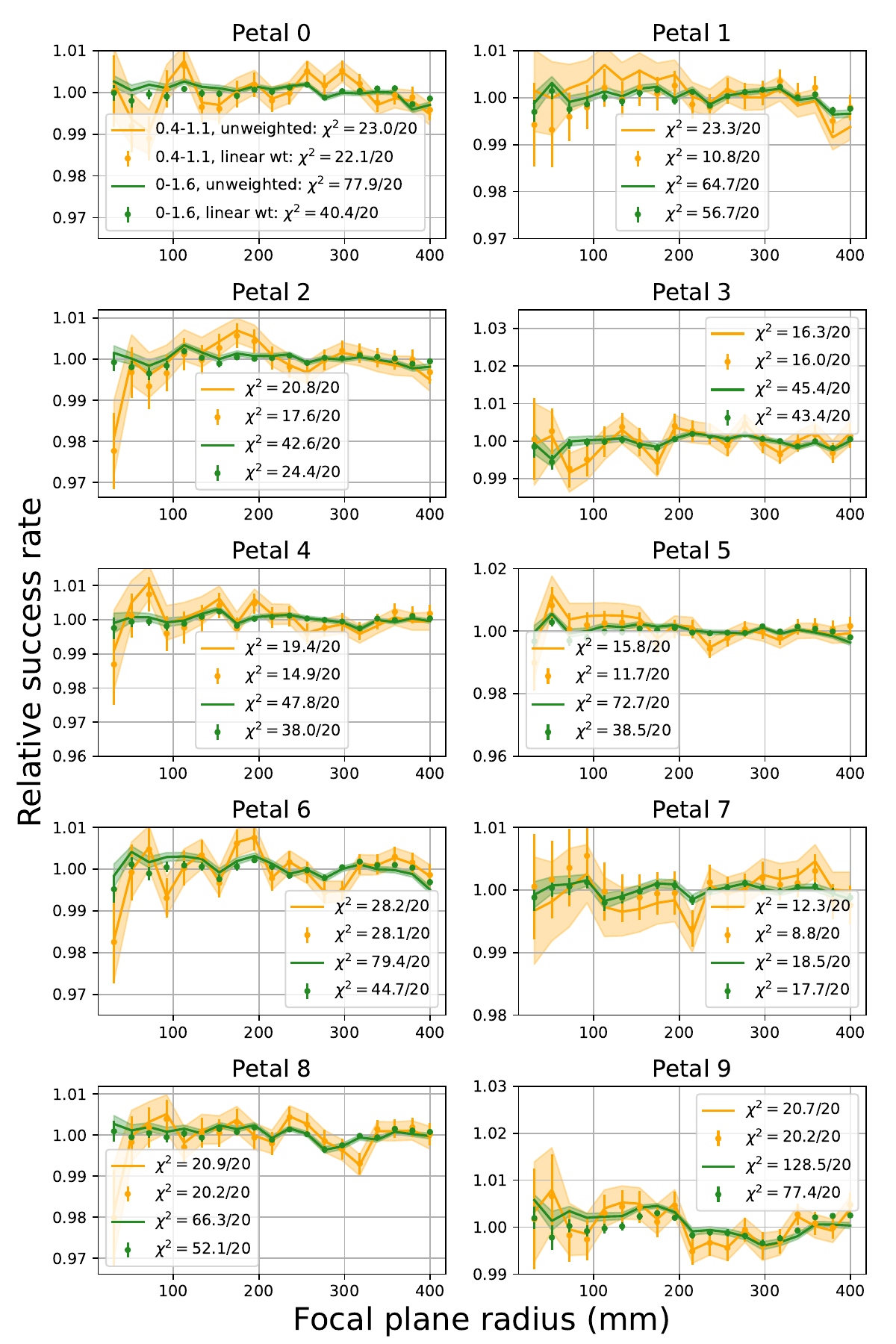}
    \caption{Redshift success rate vs.\ focal plane radius for each of the ten petals for LRG, using both the clustering redshift range $0.4 < z < 1.1$ (yellow) and the entire redshift range, $0 < z < 1.6$ (green). 
    Lines weight by the default redshift failure weight $w_{\textrm{zfail}}$, while points apply an additional linear focal plane radius weight to each petal (``linear wt''), which reduces the significance of trends but does not eliminate them.
    \label{fig:focal_plane_trends_lrg}}
\end{figure*}


\subsection{Time-dependent success rate}

\begin{figure*}
    \includegraphics[width=1.0\linewidth]{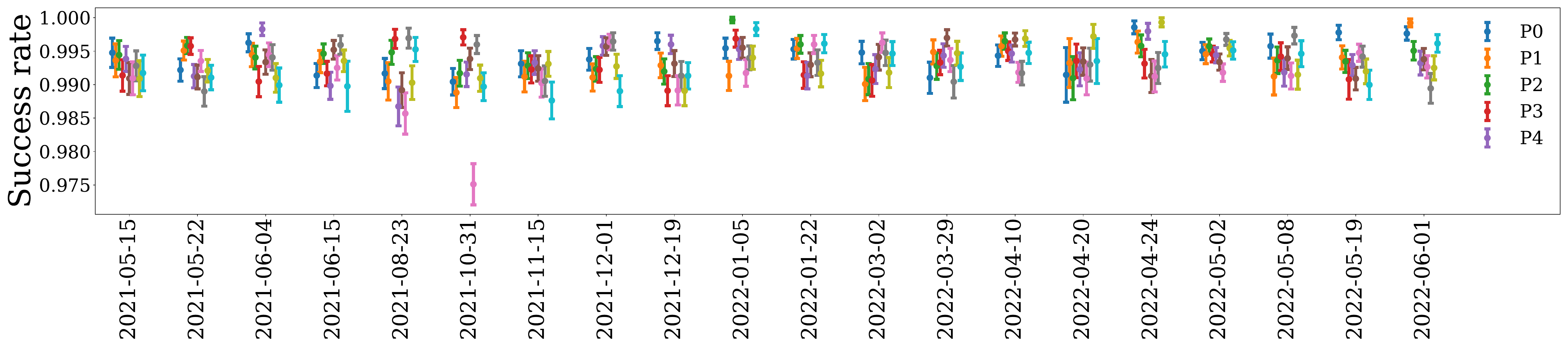}
    \includegraphics[width=1.0\linewidth]{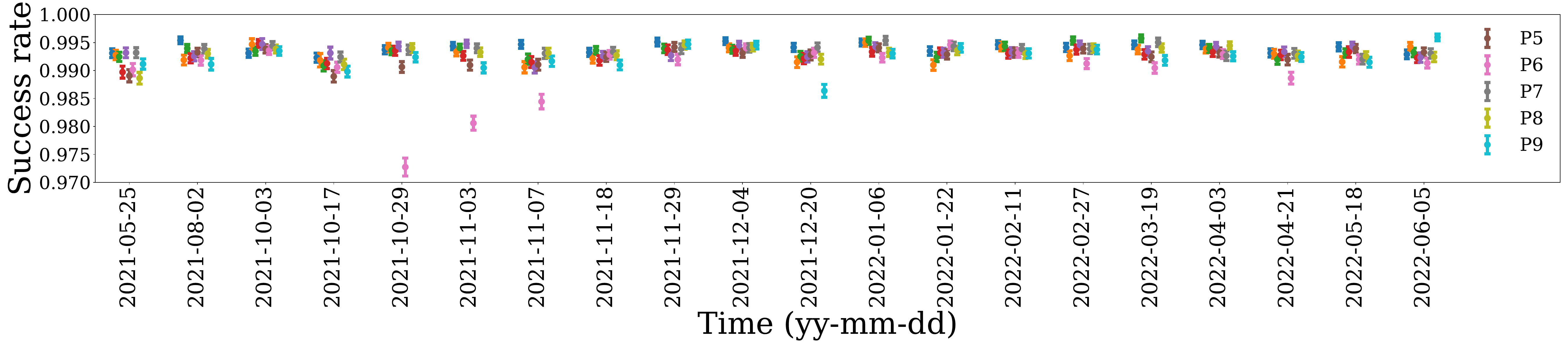}
    \caption{$z_{\textrm{fail}}$-weighted redshift success rate as a function of time for LRG in the North (top) and South (bottom). The data is also split by petal (abbreviated with P in the legend). We use a preliminary version of the catalogs to make this figure (internally labelled ``v0.6'') to show the dip in success rate in petal 9 on December 12, 2021, which was fixed
    in the final catalogs.
    \label{fig:time_dependent_success_lrg}}
\end{figure*}

We split the data into 20 bins of approximately equal sample size
using the ``LASTNIGHT'' column indicating the last night that a particular target was observed. As discussed above, some targets that are re-observed may be observed on multiple nights.
We show the dependence of LRG redshift failure weighted success rate on petal and time in Fig.~\ref{fig:time_dependent_success_lrg}.

\begin{figure*}
\centering
    \includegraphics[width=0.8\linewidth]{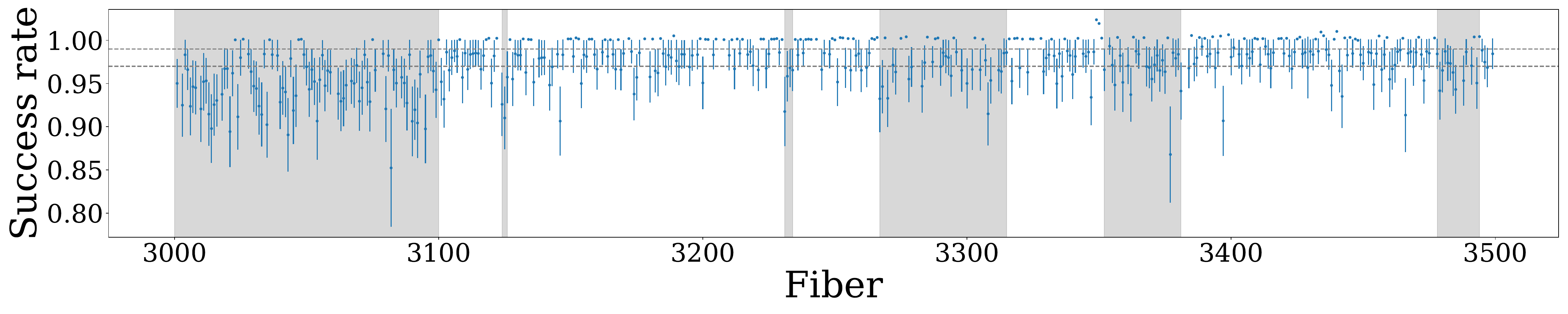}
    \includegraphics[width=0.8\linewidth]{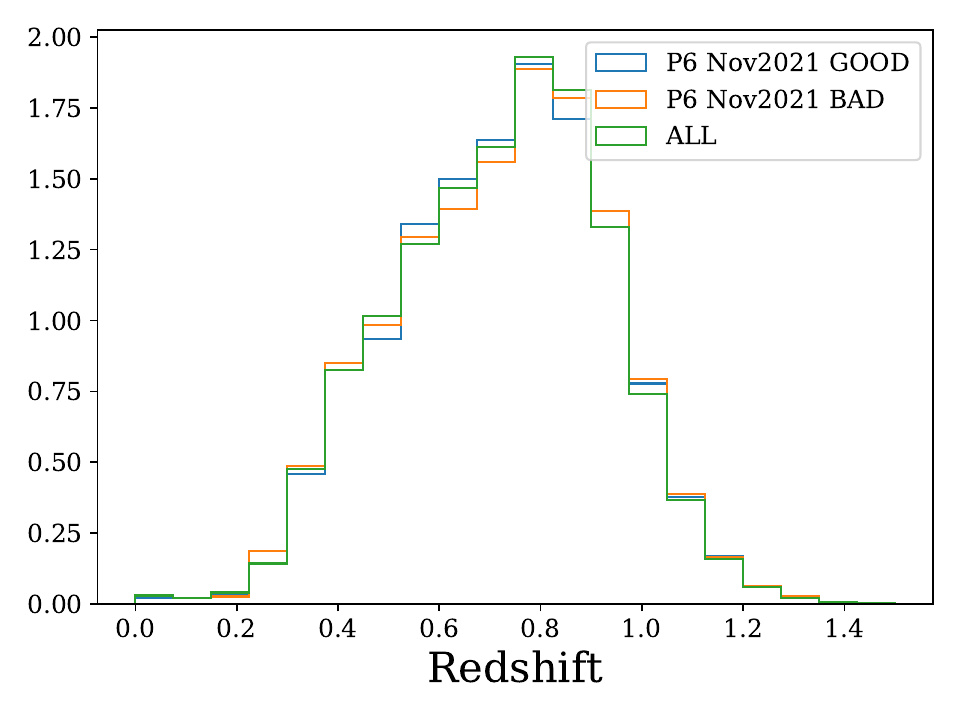}
    \caption{\textit{Top:} Per-fiber LRG success rate for petal 6 observed between October 26, 2021 and November 6, 2021 (affected by the $Z$6 amplifier instability). The two horizontal lines
    show the overall LRG success rate and the average petal 6 success rate in this time period. The shaded regions identify contiguous blocks of fibers with clearly lower success rates; the success rate elsewhere is 98.6\%, still slightly low compared to the other (normal) time-dependent fluctuations. \textit{Bottom:} Redshift distribution for both the shaded (``bad'') and unshaded (``good'') petal 6 redshifts within the $Z$6 amplitifier instability time period, and also compared to the overall LRG redshift distribution. There is no evidence for any difference in redshift distribution or spikes at particular spurious redshifts, indicating that the redshifts affected by the $Z$6 amplitifier instability are fine, just with a reduced success rate.
    \label{fig:z6_issue_perfiber}}
\end{figure*}

We find two significant dips in spectroscopic success.
First, the petal 6 success rate drops by $\sim$1\% between October 26, 2021 and November 6, 2021.  This is due to a known hardware issue: the
$Z$6 CCD amplifiers were unstable on these dates, but reverted to normal behavior after November 6. 0.95\% of the LRG sample is affected by this issue (observed on petal 6 in the specified time period).
We further measured the per-fiber redshift success rate during this time period, and found that certain fibers were more affected than others. The gray regions in Fig.~\ref{fig:z6_issue_perfiber} (identified as low success rate regions a posteriori) have 96.4\% success rate, compared to 98.6\% for the other petal 6 fibers in this date range (white region), and 99.3\% $z_{\textrm{fail}}$ success
rate in the sample as a whole.\footnote{This is higher than the unweighted success rate of 98.9\% because the redshift failure weights are always greater than one.}

Despite the lower success rate, we find no difference in redshift distribution between the lowest-success rate fibers (gray region), the other petal 6 fibers (white region), or the redshift distribution of the entire survey. We do not find any evidence for spikes or pileups at any particular redshift, and the mean redshifts are entirely consistent (Fig.~\ref{fig:z6_issue_perfiber}). The mean redshift is $0.7235 \pm 0.0018$ in the good petal 6 regions between October 26 and November 6; $0.7222 \pm 0.0022$ in the lowest success regions of petal 6; and $0.7204 \pm 0.00013$ for the entire DR1. All errorbars are standard error of the mean.
We therefore see no significant offset in the mean redshift between the white and gray regions of petal 6.
We note that the $Z$6 amplifier issue was originally identified by the redshift pileups that it created on the daily QA, but these are entirely removed by the redshift success criterion.

We therefore test the application of a uniform weight to the affected galaxies, the ``Z6\_ZFAIL'' weight.
We find that adding this weight makes no practical difference to observed galaxy clustering. We therefore do not apply the weight to the catalogs since it is somewhat arbitrary to determine
which fibers within petal 6 should be upweighted, and whether the upweighting should be the same every day that was affected, or different.

We also identify an issue with petal 9 on December 12, 2021. Fig.~\ref{fig:time_dependent_success_lrg} aggregates many nights together into the point labelled 2021-12-20, but we find that the issue is isolated to a single night and leads to a 93\% success rate in petal 9 on that night.
As described in Sec.~2.2 of \cite{DESI2024.II.KP3}, there was a bug in the calibration of the `iron' data processing affecting only December 12, 2021.
This night was reprocessed and substituted for the original `iron' reprocessing in the LSS catalog construction.

Other dips in the success rate are less significant, and we find that they are associated with success trends with number of exposures and number of nights over which a target was observed, as discussed below.

\begin{figure*}
    \includegraphics[width=1.0\linewidth]{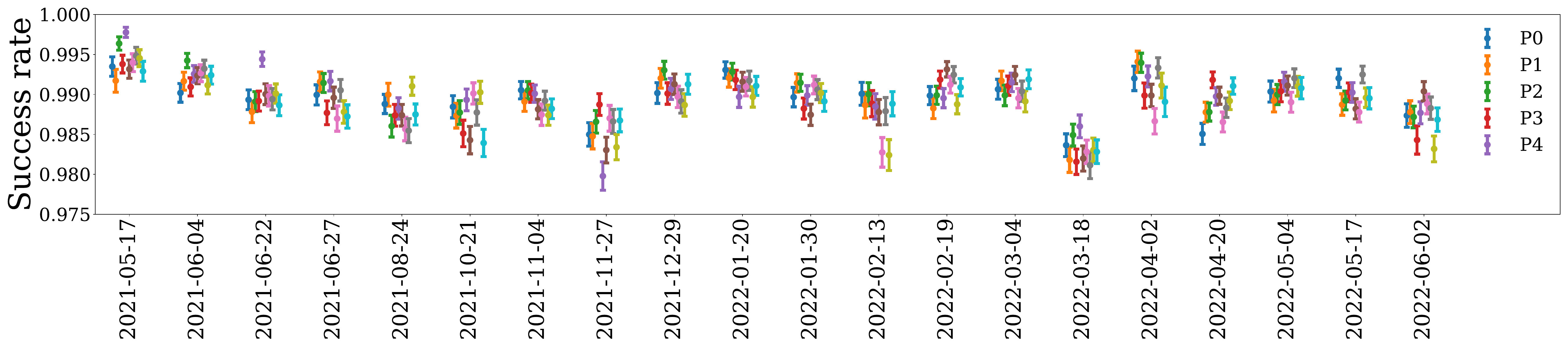}
    \includegraphics[width=1.0\linewidth]{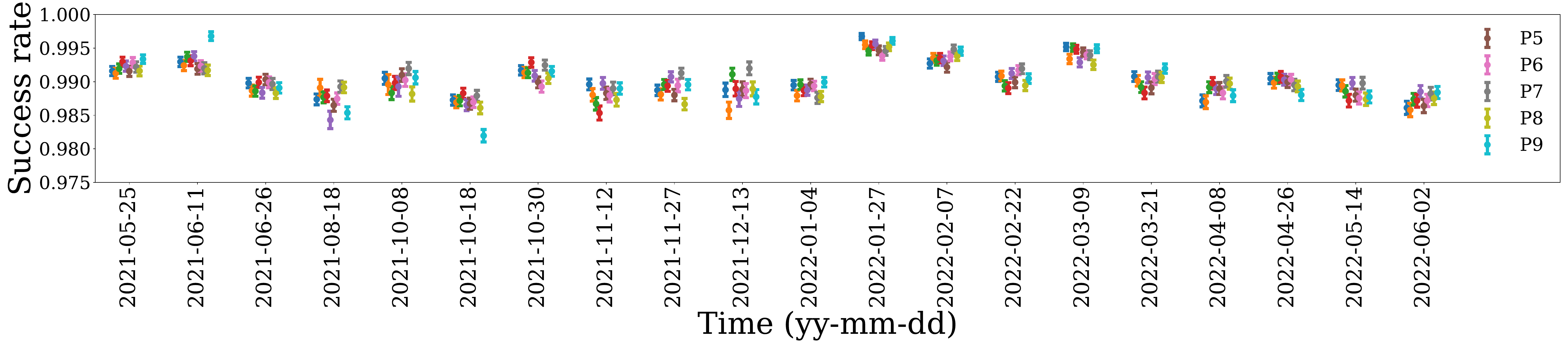}
    \caption{$z_{\textrm{fail}}$-weighted redshift success rate as a function of time for the BGS\_BRIGHT in the North (top) and South (bottom).
    \label{fig:time_dependent_success_bgs}}
\end{figure*}

We also measure the time-dependence of the BGS, ELG, and QSO success rate. We plot the BGS\_BRIGHT success rate as a function of time in Fig.~\ref{fig:time_dependent_success_bgs}. We do not see either the petal 6 or petal 9 outliers that are identified from the LRG success rate. The overall redshift success rate is less stable than for LRG, but fluctuates in a way that affects all petals equally.

\begin{figure*}
    \includegraphics[width=1.0\linewidth]{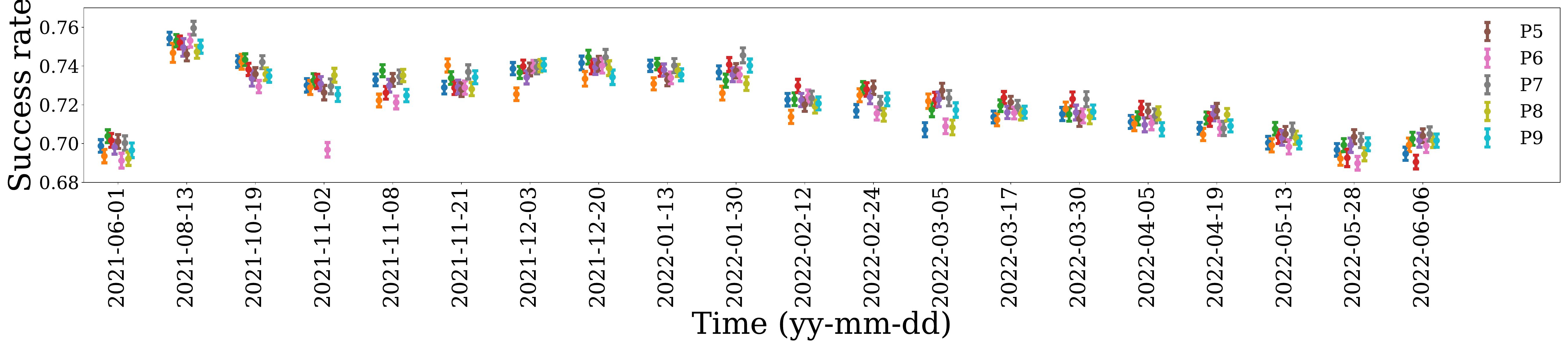}
    \includegraphics[width=1.0\linewidth]{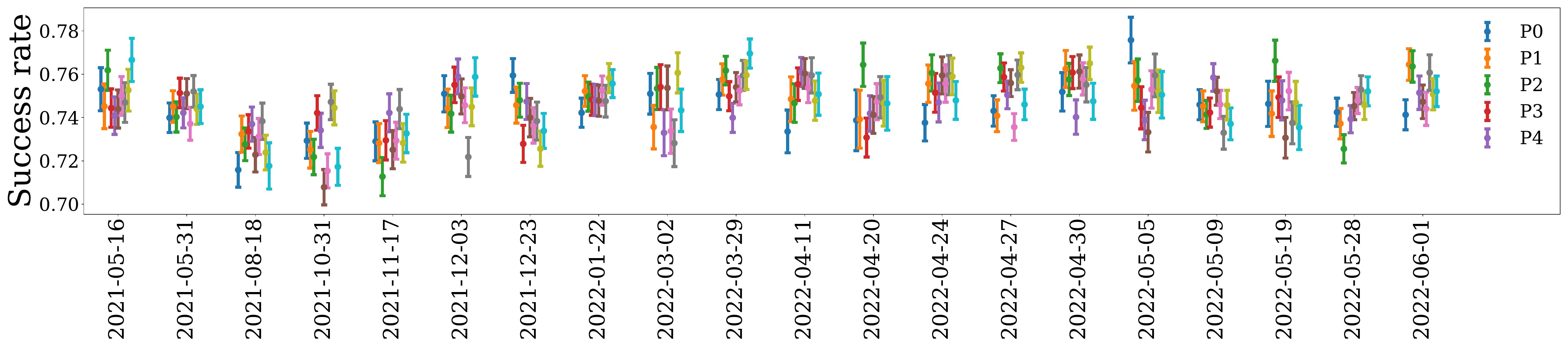}
    \caption{$z_{\textrm{fail}}$-weighted redshift success rate as a function of time for ELG in the South (top) and North (bottom).
    \label{fig:time_dependent_success_elg}}
\end{figure*}

\begin{figure*}
    \includegraphics[width=1.0\linewidth]{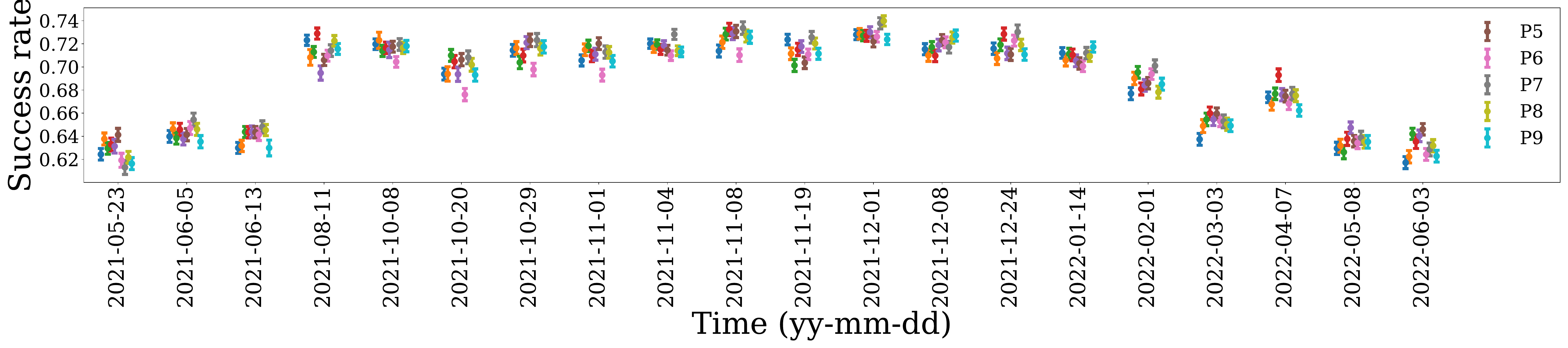}
    \includegraphics[width=1.0\linewidth]{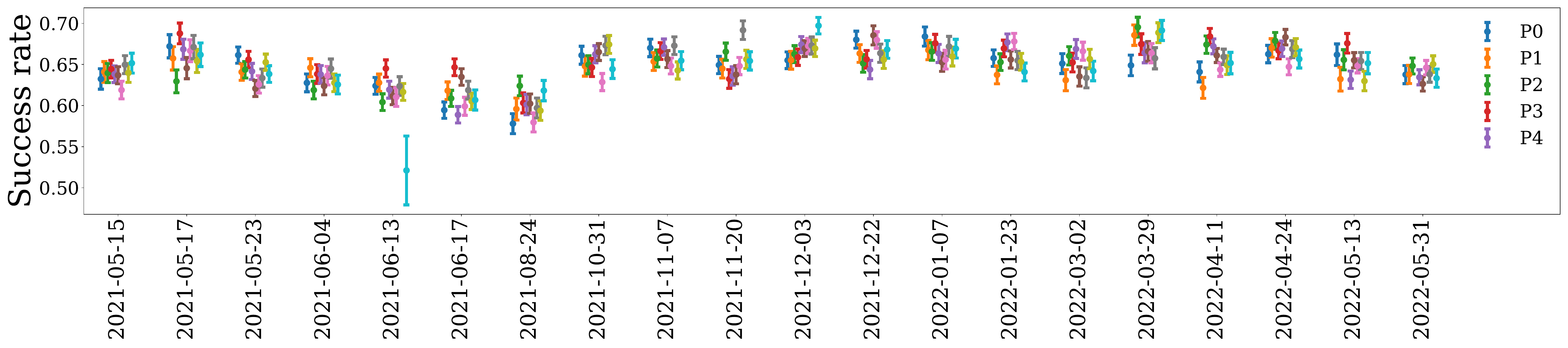}
    \caption{$z_{\textrm{fail}}$-weighted redshift success rate as a function of time for QSO in the South (top) and North (bottom).
    \label{fig:time_dependent_success_qso}}
\end{figure*}

We show the time-dependent ELG success rate in Fig.~\ref{fig:time_dependent_success_elg} and the time-dependent QSO success rate in Fig.~\ref{fig:time_dependent_success_qso}. Both ELG and QSO success rates fluctuate much more than LRG or BGS\_BRIGHT.
This is because both samples are highly impacted by inhomogeneities in the imaging; since DESI observes different areas of sky at different times, the success rate will change depending on the imaging properties of the regions that it is observing. Large variations are seen in the ELG in the South because we observe the SGC in the fall and winter and the NGC in the spring and summer; the deepest imaging is in DES in the SGC, leading to higher success rates in fall and winter.
The success rate is generally consistent between petals at a given time, except for the $Z$6 amplifier issue.
We plan to better model the imaging dependence of the success rate in future releases to allow a better study of the spectroscopic success rate stability, particularly for ELG.

We identified and investigated other fluctuations in the success rate, but these were not traceable to known hardware or processing issues and instead seemed to be the largest fluctuations from spectroscopic success rate trends with overall properties such as the survey speed. 
In LRG we found that galaxies coadded across multiple nights or exposures had a lower $z_{\textrm{fail}}$-weighted success rate. We measured 99.35\% redshift success for galaxies observed in a single exposure; 99.15\% success rate for galaxies with two exposures (16.6\% of the data); 98.95\% for galaxies with three exposures (3.2\% of the data); and 99.22\% success rate for galaxies with four exposures (0.3\% of the data). Galaxies split across multiple nights showed a similar trend: 99.32\% success for one night; 99.15\% for two nights; and 98.67\% for three nights.
We found a 0.5\% drop in success rate between galaxies in the lowest fifth percentile for survey speed (98.92\%) compared to galaxies in the top quarter of speed (99.4\%); most of this drop is concentrated in the slowest quarter of the sample, with a success rate of 99.26\% for galaxies in the 20th to 30th percentile in speed.
These trends exist after applying the TSNR2-based redshift failure weights, indicating
that there are residual trends with low-speed regions or observations requiring multiple exposures that are not captured by TSNR2 or in the model linking TSNR2 to redshift success rate.

We see similar trends in the other tracers, with QSO showing by far the largest impact for galaxies observed
across multiple exposures or multiple nights (3\% drop between 1 night or exposure and 3 nights or exposures).
Success rate showed no clear trend with speed for QSO and BGS\_BRIGHT, but the slowest ELG had a $\sim$1\% lower success rate than average. Due to the fluctuations in success rate seen with these quantities, we define
weights to correct these trends and study their impact on galaxy clustering in Section~\ref{sec:clustering}.

\begin{table*}
    \centering
    \begin{tabular}{c|cccc|cccc}
    & \multicolumn{4}{c|}{Nights} &  \multicolumn{4}{c}{Exposures} \\
    \hline
    Number & LRG & BGS\_BRIGHT & ELG & QSO & LRG & BGS\_BRIGHT & ELG & QSO \\
    \hline
    1 & 0.993 & 0.990 & 0.726 & 0.677 & 0.994 & 0.990 & 0.727 & 0.679 \\
    2 & 0.992 & 0.988 & 0.716 & 0.6453 & 0.992 & 0.989 & 0.721 & 0.659 \\
    3 & 0.987 & 0.984 & 0.727 & 0.644 & 0.990 & 0.988 & 0.712 & 0.650 \\
    4 & & & & & 0.992 & 0.995 & 0.734 & 0.673 \\
    \end{tabular}
    \caption{$z_{\textrm{fail}}$-weighted redshift success rates as a function of number of nights or number of exposures required to observe a target for sufficient effective time.
    \label{tab:coadd_numexp_numnight}}
\end{table*}










\section{Impact of residual spectroscopic systematics on galaxy clustering}
\label{sec:clustering}

\begin{figure*}
    \includegraphics[width=1.0\linewidth]{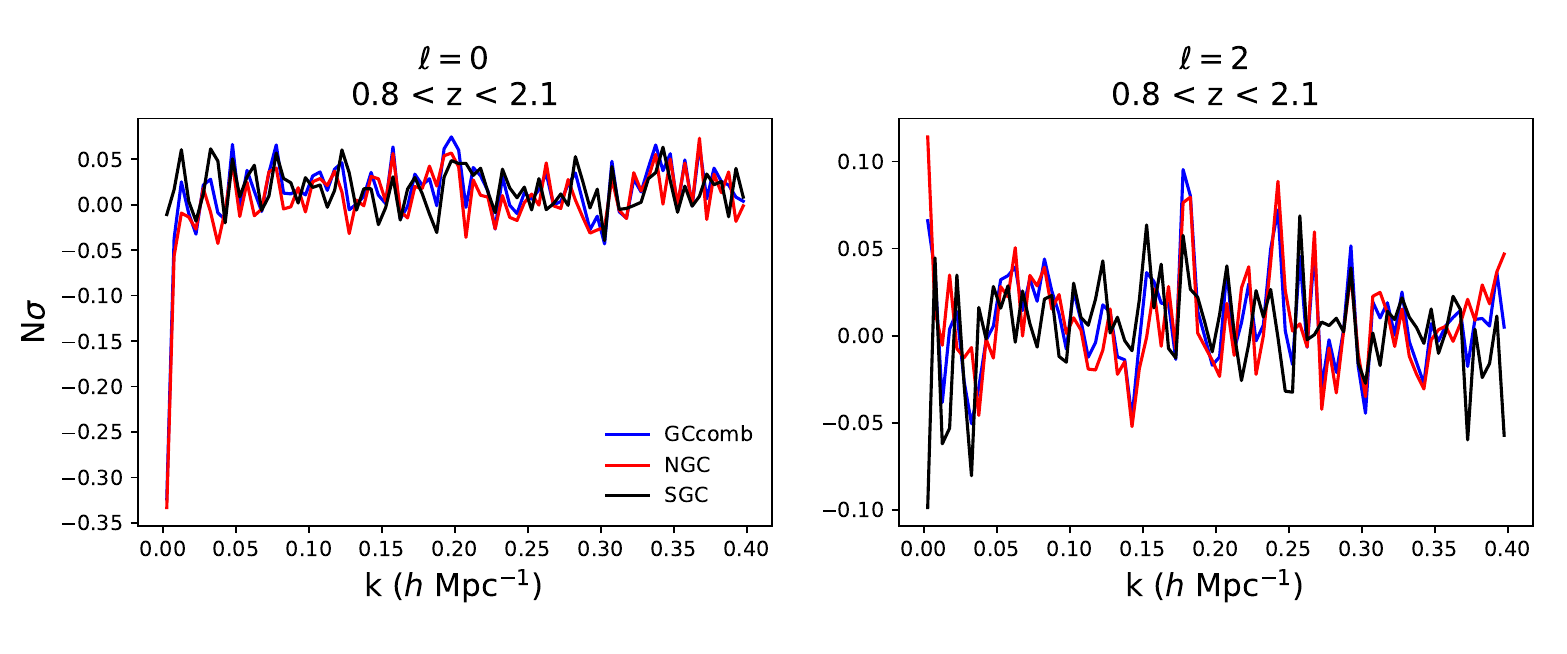}
    \caption{Deviations between the QSO power spectrum multipoles measured with the default weighting scheme, and without the redshift failure weights. We show QSO as it shows the largest impact of the redshift failure weights on clustering.
    \label{fig:remove_zfail_wts}}
\end{figure*}

The overall impact of the redshift
failure weights on clustering is small, though not necessarily negligible,
with $\chi^2_{\textrm{sys}} \sim 1$. When turning off the redshift failure weights, we find 
$\chi^2_{\textrm{sys}} = 0.26$ and 0.63 for the two ELG bins at $0.8 < z < 1.1$ and $1.1 < z < 1.6$, and 0.24 for QSO. They have a much smaller impact on LRG and BGS, with $\chi^2_{\textrm{sys}} < 0.04$ for LRG and $<0.01$ for BGS. We show the impact of turning off the redshift failure weights on the QSO in Fig.~\ref{fig:remove_zfail_wts}; the impact of turning off these weights is roughly comparable to the largest systematics arising from the additional weights.

\begin{figure*}
    \includegraphics[width=1.0\linewidth]{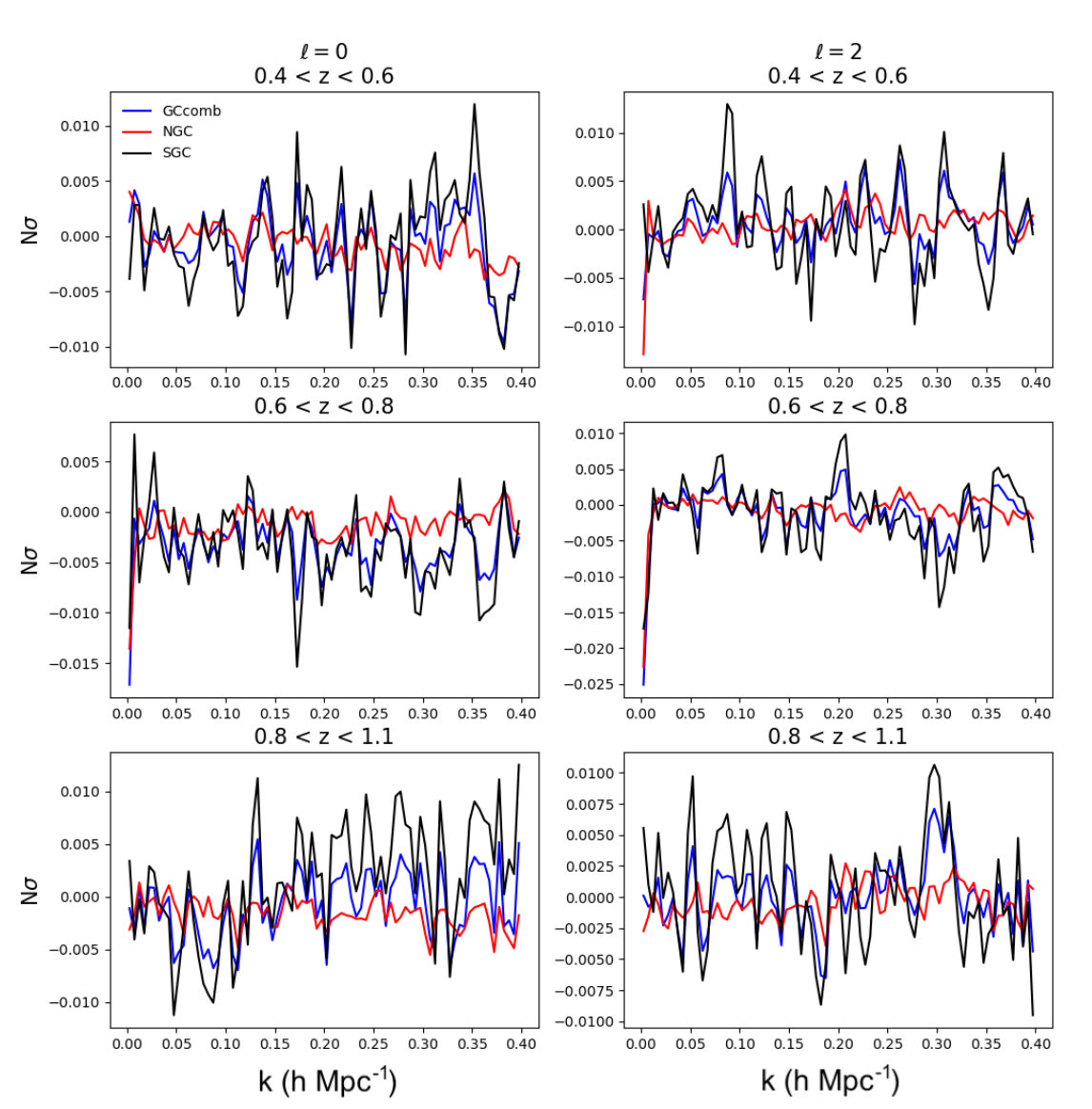}
    \caption{Deviations between the LRG power spectrum multipoles measured with the default weighting scheme, and adding an extra $Z$6 weight to fibers affected by the $Z$6 amp instability issue. Deviations are presented in units of $\sigma$, the diagonal errorbar from the analytic covariance matrix used in the DESI Y1 analysis.
    \label{fig:lrg_zfail_z6_weights}}
\end{figure*}

Next, we study the impact of the additional (optional) weights on galaxy clustering.
We find that the additional weights make a minimal difference in the measured LRG clustering.
First, we show the impact of the $Z$6 weights in Fig.~\ref{fig:lrg_zfail_z6_weights}. We consider the clustering in both the NGC and SGC, as well as the combined clustering (``GCComb''). We find
a maximum $\chi^2_{\textrm{sys}} = 0.03$, and less than a 0.03$\sigma$ offset between the default weights and the $Z$6 weights. This justifies our decision
to not include the $Z$6 weights by default, as excluding them has a completely negligible impact on galaxy clustering.

\begin{figure*}
    \includegraphics[width=1.0\linewidth]{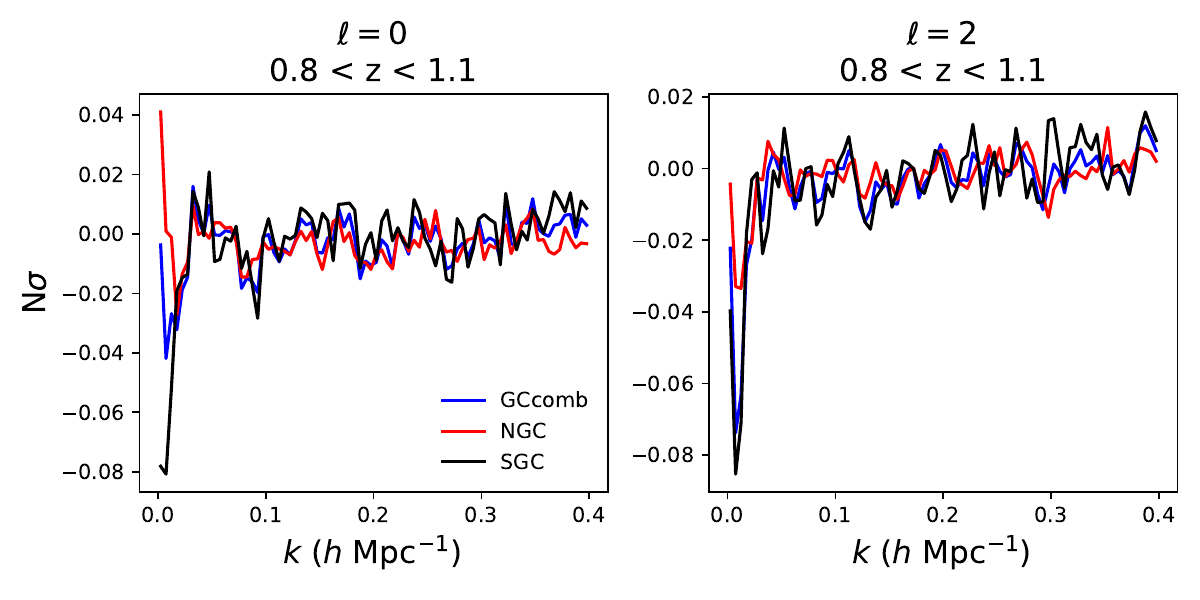}
    \caption{As in Fig.~\ref{fig:lrg_zfail_z6_weights}, but correcting the TSNR2\_LRG trend in the $0.8 < z < 0.9$ bin of LRG. We only show the $0.8 < z < 1.1$ bin since this is the only one affected by the extra weights.
    \label{fig:lrg_z0.8-0.9}}
\end{figure*}

In Fig.~\ref{fig:lrg_z0.8-0.9}, we show the impact of adding extra weights to correct the TSNR2\_LRG trend in the South at $0.8 < z < 0.9$. Adding these weights makes up to a 0.08$\sigma$ impact on the monopole and quadrupole at large scales. The largest $\chi^2_{\textrm{sys}} = 0.04$ for the SGC combination of monopole and quadrupole, and 0.02 for the combined hemisphere monopole plus quadrupole.

\begin{figure*}
    \includegraphics[width=1.0\linewidth]{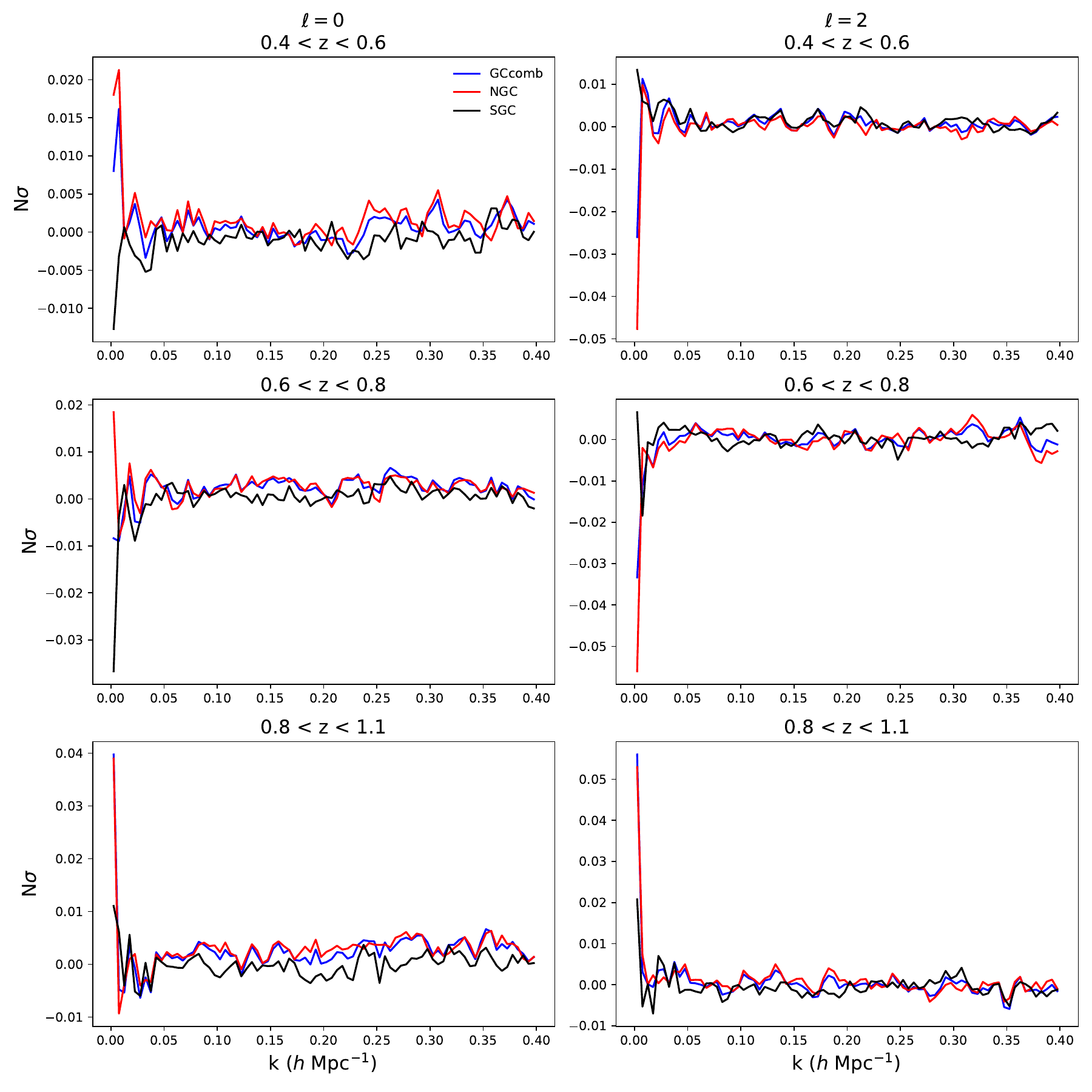}
    \caption{As in Fig.~\ref{fig:lrg_zfail_z6_weights}, but adding weights correcting the survey speed trend for LRG.
    \label{fig:lrg_speed}}
\end{figure*}

Finally, in Fig.~\ref{fig:lrg_speed}, we show the impact of adding weights for survey speed on the LRG clustering. The impact is very small at $k > 0.01$ $h$ Mpc$^{-1}$ ($<0.01\sigma$) and $<0.05\sigma$ on large scales. The largest $\chi^2_{\textrm{sys}} = 0.02$.




We also apply focal plane weights $\eta_{\textrm{zfail}}$ and weights correcting trends with number of exposures or nights.
We find that adding these weights change $P(k)$ by $<0.01\sigma$ in all $k$ bins, and $\chi^2_{\textrm{sys}}$ is less than 0.01.

We apply focal plane weights and number of night and exposure weights to BGS\_BRIGHT. We find $<0.02\sigma$ changes in all $k$ bins and $\chi^2_{\textrm{sys}} < 0.01$; i.e.\ negligible impacts on clustering.

\begin{figure*}
    \includegraphics[width=1.0\linewidth]{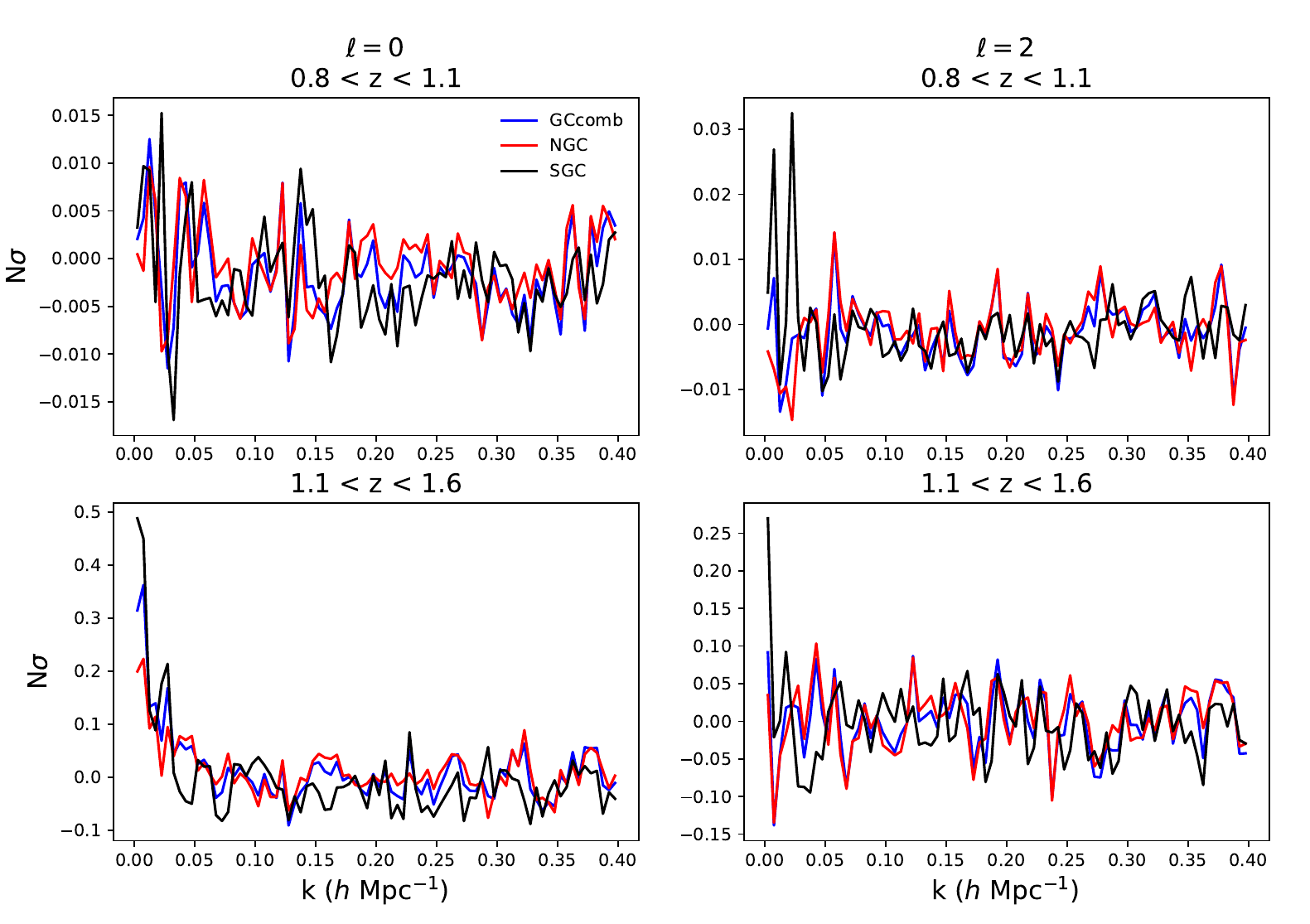}
    \caption{As in Fig.~\ref{fig:lrg_zfail_z6_weights}, but adding weights correcting the survey speed trend for ELG.
    \label{fig:elg_speed}}
\end{figure*}

We find ELG are more sensitive to applying these extra weights. The speed weights have the largest impact on ELG clustering (Fig.~\ref{fig:elg_speed}), with up to a $0.5\sigma$ impact on the large-scale monopole at $1.1 < z < 1.6$.
These translate into $\chi^2_{\textrm{sys}}$ of 0.004 and 0.47 in the two redshfit bins, when combining hemispheres. $\chi^2_{\textrm{sys}}$ is similar between both hemispheres and the combined case.

\begin{figure*}
    \includegraphics[width=1.0\linewidth]{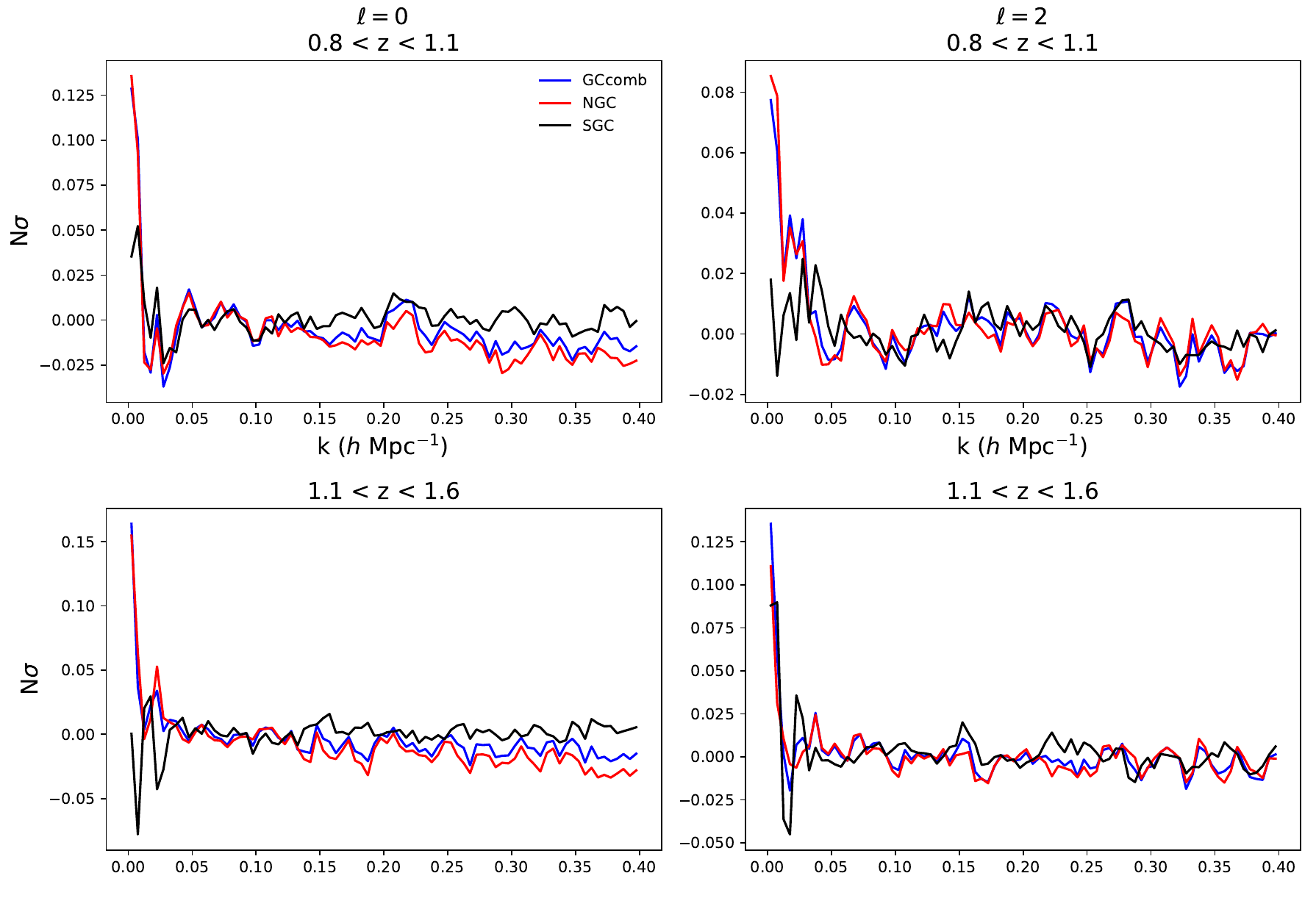}
    \caption{As in Fig.~\ref{fig:lrg_zfail_z6_weights}, but adding weights correcting the number of exposures trend for ELG.
    \label{fig:elg_numexp}}
\end{figure*}

The weights for number of nights or number of exposures have a smaller impact on ELG clustering.
We show the number of exposures weights in Fig.~\ref{fig:elg_numexp}. The impact is small ($\chi^2_{\textrm{sys}} = 0.05$ in both redshift bins) and mainly concentrated
on large scales. The impact of number of nights weight is similar, but smaller, with $\chi^2_{\textrm{sys}} = 0.05$, and a $0.05\sigma$ change in the low-$k$ monopole. The impact of focal plane weights is shown in Figs. 5 and 6 and Table 1 of the companion paper \cite{KP3s4-Yu}. The $\chi^2_{\textrm{sys}}$ is at most 0.23 and individual $k$ bins shift by $<0.1\sigma$ in an incoherent way, indicating that adding this weight has a minimal impact on the clustering measurement.

\begin{figure*}
    \includegraphics[width=1.0\linewidth]{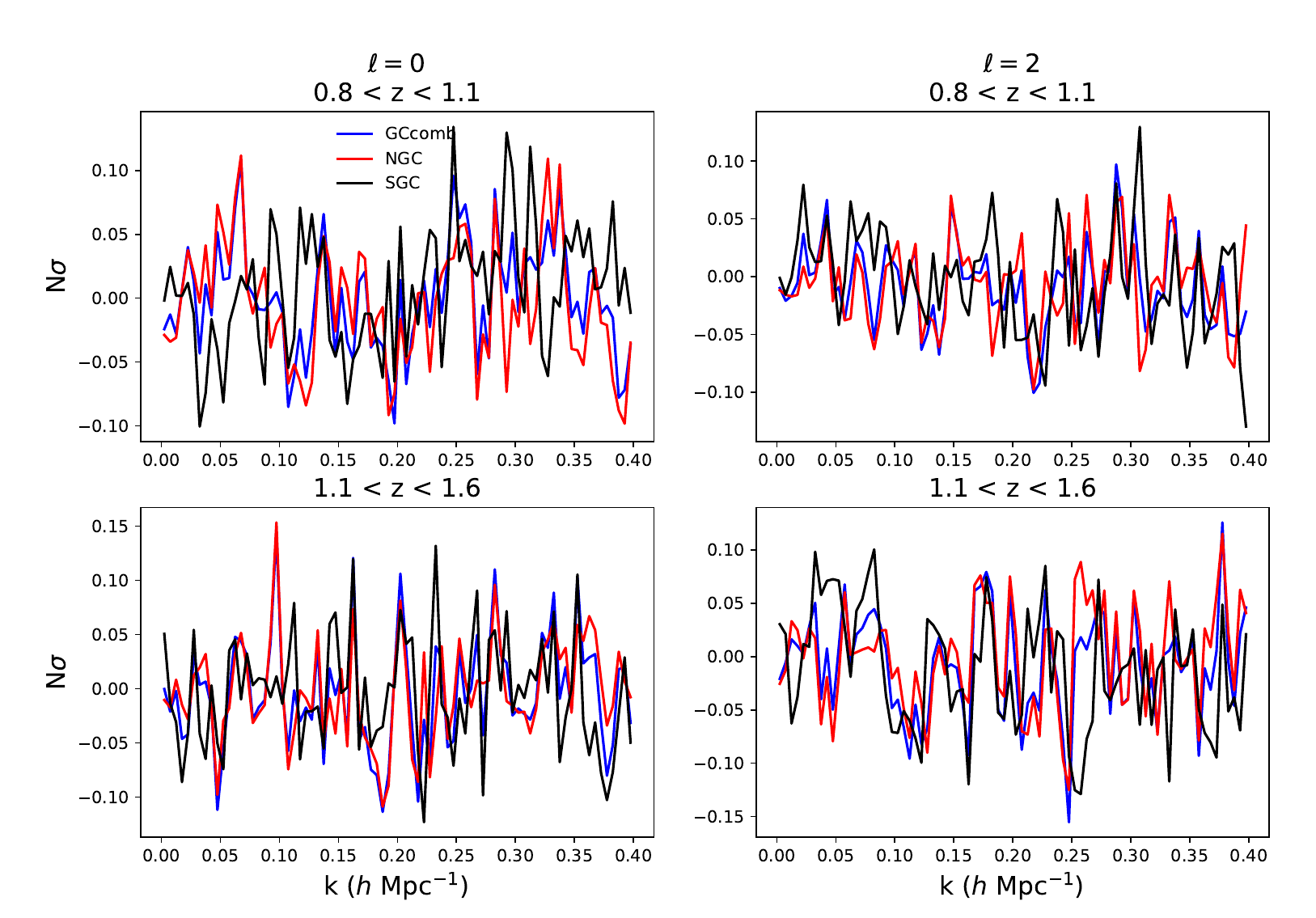}
    \caption{Impact on ELG clustering of removing fibers 1261, 1269, 1295, 1296 and 1307, visually identified as having anomalous spikes in the redshift distribution at $z=1.5$.
    \label{fig:elg_remove_spike_fibers}}
\end{figure*}

We show the impact of removing
the fibers with anomalous spikes in the redshift
distribution at $z=1.5$ (1261, 1269, 1295, 1296 and 1307).
Removing these fibers leads to random scatter
in the power spectrum multipoles with amplitude $\sim0.1\sigma$, leading to $\chi^2_{\textrm{sys}} = 0.26$ and 0.49 in the two ELG redshift bins.

\begin{figure*}
    \includegraphics[width=1.0\linewidth]{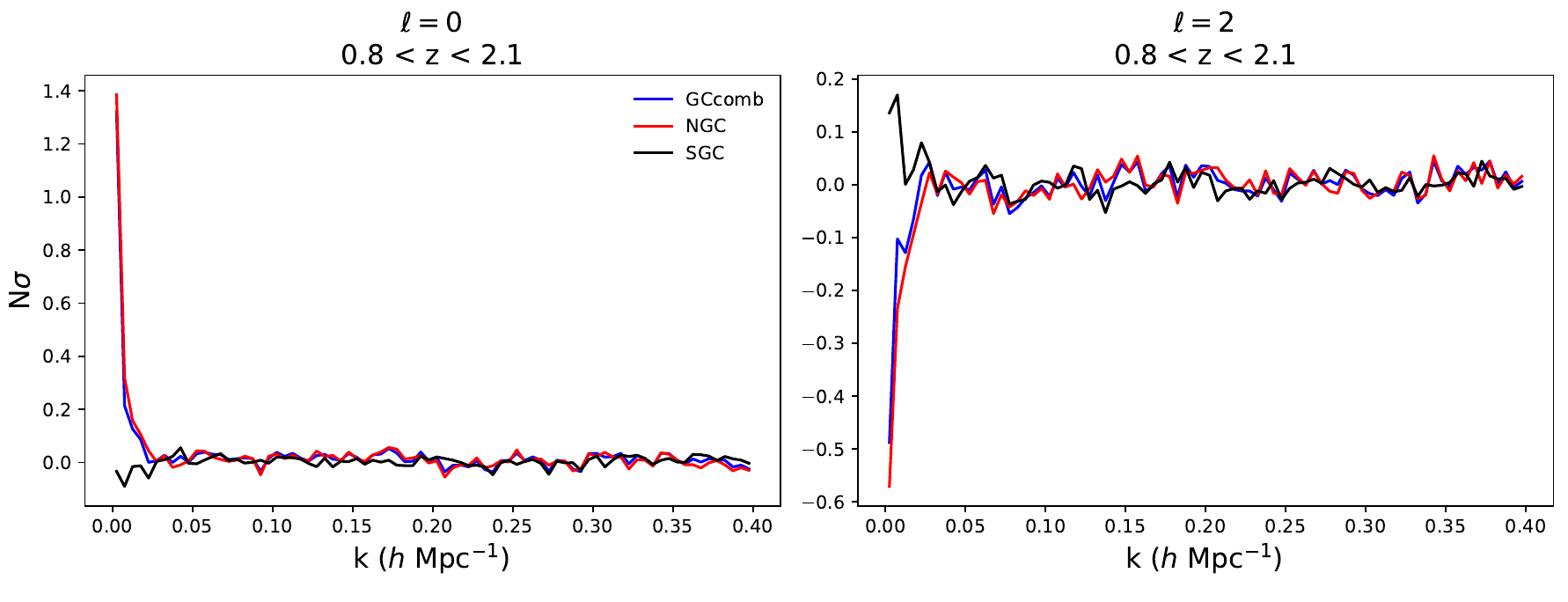}
    \caption{As in Fig.~\ref{fig:lrg_zfail_z6_weights}, but adding weights correcting the number of exposures trend for QSO.
    \label{fig:qso_numexp}}
\end{figure*}

\begin{figure*}
    \includegraphics[width=1.0\linewidth]{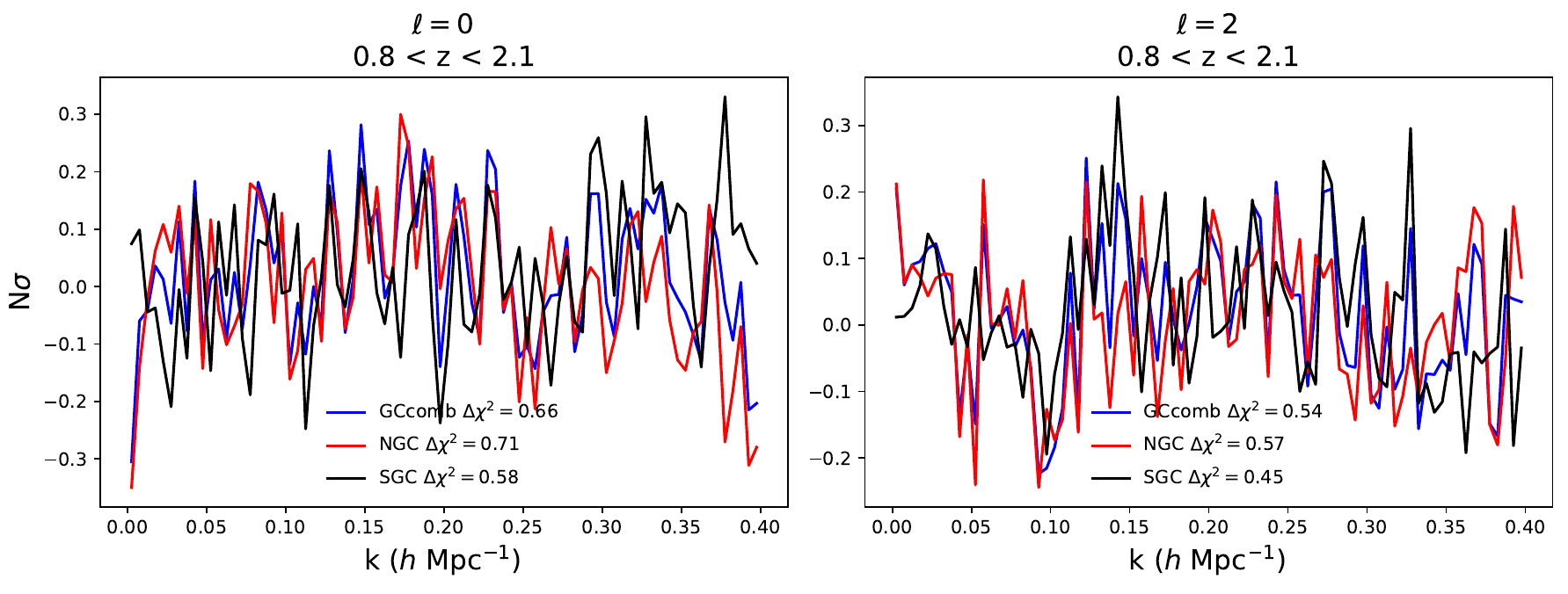}
    \caption{As in Fig.~\ref{fig:lrg_zfail_z6_weights}, but adding weights correcting the focal plane trend for QSO.
    \label{fig:qso_focal}}
\end{figure*}

These additional weights have the largest impact
on the QSO clustering. We find $\chi^2_{\textrm{sys}} = 1.9$ when applying the exposure weights (Fig.~\ref{fig:qso_numexp}), an offset that is driven by the NGC ($\chi^2_{\textrm{sys}} = 2.1$) rather than the SGC ($\chi^2_{\textrm{sys}} = 0.1$).
The offset is largest at large scales, reading 1.4$\sigma$ in the monopole and 0.6$\sigma$ in the quadrupole. The impact of number of nights weights is similarly  peaked towards large scales,
but much more modest overall, with $\chi^2_{\textrm{sys}} = 0.13$ and at most a 0.2$\sigma$ impact on the large-scale monopole.
Finally, we find a relatively large $\chi^2_{\textrm{sys}}$ impact from the focal plane weights, 1.2 in the combined hemisphere measurement (Fig.~\ref{fig:qso_focal}). However, the pattern of the impact on each $k$ bin is relatively random with a scatter of $\sim0.3\sigma$; this may simply be due to the scatter in the focal plane weights from the relatively small number of quasar targets per fiber ($\sim$400).

\subsection{Impact on cosmological parameters}
\label{sec:impact_on_params}

\begin{table*}
\small
    \centering
    \begin{tabular}{l|cc|ccc}
    Sample & $\Delta \alpha_{\parallel}$ & $\Delta \alpha_{\perp}$ & $\Delta h$ & $\Delta \Omega_m$ & $\Delta \log(A_s)$ \\
    \hline
    QSO, no.\ of exposures & +0.001 (0.059) & -0.003 (0.030) & 
    +0.001 (0.031) & -0.001 (0.020) & 0.008 (0.16) \\
    QSO, focal plane & -0.003 (0.059) & +0.003 (0.030)
    & -0.004 (0.031) & 0.0 (0.020) & +0.005 (0.16) \\
    ELG, $0.8 < z < 1.1$, speed & +0.002 (0.092) & +0.01 (0.1) & +0.002 (0.038) & -0.001 (0.034) & 0 (0.21) \\
    ELG, $1.1 < z < 1.6$, speed & +0.002 (0.080) & 0 (0.073) & +0.001 (0.020) & -0.001 (0.020) & -0.014 (0.23) \\
    ELG, $0.8 < z < 1.1$, mask $n(z)$ spike & -0.004 (0.092) & +0.01 (0.10) & +0.004 (0.038) & -0.003 (0.034) & -0.01 (0.21) \\
    ELG, $1.1 < z < 1.6$, mask $n(z)$ spike & +0.002 (0.080) & -0.003 (0.073)   & +0.002 (0.020)  & 0.0 (0.020) & -0.036 (0.23)
    \end{tabular}
    \caption{Change in parameters for BAO fits (left) and full-shape fits (right) between the default clustering measurements and clustering measurements using the weights and samples indicated in the left column. 1-$\sigma$ error for each parameter is given in parenthesis. The parameter shifts are at most 30\% of the errorbar.
    \label{tab:impact_on_parameters}}
\end{table*}

We propagate the impact of the trends with a significant $\Delta \chi^2$ onto derived
cosmological parameters (Table~\ref{tab:impact_on_parameters}).
We perform both BAO and full-shape (RSD) fits. In both cases, we fit to the blinded
data \cite{KP3s9-Andrade} and compare results with and without the weights under consideration.
For the BAO fits, we match the default settings of \cite{KP4s4-Paillas,DESI2024.III.KP4}, fitting to the un-reconstructed power spectrum.
For the full-shape fits, we use the ``velocileptors'' LPT code
\cite{Chen20,Chen21,KP5s2-Maus}
\footnote{\url{https://github.com/sfschen/velocileptors}}
with the ``physically motivated'' prior bias (all bias parameters and counterterms are multiplied by appropriate powers of $\sigma_8$) and ``maximum freedom'' in the bias parameters (i.e.\ fixing them to the coevolution relations) as implemented in the desilike code, fitting to the monopole and quadrupole.\footnote{\url{https://github.com/cosmodesi/desilike/blob/41aec182cd0d5da0657a3bb11cda74b382cad4be/desilike/theories/galaxy_clustering/full_shape.py}} We again fit to the blinded
power spectrum data, without any small-scale $\theta$-cut applied.
For the full-shape fits, we vary the cosmological parameters $h$, $\Omega_m$, $\log(A_s)$, and $\omega_b$, with the $\omega_b$ posterior dominanted by a Gaussian prior at $\omega_b = 0.02237 \pm 0.00037$.
We summarize the parameters $h$, $\Omega_m$, and $\log(A_s)$ for the full-shape fits in Table~\ref{tab:impact_on_parameters}, and $\alpha_{\parallel}$ and $\alpha_{\perp}$ for the template-based BAO fits.
While the settings in both the BAO and RSD cases do not exactly match the fiducial choices for the DESI Y1 analysis, our primary concern is the difference between the weighted and unweighted cases, not the actual values of the parameters.

We find negligible changes on the cosmological parameters when applying the
extra weights. The largest change is a 0.15$\sigma$ shift in $\log{A_s}$ when masking the 5 fibers with a spike at $z=1.5$ in ELG at $0.8 < z < 1.1$.
Apart from this shift, every other parameter shift is $<10\%$ of the errorbar.
These fits all use $k_{\textrm{min}} = 0.02$ $h$ Mpc$^{-1}$, and are thus largely unaffected by the low-$k$ uptick when applying the weights.
However, we caution that other parameter constraints which are more sensitive to large scales (e.g.\ scale-dependent bias due to primordial
non-Gaussianity) may be more impacted by these additional weights.

In the end, due to the negligible impact of the extra weights, we do not choose to use any of the extra weights
in the LSS catalogs. We only report  $w_{\textrm{zfail}}$ in the LSS catalogs and in the large-scale BAO and RSD analyses. However, we caution that other analyses of DESI data (for instance, higher-point correlation functions) may want to make different choices about what set of weights to use.






\section{Conclusions}
\label{sec:conclusions}

We develop and validate corrections for spectroscopic inhomogeneities in the DESI DR1
galaxy clustering catalogs, which consist of luminous red galaxies (LRG), bright galaxies (BGS), quasars (QSO), and emission line galaxies (ELG).
DESI tiles are observed for a uniform effective exposure time, but small variations in the actual
exposure times as well as fiber-to-fiber throughput variations lead to variations in the effective
observing time as parametrized by the template signal-to-noise ratio squared, TSNR2, for different observations.
Out of 4326 fibers with valid spectra,
we remove 60 fibers that are $>4\sigma$ outliers in success rate, compared to expectations from a Monte Carlo
simulation
using the modelled variation in success rate
with TSNR2, removing 1.2\% of the total data.
We then describe the creation of weights to mitigate
trends between redshift success rate and TSNR2 for LRG, BGS, and QSO. The process for ELG is conceptually similar but different in detail,
since ELG redshift success depends on the flux of the [OII] line and is thus very sensitive to fine
variations in the sky background due to the forest of atmospheric sky lines at the red end of DESI spectra. It is described in the companion paper, \cite{KP3s4-Yu}.

We then validate the weights in Section~\ref{sec:validate_zfail_wts}.
Without the redshift failure weights, there are significant trends for LRG and QSO; after applying the weights, their success rate is uniform.
Even after applying the weights, we find small but significant fluctuations in success rate as a function of position on the focal plane, survey speed, or number of exposures required to complete a galaxy observation.
Finally, we study the time stability of the redshift success rate and identify two periods of outlier performance. One of these is related to a
calibration error in the ``iron'' processing and was fixed for production of the clustering catalogs.
The other is related to a transient hardware
issue leading to instability in the $Z$-camera amplifier on petal 6 (lasting for 10 days).
We find this issue leads to a uniform
reduction in the spectroscopic success rate, but no change in the redshift distribution.

We propagate the impact of the remaining residual statistically significant trends
on galaxy clustering in Section~\ref{sec:clustering}.
We find that adding a weight for the  $Z$6 amplifier issue has no measurable impact on galaxy clustering, and so we continue to use redshifts
affected by this issue, without a weight to correct them.
The residual trends with the largest impact
are survey speed trends for the ELG, and number
of exposure trends for QSO. Both of them have the largest
impact on very large scales ($k < 0.03$ $h$ Mpc$^{-1}$). We find that these trends (along with other less significant trends) change cosmological parameters from BAO and RSD by $<15\%$ of their errorbar: they are negligible compared to the statistical uncertainty. However, we caution that other observables (small-scale clustering, higher-order clustering, or very large scale observations) may be affected differently by these residual trends, and their impact should be tested to ensure that it is negligible.

We aim to further develop and test these corrections for future DESI data releases.
For instance, more detailed study of the $>4\sigma$ outlier fibers may allow us to recover some of these fibers, slightly increasing the available catalogs,
or improve the modelling of success rate with TSNR2.
While the success rate is well-defined for the high-completeness LRG and BGS samples, it is harder
to define for the QSO, where a substantial fraction of QSO targets result in spectroscopically confirmed galaxies or stars.
Further study of the properties of QSO targets
that become galaxies or stars could allow for a better and more consistent definition of spectroscopic success rate for this sample. One possibility for an improvement is jointly modelling the spectroscopic and imaging systematics, which will also allow us to better distinguish chance alignments between targeting fluctuations and spectroscopic properties (as discussed in Section~\ref{sec:qso_anomaly}). 
Another possibility is to use machine learning techniques to perform a blind search for all possible
spectroscopic systematics affecting the redshift success rate--such an extension is a possibility for future DESI data releases.

The DESI instrument and survey operations continue to improve. Most notably, fiber positioning accuracy was substantially improved via a series of fiber dither tests carried out throughout the first two years of the survey \cite{Schlafly24}. Fiber positioning inaccuracies are a substantial source
 of the focal plane dependent redshift success trends,
 and the higher-order correction applied to the positioning midway through the third year of operations will substantially reduce their
 variability. Finally, we will continue to study
 whether additional weights are required to mitigate
 spectroscopic systematics trends that may be significant for other DESI analyses beyond the large-scale clustering used for BAO and RSD fits.

\section*{Data Availability}
The data used in this work will be made public as part of DESI Data Release 1 (details at \url{https://data.desi.lbl.gov/doc/releases/}). The data points corresponding to the figures
are available on Zenodo at
\url{https://doi.org/10.5281/zenodo.11277146}.

\acknowledgments

AK was supported as a CITA National Fellow by the Natural Sciences and Engineering Research Council of Canada (NSERC), funding reference \#DIS-2022-568580.
WP acknowledges support from the Natural Sciences and Engineering Research Council of Canada (NSERC), [funding reference number RGPIN-2019-03908] and from the Canadian Space Agency.
Research at Perimeter Institute is supported in part by the Government of Canada through the Department of Innovation, Science and Economic Development Canada and by the Province of Ontario through the Ministry of Colleges and Universities.
JY acknowledges the support from the SNF 200020\_175751 and 200020\_207379 ``Cosmology with 3D Maps of the Universe" research grant.

This material is based upon work supported by the U.S. Department of Energy (DOE), Office of Science, Office of High-Energy Physics, under Contract No. DE–AC02–05CH11231, and by the National Energy Research Scientific Computing Center, a DOE Office of Science User Facility under the same contract. Additional support for DESI was provided by the U.S. National Science Foundation (NSF), Division of Astronomical Sciences under Contract No. AST-0950945 to the NSF’s National Optical-Infrared Astronomy Research Laboratory; the Science and Technology Facilities Council of the United Kingdom; the Gordon and Betty Moore Foundation; the Heising-Simons Foundation; the French Alternative Energies and Atomic Energy Commission (CEA); the National Council of Humanities, Science and Technology of Mexico (CONAHCYT); the Ministry of Science and Innovation of Spain (MICINN), and by the DESI Member Institutions: \url{https://www.desi.lbl.gov/collaborating-institutions}. Any opinions, findings, and conclusions or recommendations expressed in this material are those of the author(s) and do not necessarily reflect the views of the U. S. National Science Foundation, the U. S. Department of Energy, or any of the listed funding agencies.

The authors are honored to be permitted to conduct scientific research on Iolkam Du’ag (Kitt Peak), a mountain with particular significance to the Tohono O’odham Nation.

\bibliographystyle{JHEP}
\bibliography{main,DESI2024}

\appendix

\section{Relationship between mean redshift and redshift success rate}
\label{sec:meanz_vs_success_rate}

\begin{figure*}
    \includegraphics[width=0.8\linewidth]{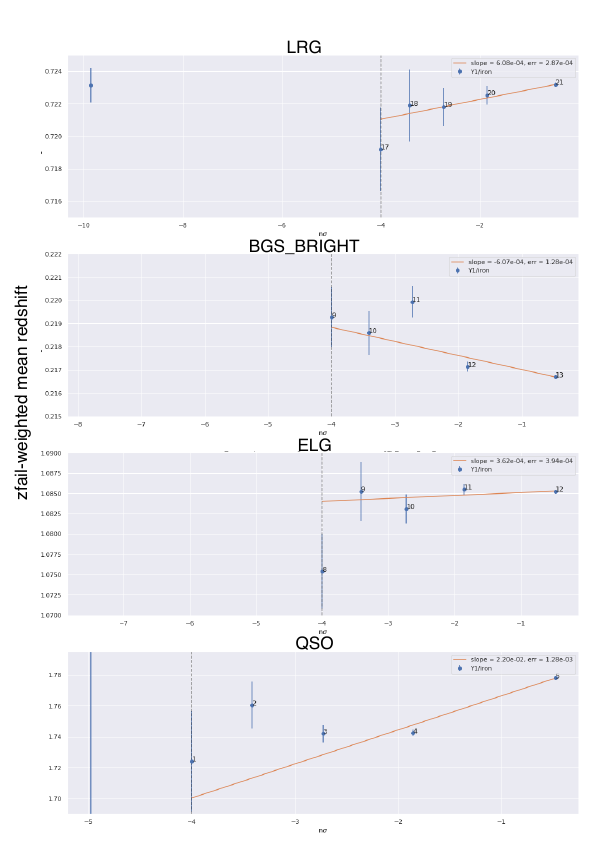}
    \caption{Relationship between success rate and redshift. On the $x$ axis, we show the number of $\sigma$ deviation from the mean success rate, measured for each fiber using Monte Carlo simulations. Fibers are then aggregated into bins in $n\sigma$. The errorbar is from the dispersion of observed redshift values. On the $y$-axis, we plot the redshift failure weighted mean redshift in each bin. The 4$\sigma$ cut used to define low success rate fibers is given by the dotted blue line.
    \label{fig:z_vs_nsig}} 
\end{figure*}

We use the Monte Carlo simulations described in Section~\ref{sec:badfibers} to remove 55 4$\sigma$ outlier
fibers from further analysis. In addition to removing targets on low success rate fibers, this analysis also enables
us to study the uniformity of galaxy properties with success rate.
In particular, the redshift failure weights are intended to ensure that galaxies are homogeneously distributed
with respect to observing conditions. One important aspect of homogeneity is the redshift distribution.
We therefore measure the mean redshift for all successful galaxies, averaged among fibers and then sorted
by each fiber's ``$n\sigma$'' deviation from the expected success rate based on the TSNR2 and fiberflux model.
When computing mean redshift, galaxies are also upweighted by their redshift failure weight (Fig.~\ref{fig:z_vs_nsig}). We fit a line to each trend, and find marginally significant relationships for 
LRG (2.1$\sigma$), and larger significances for BGS\_BRIGHT (4.7$\sigma$) and QSO (17.2$\sigma$).
The relationship is positive for LRG and QSO, and negative for BGS\_BRIGHT.
The fluctuations in redshift are very small for LRG and BGS\_BRIGHT, $\sim$0.002.
Quasars show a much larger variation, with $\Delta z$ $\sim$0.04 between the highest bin (closest success rate to the mean) and the rest. However, this is likely due to the re-observation of Ly$\alpha$ forest quasars, which were not excluded when making this plot; they lie at $z > 1.8$ and, since they are observed
multiple times, skew the distribution of quasar successes to higher redshift.
Given the small trends observed in the other tracers, we leave this for future work in the next data release.

\section{Validation of redshift failure weights in full redshift ranges and for alternative BGS samples}
\label{sec:validation_other_z_ranges}

\begin{figure*}
\vspace{-30pt}
    \includegraphics[width=0.5\linewidth]{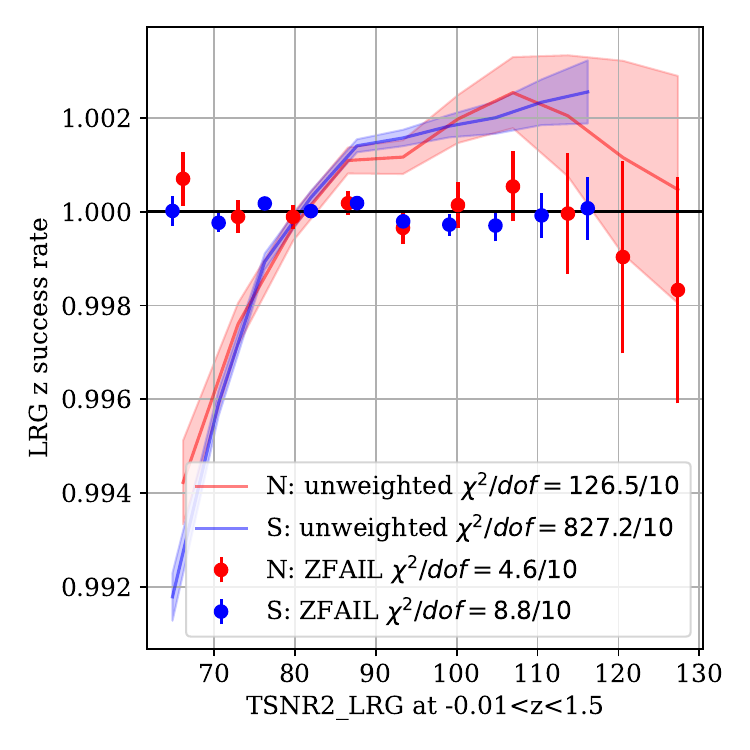}
     \includegraphics[width=0.5\linewidth]{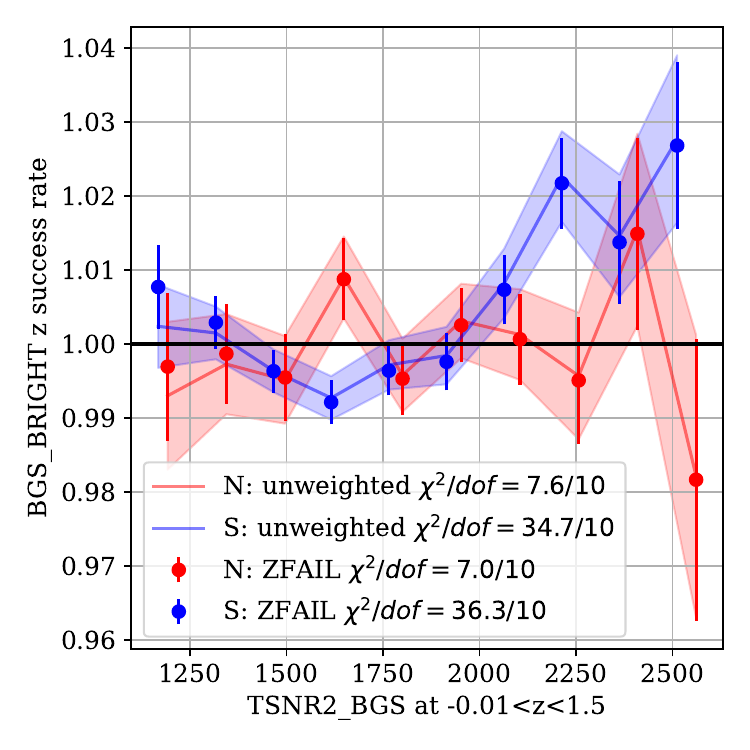}
     \includegraphics[width=0.5\linewidth]{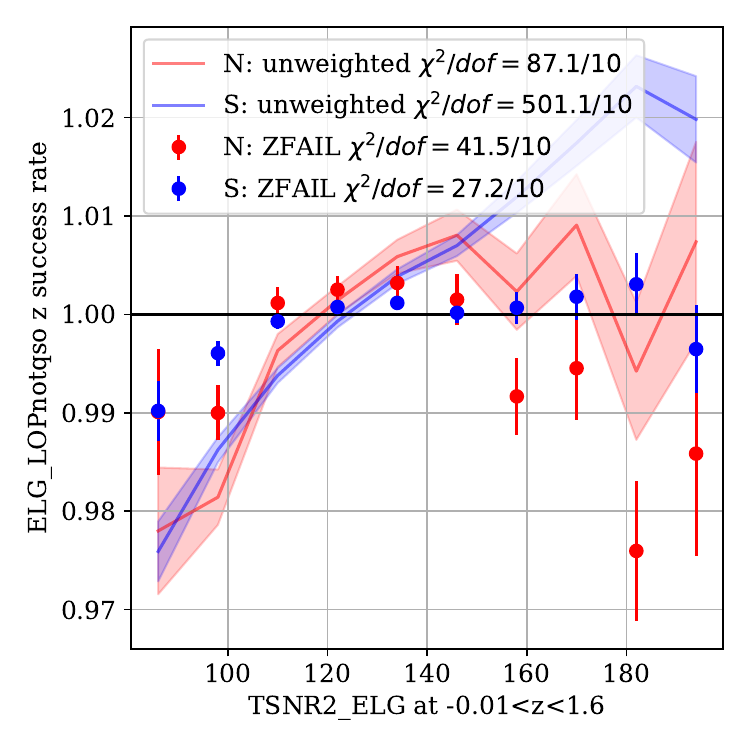}
     \includegraphics[width=0.5\linewidth]{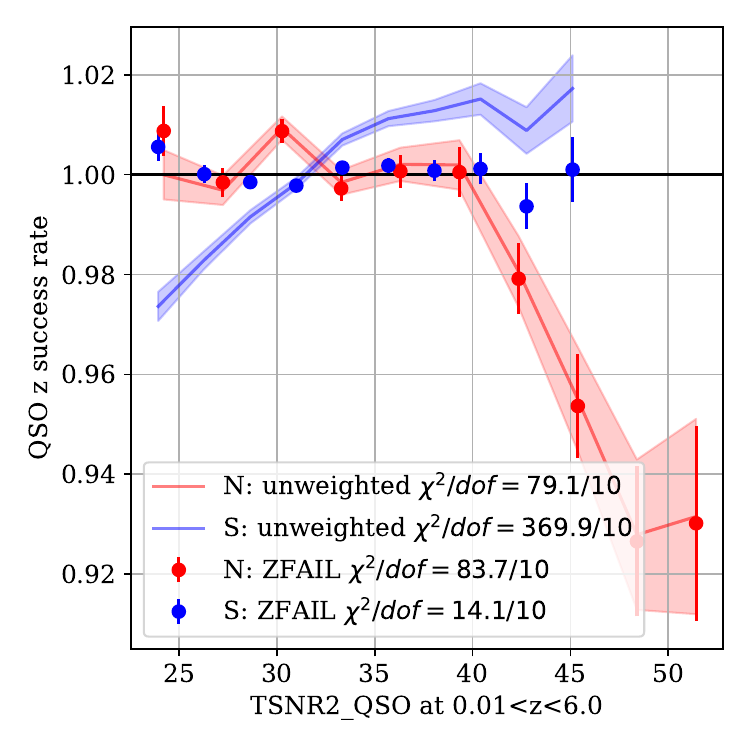}
     \includegraphics[width=0.5\linewidth]{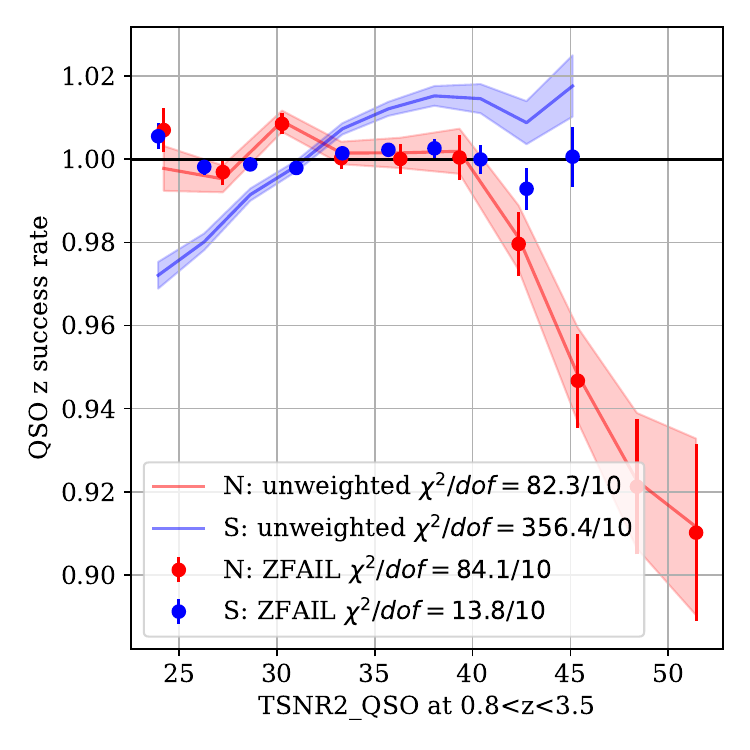}
    \caption{As in Fig.~\ref{fig:validate_overall_rate}, but using the entire redshift range of each tracer rather than the clustering redshift range. We also show the quasars using the extended redshift range ($0.8 < z < 3.5$) on the bottom.
    \label{fig:validate_overall_rate_full_z_range}}
\end{figure*}

\begin{figure*}
    \includegraphics[width=0.5\linewidth]{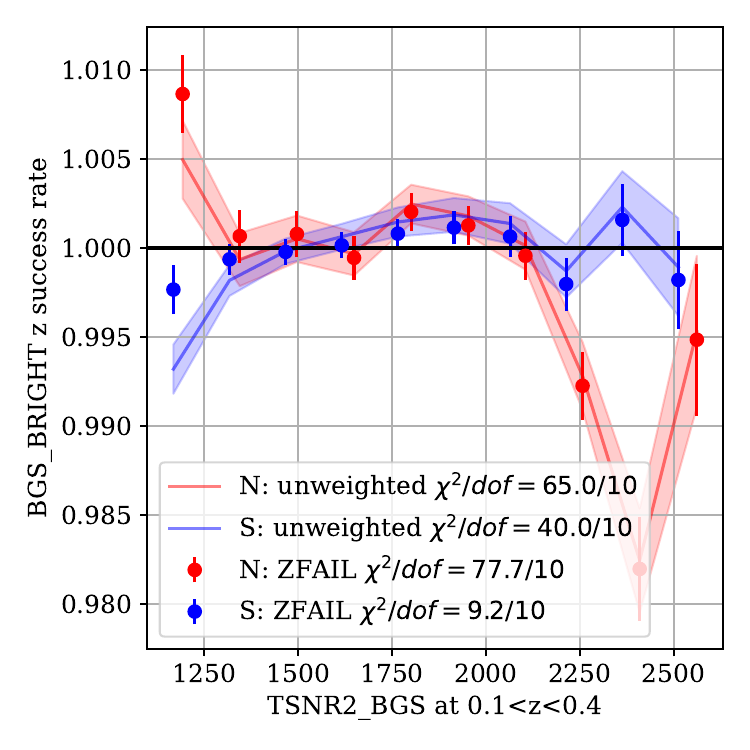}
     \includegraphics[width=0.5\linewidth]{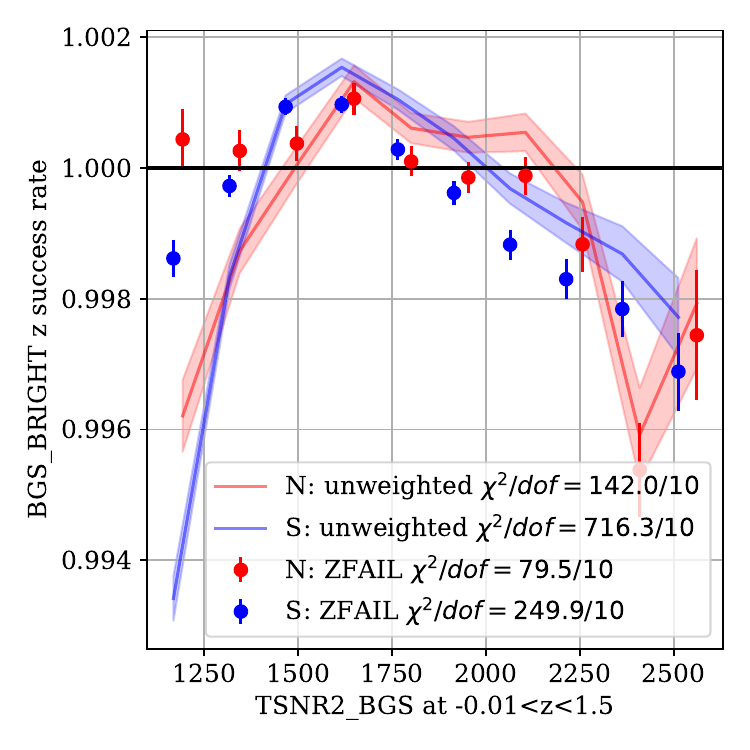}
     \includegraphics[width=0.5\linewidth]{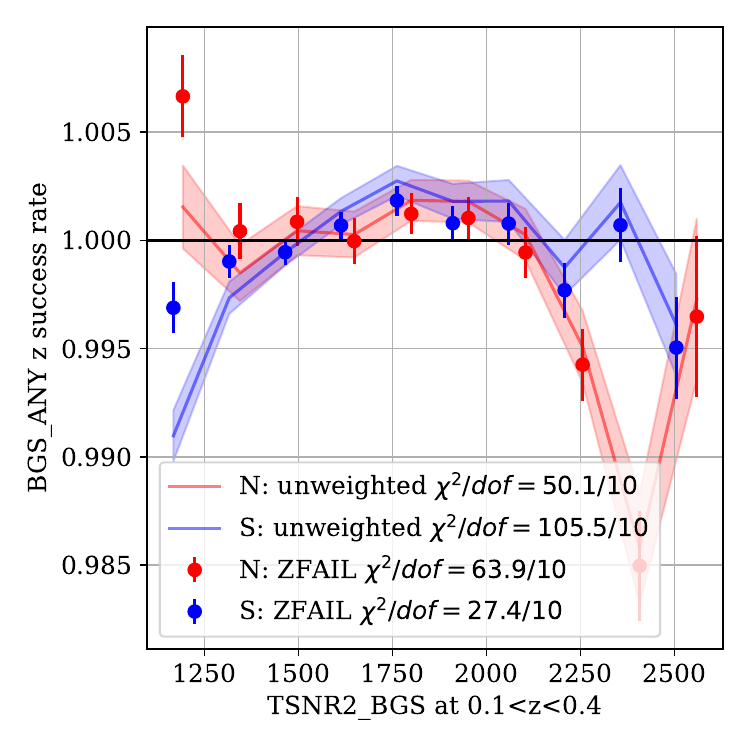}
     \includegraphics[width=0.5\linewidth]{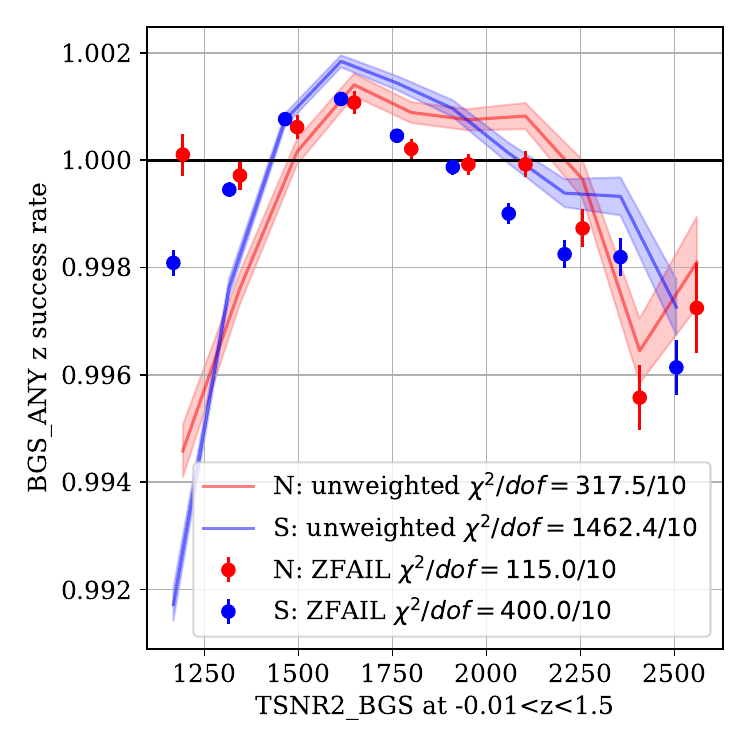}
    \caption{As in Fig.~\ref{fig:validate_overall_rate}, but for alternative definitons of the BGS sample: BGS\_BRIGHT (without the absolute magnitude cut) on top, and BGS\_ANY on bottom. We show both the clustering redshift range ($0.1 < z < 0.4$) on the left and the full redshift range on the right.
    \label{fig:validate_overall_rate_other_bgs}}
\end{figure*}

We validate the performance of the redshift failure
weights across the full redshift ranges and for alternative BGS samples in Figs.~\ref{fig:validate_overall_rate_full_z_range} and ~\ref{fig:validate_overall_rate_other_bgs}.
We find that the LRG weights perform similarly well over the full redshift range as in the clustering redshift range ($0.4 < z < 1.1$).
Likewise, the weights work well for the quasars in both the full redshift range and the extended range $0.8 < z < 3.5$. These plots show the same dip at TSNR2\_QSO $> 40$ as Fig.~\ref{fig:validate_overall_rate}, due to the 
increased galaxy fraction in quasar targets at low Galactic latitude (which aligns with the high TSNR2\_QSO regions). ELG shows a $\sim1.5\%$ drop in success rate at low TSNR2\_ELG, which is only partially corrected by the weights, leading to a significant trend in the North and a marginally significant trend in the South. As shown in \cite{KP3s4-Yu}, these trends are mitigated by also applying the focal plane weights $\eta_{\textrm{zfail}}$.

We find generally worse performance for BGS, and extending these redshift failure weights to alternative BGS
samples or redshift ranges will require a separate validation process. Considering BGS\_BRIGHT-21.5 across the entire redshift range, the trend in South is somewhat significant ($p = 7.5\times10^{-5}$, equivalent to 3.8$\sigma$ for a Gaussian). We find very large trends for BGS\_ANY and BGS\_BRIGHT (i.e.\ without the absolute magnitude cut) across the entire redshift range (Fig.~\ref{fig:validate_overall_rate_other_bgs}).
Some of these trends (e.g.\ at high TSNR2\_BGS) may be related to the change in $r_{\textrm{fiber}}$ distribution with TSNR2\_BGS (Fig.~\ref{fig:bgs_bright_fiberflux_r_distribution}). The trends with TSNR2\_BGS are much less significant when considering the clustering range $0.1 < z < 0.4$ instead of the full redshift range.
However, we find somewhat significant drops in success rate at low TSNR2\_BGS, even after applying the redshift failure weights, indicating that the redshift failure model may not adequately fit the data. In particular, the redshift success rate is a very strong function of $r_{\textrm{fiber}}$, and the simple model in Eq.~\ref{eqn:model_flux_tsnr} may not match the strong observed trends. We emphasize again that for applications of the BGS sample (``BGS\_BRIGHT-21.5'') beyond those presented here, further validation of the redshift failure weights is required.

\section{TSNR2\_QSO--failure rate trends for ``no star'' and ``high purity'' samples}
\label{sec:qso_full_nostar_high_purity}

\begin{figure*}
    \includegraphics[width=0.7\linewidth]{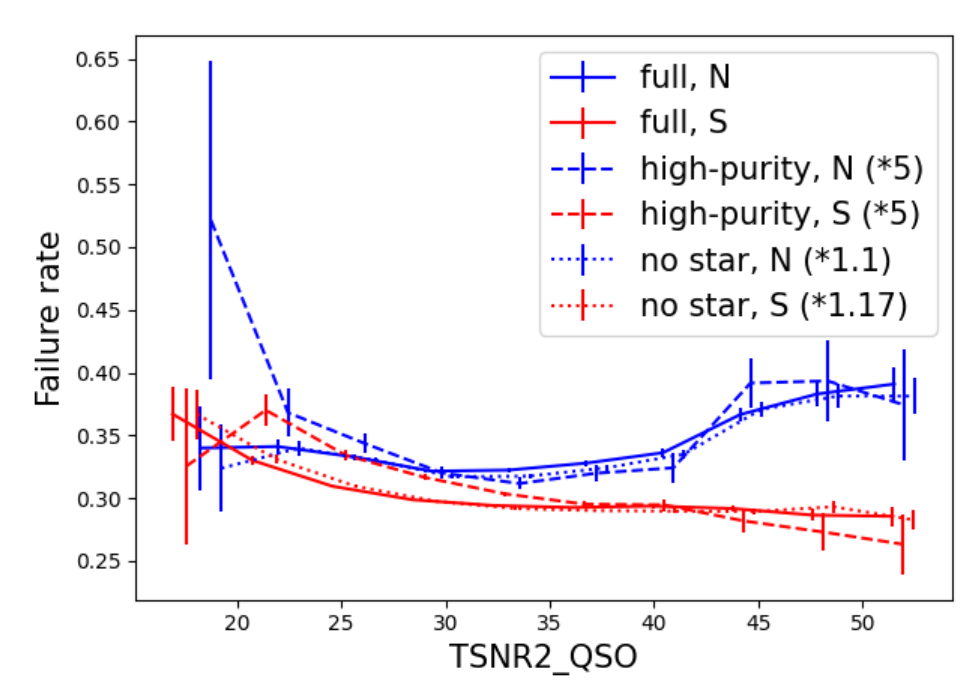}
    \caption{Failure rate as a function of TSNR2\_QSO for the full quasar sample and two alternative samples with some quasars removed: a ``no star'' sample where the star fraction is reduced from 5\% to 1\% using the photometric classification of \cite{Duncan22}; and a ``high purity'' sample where the quasar fraction is increased from 60\% to 90\% by applying the ``no star'' cut and additionally removing targets at the faint end in $r$ and W2 magnitudes. Since the failure rate (i.e.\ the fraction of quasar targets that are not spectroscopically confirmed quasars) is highest for full, the ``no star'' and ``high purity'' failure rates are scaled by the indicated factors to allow better visual comparison of the trends.
    \label{fig:qso_success_rate_subsets}}
\end{figure*}

\begin{figure*}
    \includegraphics[width=0.45\linewidth]{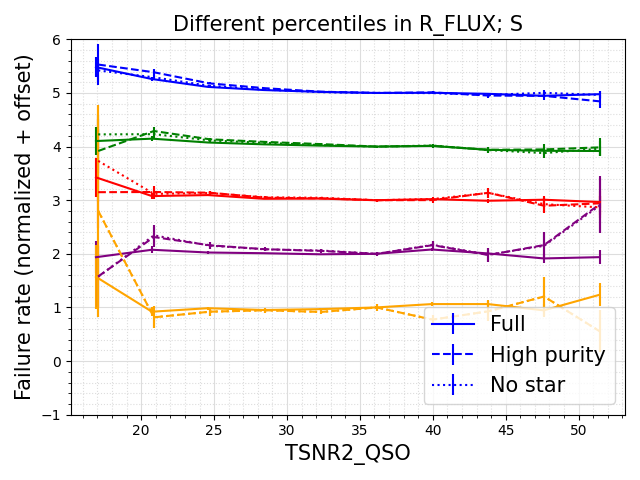}
     \includegraphics[width=0.45\linewidth]{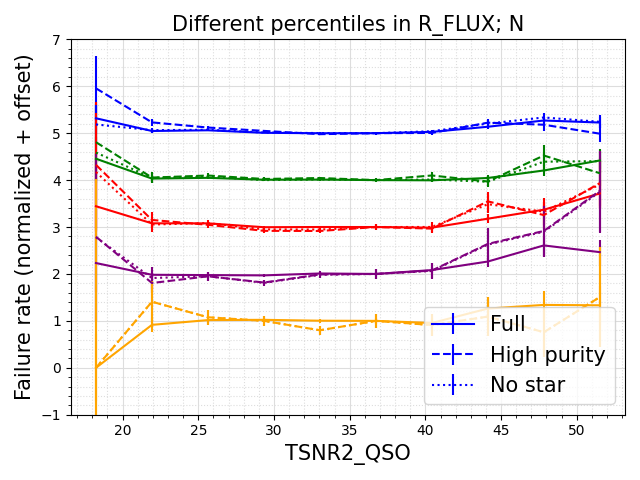}
    \caption{Failure rate as a function of TSNR2\_QSO, dividing the quasar targets into different subsets of extinction-corrected $r$ flux. The subsets are defined using the quintiles of the ``high purity'' sample
    and are the same flux cuts for all three samples. The faintest sample is at the top (blue lines) and the samples increase in brightness moving down the plot. All failure rates are normalized by the failure rate at TSNR2\_QSO = 35, and an arbitrary offset is added for clarity.
    \label{fig:qso_success_rate_vs_flux}}
\end{figure*}

We use the ``no star'' and ``high purity'' samples of \cite{Krolewski23} to assess whether the TSNR2\_QSO--failure rate trends are truly driven by fluctuations in the quasar success rate, or by fluctuations in the success rate of quasar interlopers (galaxies and stars).
We find similar behavior of the TSNR2\_QSO--failure rate trend for the ``full'' and ``no star'' sample (Fig.~\ref{fig:qso_success_rate_subsets}). The ``high purity'' sample shows a different behavior, with a more uniform rise in failure rate towards smaller TSNR2\_QSO, rather than a pronounced uptick at low TSNR2\_QSO.
However, the ``high purity'' sample is quite different from ``full'' and ``no star'' since many of the faintest quasar targets have been removed. This is important, since the failure rate strongly depends on flux\footnote{The quasar targets are selected to be point sources, so flux and fiberflux are interchangeable.} as well as TSNR2\_QSO.
We compare the quasar success rate in fixed bins of extinction-corrected $r$-band flux, defined as the quintiles of $r$ flux for the ``pure'' sample.\footnote{Since the ``pure'' sample is defined using a flux cut in $r$ magnitude, the other samples will have a considerable number of targets fainter than the faintest quintile and thus not shown in Fig.~\ref{fig:qso_success_rate_vs_flux}.}
We find that within the quintiles of $r$ flux, the success rate--TSNR2\_QSO trends are indistinguishable within the errorbars, both in the North and the South (Fig.~\ref{fig:qso_success_rate_vs_flux}).
We therefore conclude that the different behavior seen between the ``high purity'' and ``full'' quasar samples in Fig.~\ref{fig:qso_success_rate_subsets} is driven by the different $r$ flux distribution, rather than actual differences in the relative success rate.

Thus we conclude that the measured quasar redshift success rate is representative of the ``true'' redshift success rate, defined as the ratio of spectroscopically-classified quasars to true quasars within the target sample, and is robust to the presence of interlopers (stars and galaxies) within the quasar sample.

\section{Redshift success vs.\ focal plane radius for BGS, ELG, and QSO}
\label{sec:focal_plane_trends}

\begin{figure*}
    \includegraphics[width=0.8\linewidth]{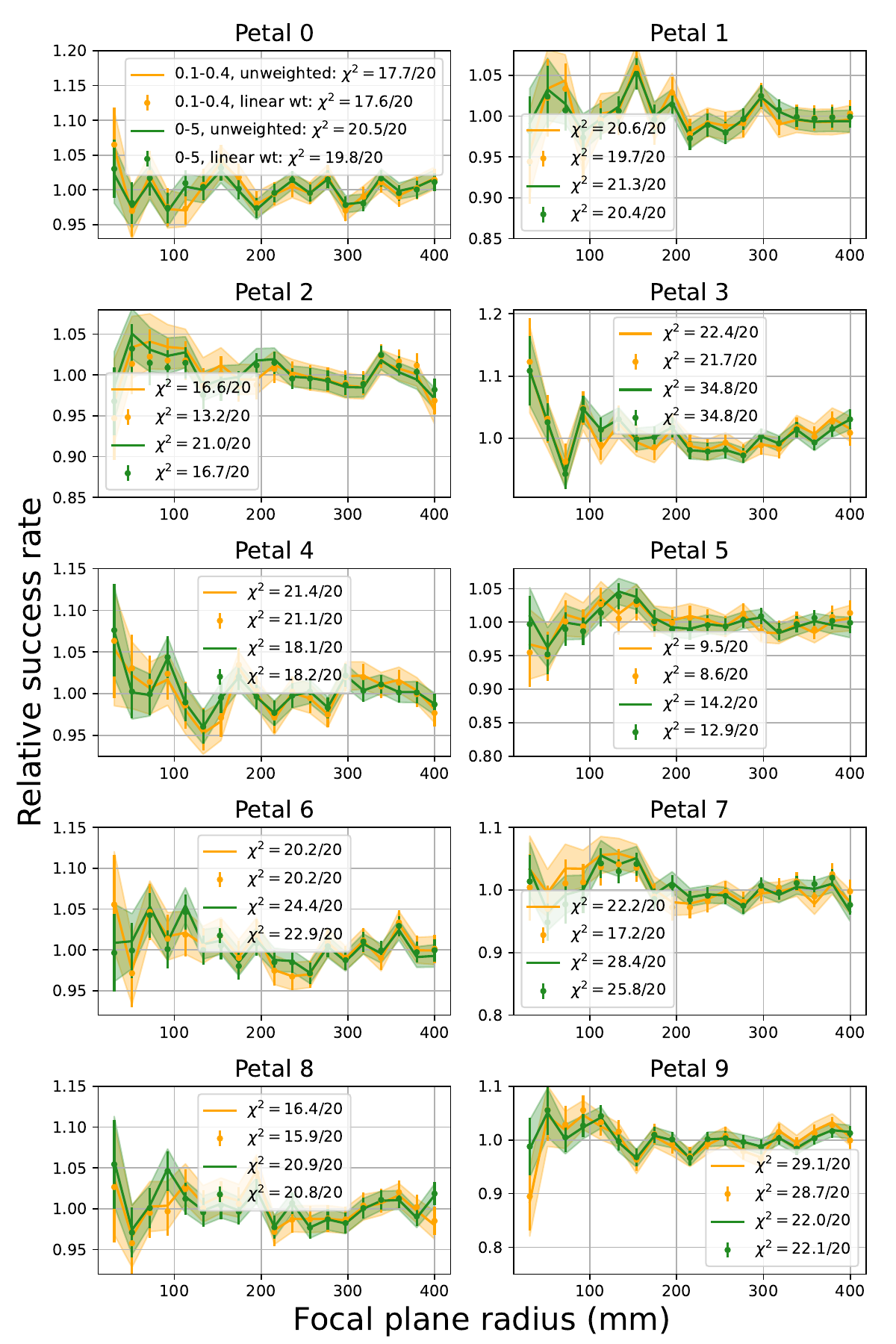}
    \caption{As in Fig.~\ref{fig:focal_plane_trends_lrg}, but for BGS\_BRIGHT-21.5.
    \label{fig:focal_plane_trends_bgs}}
\end{figure*}

\begin{figure*}
    \includegraphics[width=0.8\linewidth]{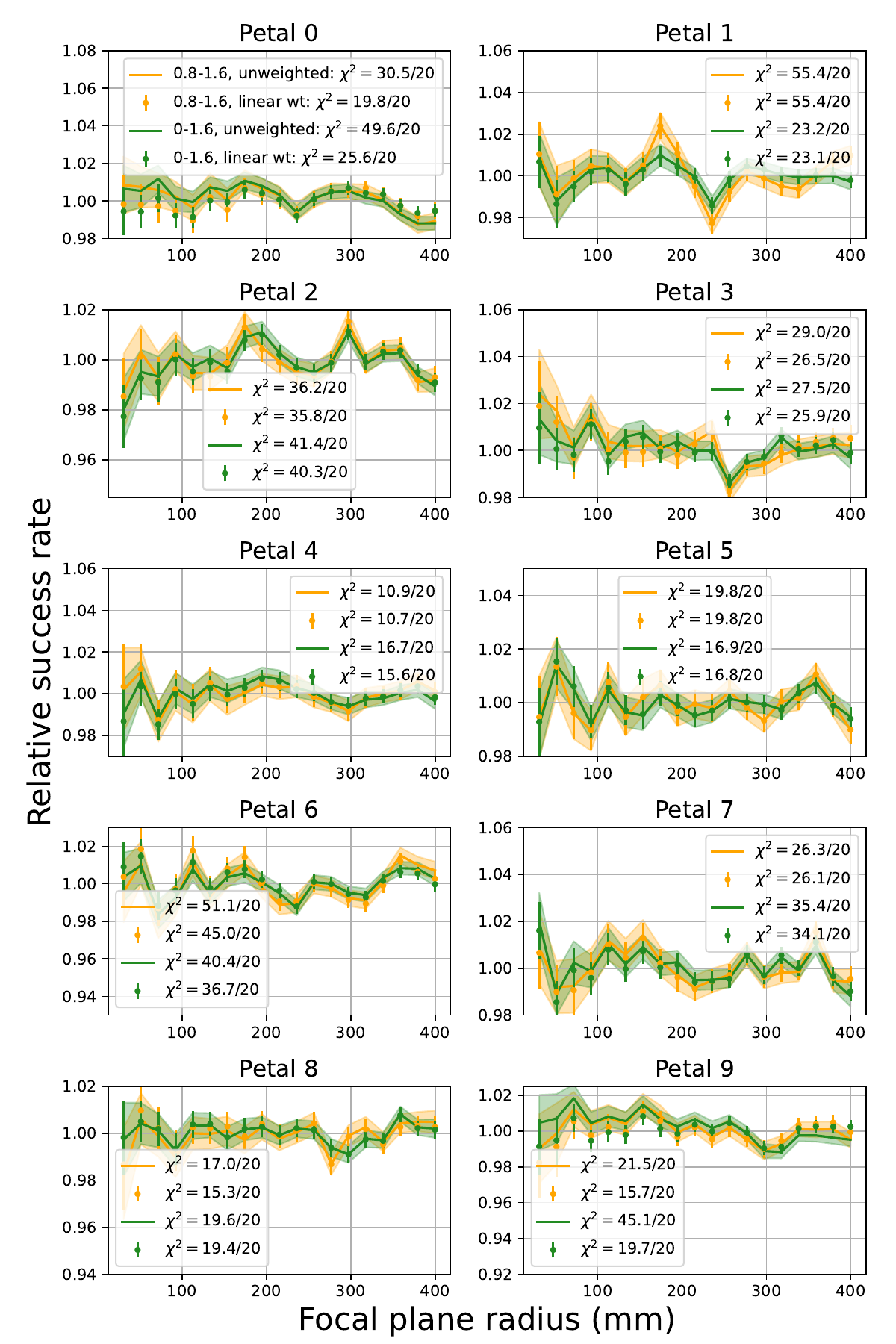}
    \caption{As in Fig.~\ref{fig:focal_plane_trends_lrg}, but for ELG.
    \label{fig:focal_plane_trends_elg}}
\end{figure*}

\begin{figure*}
    \includegraphics[width=0.8\linewidth]{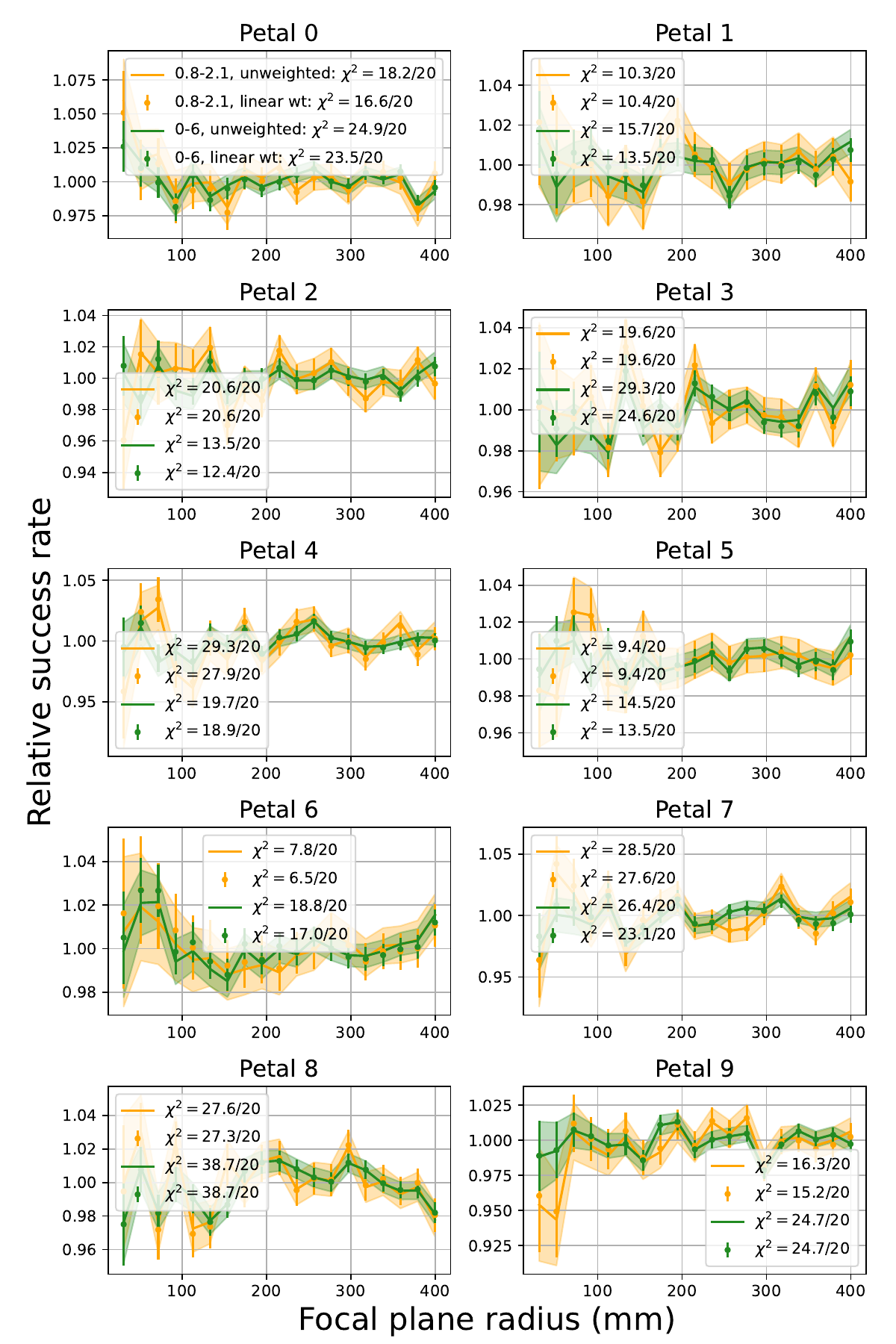}
    \caption{As in Fig.~\ref{fig:focal_plane_trends_lrg}, but for QSO.
    \label{fig:focal_plane_trends_qso}}
\end{figure*}

We show trends of redshift success vs.\ focal plane radius for BGS\_BRIGHT-21.5, ELG, and QSO in Figs.~\ref{fig:focal_plane_trends_bgs},~\ref{fig:focal_plane_trends_elg},~\ref{fig:focal_plane_trends_qso}.
We only find a $>3\sigma$ trend in petals 1 and 6 for ELG using the clustering redshift range. We find significant trends in petals 0, 2, 6 and 9 for ELG using the full redshift range. For petals 0 and 9, applying the linear focal plane radius weights improves the trends to be statistically consistent with no trend. We do not find significant trends for QSO and BGS when considering the entire redshift range.


\end{document}